\documentclass[english]{iopart}

\usepackage{graphicx}
\usepackage{url}
 \usepackage{amssymb}

\newcounter{fig}
\begin{document}

\title[Symmetries of differential equations]
{\Large Modular forms, Schwarzian conditions, and symmetries of differential equations in physics}

\vskip .3cm 


\author{Y. Abdelaziz, J.-M. Maillard$^\pounds$}
\address{$^\pounds$ LPTMC, UMR 7600 CNRS, 
Universit\'e de Paris 6, Tour 24,
 4\`eme \'etage, case 121, 
 4 Place Jussieu, 75252 Paris Cedex 05, France}

\ead{maillard@lptmc.jussieu.fr}

\begin{abstract}

We give examples of infinite order rational 
transformations that leave  linear differential 
equations covariant. These examples are
non-trivial yet simple enough illustrations
of exact representations of the renormalization group.
We first illustrate covariance properties on order-two 
linear differential operators associated with identities 
relating the same $\, _2F_1$ hypergeometric function with different
rational pullbacks. These rational transformations 
are solutions of a differentially algebraic 
equation that already emerged in a paper by Casale on the
Galoisian envelopes. We provide two new and more general results of 
the previous covariance by rational functions: a new Heun 
function example and a higher genus $\, _2F_1$ hypergeometric 
function example. We then focus on identities relating 
the same $\, _2F_1$ hypergeometric function with two different 
algebraic pullback transformations: such remarkable identities 
correspond to modular forms, the algebraic transformations
being solution of another differentially algebraic 
Schwarzian equation that also emerged in Casale's paper. Further, 
we show that the first differentially algebraic 
equation can be seen as a subcase of the last Schwarzian differential 
condition, the restriction corresponding to a factorization condition 
of some associated order-two linear differential operator.
Finally, we also explore generalizations of these results, for instance, 
to $\, _3F_2$, hypergeometric functions, and show that 
one just reduces to the previous 
$\, _2F_1$ cases through a Clausen identity. The question of the reduction 
of these Schwarzian conditions to 
modular correspondences remains an open question.
In a $ \, _2F_1$ hypergeometric framework the Schwarzian condition
encapsulates all the modular forms 
and modular equations of the 
theory of elliptic curves, but these two conditions are actually richer than elliptic 
curves or $\, _2F_1$ hypergeometric functions, as can be seen 
on the Heun and higher genus example. 
This work is a strong incentive to develop more differentially algebraic 
symmetry analysis in physics. 

\end{abstract}

\vskip .3cm


\noindent {\bf AMS Classification scheme numbers}: 34M55, 47Exx, 32Hxx, 32Nxx, 34Lxx, 34Mxx, 14Kxx, 14H52 

\vskip .3cm


\vskip .1cm
 
\section{Introduction: infinite order symmetries.}
\label{int}

In its simplest form, the concept of symmetries in physics 
corresponds to a (univariate) transformation 
$\, x \, \rightarrow \, \, R(x)$ preserving some structures. Whether
these structures are linear differential equations, or more complicated 
mathematical objects (systems of differential equations, functional 
equations, etc ...), they must be {\em invariant} or  {\em covariant} under 
the previous transformations $\, x \, \rightarrow \, \, R(x)$. Of course, 
these transformation symmetries  can be studied, per se, 
in a discrete dynamical perspective\footnote[1]{In their 
pioneering work Julia, Fatou and Ritt the theory of iteration 
of rational functions was seen as a method for investigating 
functional equations~\cite{Fatou,Ritt,Fatou2}. More generally, 
one can try to find all pairs of {\em commuting rational functions}, 
see~\cite{Eremenko}.}. Along this iteration line, or more generally, 
{\em commuting transformations} line, there is no need 
to underline  the success of the renormalization group revisited
 by Wilson~\cite{Migdal,Fisher} seen as a fundamental
 symmetry in lattice statistical mechanics or field theory.  

\vskip .2cm

The renormalization of the one-dimensional Ising model without 
a magnetic field (even if it can also be performed with 
a magnetic field~\cite{Hindawi}), which corresponds to the simple
(commuting) transformations $\, x \, \rightarrow \, x^n$ 
(where $\, x\, = \,\, \tanh(K)$), is usually seen as the heuristic 
``student'' example of {\em exact} renormalization in physics, but it is
trivial being one-dimensional. For less academical models one 
could think that no exact\footnote[2]{For instance, a Migdal-Kadanoff 
decimation can introduce, in a finite-dimensional parameter space 
of the model, rational transformations that can be seen as efficient 
approximations of the generators of the renormalization group, hoping  
that the basin of attraction of the fixed points of the 
transformation is ``large enough''.} closed form representation of the 
renormalization group exists, but can one hope to find anything better ?
For Yang-Baxter integrable models~\cite{broglie,bo-ha-ma-ze-07b} 
with a canonical genus-one 
parametrization~\cite{Automorphisms,Baxterization,BeMaVi92} (elliptic 
functions of modulus $\, k$) {\em exact} representations
of the  generators of the renormalization group 
happen to exist. Such exact symmetry transformations must have
$\, k \, = \, 0$ and $\, k \, = \, 1$ as a fixed point,  be 
compatible with the Kramers-Wannier duality 
$\, k \, \leftrightarrow  \, 1/k$, and, most importantly,
be compatible with the {\em lattice of periods} of the elliptic 
functions parametrizing the model. Thus, these exact generators
must be the {\em isogenies}~\cite{Heegner,buium} of the elliptic functions 
(of modulus $\, k$).
The simplest example of a transformation carrying these properties
is the {\em Landen transformation}~\cite{bo-ha-ma-ze-07b,Heegner} 
\begin{eqnarray} 
\label{Landen}
\hspace{-0.95in}&& \quad \quad \quad \quad \quad \quad
k \, \quad  \longrightarrow \, \quad k_L \, = \, \, 
{{2 \sqrt{k}} \over {1+k}},
\end{eqnarray} 
with the {\em critical point} of the 
square Ising model (resp. Baxter model) given by the fixed 
point of the transformation: $\, k\,= \, 1$.

This algebraic transformation corresponds 
to multiplying 
({\em or dividing} because of the 
modular group symmetry $\tau \,\leftrightarrow \, 1/\tau$)
 the ratio $\tau$ of the two periods of the elliptic curves
 $ \, \,\tau \, \, \longleftrightarrow \, \, 2\, \tau$.  
The other (isogeny) transformations\footnote[9]{See 
for instance (2.18) in~\cite{Canada}.} 
correspond to 
$\tau \,\leftrightarrow \, N \cdot \tau$,
for various integers $\, N$.

Setting out to find the precise covariance of some of the physical quantities
related to the 2-D Ising model, like the partition function 
per site, the correlation functions, the $\, n$-fold correlations 
$\, \chi^{(n)}$ associated with the full 
susceptibility~\cite{High,ze-bo-ha-ma-05b,bo-gu-ha-je-ma-ni-ze-08,higher3}, 
with respect to transformations of the Landen type (\ref{Landen}), is a 
difficult task. An easier goal would be to find a covariance, not 
on the selected\footnote[5]{They are not only Fuchsian, the 
corresponding linear differential operators are globally nilpotent 
or $\, G$-operators~\cite{bo-bo-ha-ma-we-ze-09,Andre,Andre2}.} linear 
differential operators that these quantities satisfy, 
but on the different {\em factors} of these operators. Luckily the factors 
of the operators associated with these physical quantities are linear 
differential operators whose solutions can be expressed in terms of 
{\em elliptic functions, modular forms}~\cite{bo-bo-ha-ma-we-ze-09} 
(and beyond $\, _4F_3$ hypergeometric functions associated with 
{\em Calabi-Yau ODEs}~\cite{IsingCalabi,IsingCalabi2}, etc ...).

\vskip .2cm

Let us give an illustration of the precise action of non-trivial symmetries 
like (\ref{Landen})
on some elliptic functions that actually occur in the 2-D Ising 
model~\cite{IsingCalabi,IsingCalabi2,Christol}: weight-one {\em modular forms}. 

\vskip .2cm 

Let us introduce the $\, j$-invariant\footnote[8]{The $j$-invariant~\cite{Heegner,Canada} 
(see also Klein's modular invariant) regarded as a function of a complex 
variable $\, \tau$, is a modular function of weight zero for $\, SL(2,\, \mathbb{Z})$.} 
of the elliptic curve and its 
transform by the Landen transformation
\begin{eqnarray} 
\label{jjprime}
\hspace{-0.95in}&& \quad \quad \quad
 j(k) \, = \, \, \, \, 256
\cdot {{(1-k^2+k^4)^3} \over {k^4 \cdot (1-k^2)^2}},
\quad \quad  j(k_L) \, = \, \, \, \, 
16 \cdot {\frac { (1+14\,{k}^{2}+{k}^{4})^3}{
 (1-{k}^{2})^{4} \cdot {k}^{2} }}.
\end{eqnarray}
and let us also introduce the two corresponding 
{\em Hauptmoduls}~\cite{Heegner}
\begin{eqnarray} 
\label{Haupt}
\hspace{-0.95in}&& \quad \quad \quad\quad \quad
x \, \, = \, \, \, {{1728} \over {j(k)}}, 
\quad \quad \quad \, 
y \, \, = \,\, \,  {{1728} \over {j(k_L)}}, 
\end{eqnarray}
with the two Hauptmoduls being related by the 
{\em modular equation}~\cite{Andrews,Atkin,Hermite,Hanna,Morain,Weisstein}:
\begin{eqnarray}
\label{modularcurve}
\hspace{-0.95in}&& \quad 
1953125\,{x}^{3}{y}^{3} \, \, -187500\,{x}^{2}{y}^{2} \cdot \, (x+y) \, \, 
+375\, xy \cdot \, (16\,{x}^{2}-4027\,xy+16\,{y}^{2})
\nonumber \\ 
\hspace{-0.95in}&& \quad \quad  \quad 
\quad  
 \,  -64\, \, (x+y)  \cdot \, ({x}^{2}+1487\,xy+{y}^{2}) 
\,\,  +110592\,xy
 \, \,  \,= \, \,\,  \, \, 0. 
\end{eqnarray}
The transformation 
$\, x \, \rightarrow  \, \, y(x) \, = \, y$, where $\, y$ 
is given by the modular equation (\ref{modularcurve}), 
is an {\em algebraic} transformation 
{\em which corresponds to the Landen transformation} 
(as well as the inverse Landen transformation: it is {\em reversible}
because of the $\,x \, \leftrightarrow \, y$ symmetry 
of (\ref{modularcurve})).
The emergence of a {\em modular form}~\cite{IsingCalabi,IsingCalabi2,Christol}
corresponds to the remarkable identity on the {\em same}
hypergeometric function but where the pullback $\, x$ is changed 
$\, x \, \rightarrow  \, \, y(x) \, = \, y$ according to the 
modular equation (\ref{modularcurve}) corresponding to the 
Landen transformation, or inverse Landen transformation
\begin{eqnarray}
\label{modularform2explicit}
\hspace{-0.95in}&& \quad \quad  \quad  \quad  \quad 
 _2F_1\Bigl([{{1} \over {12}}, \, {{5} \over {12}}], \, [1], \, y  \Bigr)
\, = \, \, \, \,\, 
 {\cal A}(x) \cdot \,
 _2F_1\Bigl([{{1} \over {12}}, \, {{5} \over {12}}], \, [1], \,  x  \Bigr), 
\end{eqnarray}
where $\, {\cal A}(x)$ is an algebraic function given by:
\begin{eqnarray}
\label{wherecalA}
\hspace{-0.95in}&& \quad \quad  \quad 
1024\,\,{\cal A}(x)^{12} \,  \, 
-1152\,\,{\cal A}(x)^{8}  \, \, +132\,\,{\cal A}(x)^{4}
 \, \,  +125\,x \, \,  -4 \, \,  \,  \,= \, \, \, \, \,  0. 
\end{eqnarray}
The emergence of a modular form is thus associated with 
 a selected hypergeometric function having an 
{\em exact covariance property}~\cite{Stiller,Zudilin} 
{\em with respect to an infinite order algebraic transformation}, 
corresponding here to the Landen transformation, which is 
precisely what we expect for an exact representation
of the renormalization group of the square Ising 
model~\cite{Hindawi,Heegner}.

With the example of the Ising model one sees that
the exact representation of the 
renormalization group immediately requires considering the 
isogenies of {\em elliptic curves}~\cite{Heegner}, and 
thus transformations, corresponding to the {\em modular equations}, 
$\, x \, \rightarrow \, y(x)$ which are (multivalued) 
{\em algebraic} functions. 

In a previous paper~\cite{Hindawi}, we studied simpler 
examples of identities on $\, _2F_1$ hypergeometric functions 
where the transformations $\, x \, \rightarrow \, y(x)$ were
 {\em rational functions}. In that paper we found 
that the rational functions $\, y(x)$ 
are {\em differentially algebraic}\footnote[2]{All the non-linear 
differential equations we consider in this paper (see (\ref{mad}),
see the Schwarzian equations (\ref{condition1}), (\ref{condition3F2}), ... 
 below) are differentially algebraic~\cite{Selected,IsTheFull}, i.e.
$\, y(x)$  is a solution of a polynomial equation $P(x,y,y',y'',...)$, 
as a consequence of the fact that the functions $\, A_R(x)$
or $\, W(x)$ in these equations are {\em rational functions} 
instead of general meromorphic functions (see (\ref{cas2}), 
(\ref{Casale}) below).
 }~\cite{Selected,IsTheFull}: they 
verify a (non-linear) differential equation 
\begin{eqnarray}
\label{mad}
\hspace{-0.95in}&& \quad \quad  \quad \quad 
A(y(x)) \cdot \, y'(x)^2 \,\,     = \,\,  \,  \,\,
     A(x) \cdot \, y'(x)  \, \, \,  + y''(x). 
\end{eqnarray}
where $\, A(x)$ is a rational function (which is in 
fact a log-derivative~\cite{Hindawi}).
This non-trivial condition coincides exactly with 
one of the conditions G. Casale 
obtained~\cite{Casale,Casale2,Casale3,Casale4,Casale5,Casale6,Casale7}
 in a classification of 
Malgrange's $\, {\cal D}$-envelope and $\, {\cal D}$-groupoids  on 
$\mathbb{P}_1$. Denoting $\, y'(x)$, $\, y''(x)$ and $\, y'''(x)$ the 
first, second and third derivative of $\, y(x)$ with respect to $\, x$, 
these conditions read respectively\footnote[5]{More generally see 
the concept of differential algebraic invariant of 
isogenies in~\cite{buium}.} 
\begin{eqnarray}
\label{cas2}
\hspace{-0.95in}&& \quad \quad  \quad \quad 
\mu(y) \cdot \, y'(x) \,\,  -\mu(x) \,\,  + \, {{y''(x)} \over{ y'(x)}}
 \, \,\,  = \, \, \, \, 0, 
\\
\label{Casale}
\hspace{-0.95in}&& \quad \quad  \quad \quad 
\nu(y) \cdot \, y''(x)^2 \,  \, -\nu(x) \, \, \, 
+ \, {{y'''(x)} \over{ y'(x)}}  \,\, 
 -{{3} \over {2}} \cdot \,  \Bigl({{y''(x)} \over{ y'(x)}}\Bigr)^2 
 \, \, = \, \,\,  \, 0, 
\end{eqnarray}
together with $\, \gamma(y) \cdot \,  y'(x)^n \, - \, \gamma(x)
\, \, = \, \, \, 0$ 
and $\, h(y) \, \, = \, \, \, h(x)$, 
corresponding respectively to rank two, rank three, 
together with rank one and rank nul groupo\"ids, 
where $\, \nu(x)$, $\, \mu(x)$, $\, \gamma(x)$ 
are {\em meromorphic} functions ($h(x)$ is holomorph).
Clearly Casale's condition (\ref{cas2}) is 
{\em exactly the same condition as} the one we already 
found in~\cite{Hindawi}, and this is not a coincidence !
In this paper we will refer to Casale's first condition (\ref{cas2})
as the ``rank-two condition'', and to the 
Casale's second condition (\ref{Casale}) as 
the ``rank-three condition'', or the 
``Schwarzian condition''. When our paper~\cite{Hindawi} was 
published we had no example corresponding to a 
{\em Schwarzian condition} like (\ref{Casale}).

Without going into the details of 
Malgrange's pseudo-groups~\cite{Casale2,Casale7}, Galoisian envelopes, 
$\, {\cal D}$-envelopes of a germ of foliation~\cite{Casale6}, 
and $\, {\cal D}$-groupo\"ids, let us 
just say that these concepts are built in order to generalize
the idea of differential Galois groups to {\em non-linear}~\cite{Malgrange} 
ODEs\footnote[2]{In the case of linear ODEs the $\, {\cal D}$-envelope
gives back the differential Galois group of the linear ODEs.} or 
{\em non-linear} functional equations\footnote[1]{The typical example is  
the (non-linear) functional equation $\, f(x+1) \, = \, \, y(f(x))$, which is 
such that its Malgrange pseudo-group (generalization of the 
Galois group) will be ``small enough'' if and only if, there 
exists a rational function $\, \nu(x)$, such that the 
Schwarzian condition (\ref{Casale}) is satisfied.}
(see~\cite{Paul}). In an experimental 
mathematics pedagogical approach, we will provide more 
examples of {\em rational} transformations verifying rank-two
condition (\ref{cas2}), and new pedagogical examples of {\em algebraic} 
transformations verifying  {\em Schwarzian conditions} like 
in (\ref{Casale}). We hope that these (slightly obfuscated for 
physicists) Galoisian envelope conditions will become clearer 
in a framework of {\em identities on hypergeometric functions}. In 
a {\em modular form} perspective, we will show that the infinite 
number of algebraic transformations corresponding to the 
{\em infinite number of the modular equations}, are
solutions of a {\em unique} Schwarzian condition 
(\ref{Casale}) with $\, \nu(x)$ a {\em rational function}. 
 
\vskip .1cm

The paper is organized as follows. We first recall the $\, _2F_1$ 
results in~\cite{Hindawi} which correspond to rational transformations 
and rank-two condition (\ref{cas2}) on these rational 
transformations. We then display a set of new results also
corresponding  to rational transformations with condition (\ref{cas2}).
Then focusing on a modular form hypergeometric identity, we 
show that it actually 
provides a first heuristic example of a Schwarzian condition 
(\ref{Casale}) where $\, \nu(x)$ is a rational function 
and analyze them in detail. 
We then show that the rank-two condition (\ref{cas2}) 
is a subcase of the rank-three Schwarzian condition 
(\ref{Casale}), the restriction corresponding to a 
{\em factorization condition} of some associated order-two 
linear differential operator.
We then explore generalizations of the hypergeometric identity 
to $\, _3F_2$, $\, _2F_2$ and $\, _4F_3$ hypergeometric functions, 
and show that the $\, _3F_2$ attempt, in fact, just reduces to
the previous $\, _2F_1$ cases through a Clausen identity.

\vskip .1cm

\vskip .2cm
 
\section{Recalls: rational transformation and $\, _2F_1$ hypergeometric functions } 
\label{recalls}

We recall a few examples and results from~\cite{Hindawi} on the 
hypergeometric examples displayed in~\cite{Vidunas}.
The hypergeometric function
\begin{eqnarray}
\label{vid}
\hspace{-0.95in}&& \quad \quad  \quad 
Y(x) \,\, = \, \,\,\,  x^{1/4} \cdot  \,
 _2F_1\Bigl( [{{1} \over {2}}, {{1} \over {4}}],
 [{{5} \over {4}}];\,  x\Bigr) 
\,\, = \, \,  \, \,
{{1} \over {4}} \cdot \, 
 \int_0^{x} \, t^{-3/4} \cdot \,  (1-t)^{-1/2}  \cdot \,  dt 
 \nonumber \\
\hspace{-0.95in}&& \quad \quad \quad  \quad \quad \quad
  \,  = \, \, \,
 x^{1/4} \cdot  \, (1-x)^{-1/2} \cdot \, \, \, 
_2F_1\Bigl( [{{1} \over {2}}, {{1} \over {4}}], [{{5} \over {4}}];
 \, {{-4\, x} \over {(1-x)^2}} \Bigr), 
\end{eqnarray}
is the integral of an algebraic function.  
It has a simple covariance property
 with respect to the {\em infinite order
rational} transformation 
$\, x \, \rightarrow \, -4\, x/(1-x)^2$:
\begin{eqnarray}
\label{F}
Y\Bigl({{-4\, x} \over {(1-x)^2}} \Bigr)
\,\, \,  = \, \, \,  \,\,\, (-4)^{1/4} \cdot Y(x). 
 \end{eqnarray}
This  hypergeometric  function can be seen 
as an 'ideal' example of physical functions,
 covariant by an exact (rational)
transformation. Three other hypergeometric functions
with similar  covariant properties were
analyzed in~\cite{Hindawi}:
\begin{eqnarray}
\label{YM3first}
\hspace{-0.95in}&&    
Y(x) \, \, = \, \, \, 
x^{1/3} \cdot \, 
_2F_1\Bigl([{{1} \over {3}}, \, {{2} \over {3}}], \, [{{4} \over {3}}], \, x\Bigr)
\, \, = \, \, \,  
{{1} \over {3}} \cdot \, 
 \int_0^{x} \, t^{-2/3} \cdot \,  (1-t)^{-2/3}  \cdot \,  dt 
 \\
\hspace{-0.95in}&& 
  \,  = \, \, \,  (-8)^{-1/3} \cdot  \,
 R(x)^{1/3} \cdot  \, 
_2F_1\Bigl( [{{1} \over {3}}, \, {{2} \over {3}}], \, [{{4} \over {3}}],
 \,  R(x)\Bigr), 
\quad \, \hbox{with:}  \quad \quad 
R(x) \, = \, \,  {{x \cdot \, (x \, -2)^3} \over {(1 \, -2 \, x)^3}}, 
 \nonumber 
\end{eqnarray}
as well as
\begin{eqnarray}
\label{YM6first}
\hspace{-0.95in}&& 
Y(x) \, \, = \, \, \, \,
x^{1/6} \cdot \, 
_2F_1\Bigl([{{1} \over {2}}, \, {{1} \over {6}}], \, [{{7} \over {6}}], \, x\Bigr) 
\, \, = \, \, \, \, 
{{1} \over {6}} \cdot \, 
 \int_0^{x} \, t^{-5/6} \cdot \,  (1-t)^{-1/2}  \cdot \,  dt 
 \\
\hspace{-0.95in}&& \quad 
  \,  = \, \, \,  (-27)^{-1/6} \cdot  \,
 R(x)^{1/6} \cdot  \, 
_2F_1\Bigl( [{{1} \over {2}}, \, {{1} \over {6}}], \, [{{7} \over {6}}],
 \,  R(x)\Bigr), 
\quad \hbox{with:}  \quad \quad 
R(x) \, = \, \,  {{ -27 \, x} \over { (1\, -4 \, x)^3}},
 \nonumber 
\end{eqnarray}
which can be seen as a particular subcase ($\alpha \, = \, \, 1/2$) 
of the identity on hypergeometric functions:
\begin{eqnarray}
\hspace{-0.95in}&& \quad \quad \quad \quad 
_2F_1\Bigl([\alpha, \, {{1-\alpha} \over {3}}], \, [{{4\, \alpha \, +5} \over {6}}], \, x\Bigr)
\nonumber \\
\hspace{-0.95in}&& \quad \quad  \quad \quad \quad \quad \quad 
\, \, = \, \, \, (1\, -4\, x)^{-\alpha} \cdot \, 
_2F_1\Bigl([{{\alpha} \over {3}}, \, {{\alpha+1} \over {3}}], \, [{{4\, \alpha \, +5} \over {6}}], \, 
{{-\, 27 \, x} \over {(1\, -4\, x)^3}}\Bigr),
\end{eqnarray}
and, finally, the simple function $\, Y(x) \, = \, \, tanh^{-1}(x^{1/2})$ 
that one represents as a hypergeometric function:
\begin{eqnarray}
\label{zero}
\hspace{-0.95in}&& \quad 
Y(x)\,\, = \, \,\,\, \,  x^{1/2} \, \cdot \, 
_2F_1\Bigl([1,\, {{1} \over {2}}],[{{3} \over {2}}], \, x\Bigr)
\,\, = \, \,\,\,
{{1} \over {2}} \cdot \, 
 \int_0^{x} \, t^{-1/2} \cdot \,  (1-t)^{-1}  \cdot \,  dt 
 \\
\hspace{-0.95in}&& \quad \quad 
  \,  = \, \, \,  (4)^{-1/2} \cdot  \,
 R(x)^{1/2} \cdot  \, 
_2F_1\Bigl( [1,\, {{1} \over {2}}],[{{3} \over {2}}], \,
 \,  R(x)\Bigr), 
\quad \hbox{with:}  \quad \quad 
R(x)  \,\,  = \, \,\,\,  \, 
{{4 \, x } \over {(1\, +x)^2}}.
 \nonumber 
\end{eqnarray}

Though not mentioned in~\cite{Hindawi}, two other hypergeometric 
functions, also covariant under a rational transformation, could 
have been deduced from the previous hypergeometric examples
 using Goursat and Darboux 
identities (see \ref{more2F1GoursatDarboux}, and 
especially (\ref{newidR1R2R3})):
\begin{eqnarray}
\label{more1}
\hspace{-0.95in}&& \quad 
Y(x) \, = \, \, \, x^{1/4} \cdot \, (1\, -x)^{1/4} \cdot \, 
 _2F_1\Bigl([{{1} \over {2}}, \, 1], \, [ {{5} \over {4}}], \, x\Bigr) 
\,\, = \, \,\,\,
{{1} \over {4}} \cdot \, 
 \int_0^{x} \, t^{-3/4} \cdot \,  (1-t)^{-3/4}  \cdot \,  dt 
\nonumber \\
\hspace{-0.95in}&& \quad \quad  
\, \, = \, \, \,  (-4)^{-1/4} \cdot \, 
R(x)^{1/4} \cdot \, (1\, -R(x))^{1/4} \cdot \, 
_2F_1\Bigl([{{1} \over {4}}, \, {{1} \over {2}}], \, [ {{5} \over {4}}], \, 
R(x) \Bigr)
\\
\label{more1pull}
\hspace{-0.95in}&& \quad \quad 
 \quad \hbox{where:} \quad  \quad  \quad  \quad  \quad  
R(x) \, \, = \, \, \, 
{{ -4 \cdot \, x \cdot \, (1\, -x)} \over { (1 \, -2 \, x)^2}}, 
\end{eqnarray}
and 
\begin{eqnarray}
\hspace{-0.96in}&& 
\label{QQQ8first}
Y(x) \, = \, \, x^{1/6} \cdot \, 
_2F_1\Bigl([{{1} \over {6}}, \, {{2} \over {3}}], \, [{{7} \over {6}}], \, x\Bigr) 
\,\, = \, \,\,\,
{{1} \over {6}} \cdot \, 
 \int_0^{x} \, t^{-5/6} \cdot \,  (1-t)^{-2/3}  \cdot \,  dt 
 \\
\hspace{-0.96in}&&     
\, = \, \,   (64)^{-1/6}  \cdot  R(x)^{1/6} \cdot \, 
_2F_1\Bigl([{{1} \over {6}}, \, {{2} \over {3}}], \, [{{7} \over {6}}], \, R(x)\Bigr)
 \quad \hbox{with:} \,   \quad R(x) \,  = \, \,  
{{64 \, x } \over{ (1\, +18 \, x \, -27 \, x^2)^2}}.
\nonumber
\end{eqnarray}

\vskip .1cm 

These six  hypergeometric functions are incomplete integrals that 
are canonically associated with an algebraic curve 
$\,\, u^N \, - \, P(t) \, = \, \, 0\,\,$ of {\em genus one} 
for (\ref{vid}), (\ref{YM3first}), (\ref{YM6first}), (\ref{more1})
(\ref{QQQ8first}), and genus zero for  (\ref{zero})
\begin{eqnarray}
\label{form}
\hspace{-0.95in}&&  
Y(x)\,\, = \, \,\, {{1} \over {N}} \cdot \,  \int_0^{x} \, {{dt} \over { u(t) }} 
 \,\, = \, \,\,
{{1} \over {N}} \cdot \,  \int_0^{x} \, {{dt} \over {\sqrt [N]{P(t)} }}, 
\quad \, \,  \hbox{or:} \quad \quad  
N \cdot \,  Y'(x) \, = \, \, {{1} \over { u(x)}},
\end{eqnarray}
and are solutions of a 
second order linear differential operator:
\begin{eqnarray}
\label{Omega}
\hspace{-0.95in}&& \quad \quad \quad \,\,
\Omega 
\,\,\, = \,\,\, \,\,  \omega_1 \cdot  D_x, 
\qquad  \,\,\quad \hbox{with:} \quad \quad  \quad \quad \,
   \omega_1  \, = \, \,\, \,\,
 D_{x} \,\, +  A_R(x), 
 \end{eqnarray}
where $\, D_x$ denotes $\, d/dx$, 
where a rational function $\, A_R(x)$
is\footnote[1]{The fact that $\, A_R(x)$ is 
the log-derivative of the $\, N$-th root of a rational function, here 
a polynomial, is a consequence of the fact that $\, \Omega$ 
is a globally nilpotent linear differential operator~\cite{bo-bo-ha-ma-we-ze-09}.}  
the logarithmic derivative of a simple algebraic\footnote[2]{Note that 
$\, u(x)$ being an algebraic function, 
these examples are such that $\, N \cdot Y'(x) \, = \, 1/u(x)$, 
$\, Y'(x)$ is holonomic but also its reciprocal $\, 1/Y'(x)$.} function 
$\, u(x) \, = \, \, \sqrt [N]{P(x)}$.
The expressions of the rational functions $\, A_R(x)$ read 
respectively for the four
 hypergeometric examples (\ref{vid}), (\ref{YM3first}), (\ref{YM6first}) 
and (\ref{zero})
\begin{eqnarray}
\label{respecAR}
\hspace{-0.95in}&& \quad \,\,
 {{1} \over {4}} \, {{3 \, -5\,x} \over {x \cdot \, (1\,-x)}}, \quad \quad
{{2} \over {3}} \cdot {\frac {1-2\,x}{ x \cdot \,(1\, -x) }}, \quad
\quad {{1} \over {6}} \cdot {\frac {5-8\,x}{ x \cdot \,(1\,-x) }}, \quad \quad
 \, {{1} \over {2}} \cdot {\frac {1 \, -3\,x}{ x \cdot \, (1\,-x) }}.
\end{eqnarray}
and for the two new  examples (\ref{more1}) and (\ref{QQQ8first}):
\begin{eqnarray}
\label{respecARmore}
\hspace{-0.95in}&& \quad \quad \quad \quad \quad \quad  \quad \quad 
{{3} \over { 4}} \cdot \, {{1 \, -2 \, x} \over {x \cdot \, (1 \, -x) }}, 
 \,\, \quad \quad \quad \quad 
{{1} \over { 6}} \cdot \, {{5 \, -9 \, x} \over {x \cdot \, (1 \, -x) }}.
\end{eqnarray}

\vskip .1cm 

In the interesting cases emerging 
in physics~\cite{bo-ha-ma-ze-07b,IsingCalabi,IsingCalabi2,Christol}, 
the operator  $\, \Omega$ happens to be globally 
nilpotent~\cite{bo-bo-ha-ma-we-ze-09}, in which case
$\, A_R(x)$ is the log-derivative 
of the $\, N$-th root of a rational function. At first 
we do not require $\, \Omega$ to be globally 
nilpotent\footnote[9]{Imposing the 
global nilpotence generates additional relations 
(see section (\ref{assuminglob}) below).}, then we will see 
what this assumption entails.

\vskip .2cm  

Let us consider a rational transformation 
$\, x \, \rightarrow \, R(x)$ and the order-one operator 
$\, \, \omega_1 \, = \, \, D_x \, \, + \, A_R(x)$. 
The change of variable $\, x \, \rightarrow \, R(x) \, $
on the order-one operator  $\, \omega_1$  reads:
\begin{eqnarray}
\label{RHSL1}
\hspace{-0.95in}&& \quad \,  \,  \quad  \quad 
D_x \, + \, \, A_R(x) \quad \longrightarrow \quad    \, \, \, \,
 1/R'(x) \cdot  D_x \, + \, \, A_R(R(x)) 
\nonumber \\
\hspace{-0.95in}&& \quad \, \quad   \,  \quad  \quad  \quad  \quad 
 \, \, = \, \, \, \,
\gamma(x) \cdot {\cal L}_1 
\, \, = \, \, \, \, 
\gamma(x) \cdot \Bigl( D_x \, + \, \, A_R(R(x)) \cdot \, R'(x) \Bigr),  
\end{eqnarray}
with $\, \gamma(x) \, = \, \,  1/R'(x)$.
Now  imposing the order-one operator $\, {\cal L}_1$ of the RHS expression (\ref{RHSL1}) 
to be equal to the conjugation by $\, \gamma(x)$ of 
$\, \, \omega_1 \, = \, \, D_x \, \, + \, A_R(x)$, namely 
\begin{eqnarray}
\hspace{-0.95in}&& \quad \quad  \quad 
\gamma(x) \cdot 
\Bigl(D_x \, + \, \, A_R(x)\Bigr) \cdot {{1} \over {\gamma(x)}}
\, \, \, = \, \, \,\, D_x \, \, \, - \, {{ d\ln(\gamma(x))} \over {dx}} \,\,  + \, A_R(x),
 \nonumber 
\end{eqnarray}
one deduces a rank-two functional equation~\cite{Hindawi}
 on $\, A_R(x)$ and $\, R(x)$:
\begin{eqnarray}
\label{mad}
\hspace{-0.95in}&& \quad \quad  \quad \quad  \quad \quad 
A_R(R(x))  \cdot \, R'(x)^2 \,\,     = \,\,  \,  \,\,
   A_R(x) \cdot \, R'(x) \, \, \,  + \, R''(x) . 
\end{eqnarray}
This condition is {\em exactly} the first rank-two condition 
of Casale given by (\ref{cas2}). Using the chain rule formula of 
derivatives for the composition of functions, one can show, 
for a given rational function $\, A_R(x)$, that  the composition 
$\, R_1(R_2(x))$ {\em verifies condition} (\ref{mad}) 
if two rational functions $\, R_1(x)$ 
and  $\, R_2(x)$ verify condition (\ref{mad}). In 
particular if $\, R(x)$ verifies condition (\ref{mad}), 
all the iterates of $\, R(x)$ also verify that 
condition\footnote[5]{This is in agreement with the fact 
that (\ref{mad})  is the condition for $\, \Omega\, =\,\,$
$ (D_x\, + \, A_R(x)) \cdot D_x$ to be 
covariant by $x \, \rightarrow \, R(x)$: this condition is
obviously preserved by the composition of $\, R(x)$'s
 (for $\, A(x)$ fixed).}: $\, R(x) \, $
$\longrightarrow \,\, \, R(R(x))$,
 $\, R(R(R(x))), \,\, \cdots$ 

\vskip .2cm

 Keeping in mind the well-known example of the
parametrization of the standard map
 $\, x \, \rightarrow \,4\, x \cdot (1-x)$ with 
$\, x \, = \, \sin^2(\theta)$, yielding
 $\, \theta \, \rightarrow \, 2 \, \theta$,
let us seek a ({\em transcendental}) 
parametrization $\, \,x \, = \, P(u)\,$
such that\footnote[2]{This is the idea of Siegel's 
linearization~\cite{Siegel,Siegel2,Almost} 
(or Koenig's linearization theorem see~\cite{Milnor}).} 
\begin{eqnarray}
\hspace{-0.95in}&& \quad \quad  \quad \quad 
R_{a_1}\Bigl(P(u) \Bigr) \, = \, \, P(a_1\, u) 
\qquad \hbox{or:}   \quad \quad \quad \, \,\, \,
R_{a_1}\, = \, \,  P \circ  H_{a_1} \circ \, Q,
\end{eqnarray}
where $\, H_{a_1}$ denotes the scaling
 transformation $\, \,x \, \rightarrow \, \, a_1 \cdot x\,$
and $\, Q \, = \, \, P^{-1}$ denotes the composition inverse  
of $\, P$. One can also 
verify an essential property that we expect to be true 
for a representation of the renormalization group,
 namely that two  $\, R_{a_1}(x)$ for different 
values of $\, a_1$  commute, 
the result corresponding to the product of these two  $\, a_1$:
\begin{eqnarray}
\label{commute}
\hspace{-0.95in}&& \quad \quad  \quad \quad  \quad \quad 
R_{a_1}\Bigl(R_{b_1} (x) \Bigr) \,\,  \, = \, \, \, \, 
R_{b_1}\Bigl(R_{a_1} (x) \Bigr) 
\,\,  \, = \, \, \, \, R_{a_1 \cdot b_1} (x).
\end{eqnarray}
The neutral element of this abelian group
corresponds to $\, a_1 \, = \, 1$,
giving the identity transformation 
$\, R_{1}(x) \, = \, \, x$.
Performing the composition inverse  of $\, R_{a_1}(x)$
 amounts to changing 
$\, a_1$ into its inverse $\, 1/a_1$. 
The structure of the (one-parameter)  group and 
the extension of the composition of $\, n$ times a rational 
function $\, R(x)$ (namely $\, R(R( \cdots R(x) \cdots ))$)
 to $\, n$ {\em any complex number},
is a straight consequence of this relation.  For example, 
in the case of the $\, _2F_1$ hypergeometric 
function (\ref{YM3first}), the one-parameter series 
expansion of $\, R_{a_1}(x)$ reads:
\begin{eqnarray}
\label{mad2oneseriesfirst}
\hspace{-0.95in}&&   \quad \quad 
R(a, \, x) \,\,     = \,\,  \,  \,\,
a \cdot \, x \, \, \, + \, a \cdot \, (a-1) \cdot \, S_a(x) 
\quad  \quad \quad  \quad \quad \hbox{where:} 
\\
\hspace{-0.95in}&& \quad \quad 
S_a(x)   \,\,     = \,\,  \,  \,\,
 - {{1} \over {2}}  \cdot \, {x}^{2} \, \, \, 
+ {{1} \over {28}} \cdot  \, (5\,a-9) \cdot \, {x}^{3}
\,\, -{\frac { 
\, (3\,{a}^{2}-12\,a+13) }{56}} \, \cdot {x}^{4}
 \, \,  \,  \, + \, \,\,  \cdots 
 \nonumber
\end{eqnarray} 
This one-parameter series (\ref{mad2oneseriesfirst})
is a  family of commuting one-parameter series solution of 
the rank-two condition (\ref{mad}), and these solution series
have {\em movable singularities} (more details 
in \ref{more2F1examples}). Defining some 
``infinitesimal composition'' 
($Q \, = \, P^{-1}$, $\epsilon \, \simeq \, 0$)
\begin{eqnarray}
\label{R1pluseps}
\hspace{-0.95in}&& \quad  \quad \quad  \quad \quad  \,
R_{1\, + \, \epsilon}(x)\, \, = \,  \, \,  \, \, 
 P \circ \,\, H_{1\, + \, \epsilon} \circ \, Q(x)
\,\,  = \, \, \,\, \,   x \,\, \, + \, \epsilon \cdot F(x)
 \,\,\, \,\,  + \,\,  \cdots
\end{eqnarray}
 we see, from (\ref{commute}), that $\, R_{a_1}(R_{1\, + \, \epsilon}(x)) \,= \, \, 
R_{1\, + \, \epsilon}(R_{a_1} (x))$. Using (\ref{R1pluseps}) and Taylor expansion 
one gets the following relations
between $\, R_{a_1}(x)$ and the function\footnote[1]{Generically, 
$ \, F(x)$  is a transcendental function, 
not a rational nor an algebraic function. } $\, F(x)$:
\begin{eqnarray}
\label{commuteinf}
\hspace{-0.95in}&&  
R_{a_1}\Bigl(R_{1\, + \, \epsilon}(x) \Bigr) \,   = \, \,
R_{a_1}\Bigl (x \, + \, \epsilon \cdot F(x)  + \,  \cdots \Bigr)
 \, \,  = \, \,  \, R_{a_1}(x) 
\, \,  + \, {{dR_{a_1}(x) } \over {dx}} \cdot  \epsilon \cdot  F(x)
 \, \,  + \,\,  \cdots
\nonumber \\ 
\hspace{-0.95in}&&  \quad  \quad  \quad 
= \,\,  R_{1\, + \, \epsilon}\Bigl(R_{a_1} (x) \Bigr) \,\, \,   = \, \, \,
R_{a_1} (x) \, \, + \, \, \epsilon \cdot  F(R_{a_1} (x)) \, \, + \, \, \,  \cdots
\end{eqnarray}
which gives at the first order in $\, \epsilon$:
\begin{eqnarray}
\label{Ra1}
\hspace{-0.95in}&&  \quad \quad \quad  \quad \quad 
 {{dR_{a_1}(x) } \over {dx}}
 \cdot \, F(x) \,\, = \, \,\,\, F(R_{a_1}(x)). 
\end{eqnarray}
For  $\, R(x)$ and for the $\, n$-th iterates of the rational function 
$\, R(x)$ (which are in the one-parameter family $\, R_{a_1}(x)$) 
relation (\ref{Ra1}) reduces to:
\begin{eqnarray}
\hspace{-0.95in}&& \quad  \, \,\, 
  R'(x)
 \cdot \, F(x) \,\, = \, \,\, F(R(x)),   
\quad \quad \quad  {{dR^{(n)}(x) } \over {dx}}
 \cdot \, F(x) \,\, = \, \,\, F(R^{(n)}(x)), 
 \\
\hspace{-0.95in}&& \quad  \quad \quad  \quad  \, \,\, 
\hbox{where:} \qquad \quad  \quad  \quad \quad \, \,\, 
 R^{(n)}(x) \, = \, \, \, R(R(\cdots R(x)) \cdots ). 
\nonumber
\end{eqnarray}
From (\ref{R1pluseps}) one gets 
$\, P(Q(x) \, + \, \epsilon \cdot \, Q(x)) \, = 
\,  x \, + \,  P'(Q(x))  \cdot \, \epsilon \cdot \, Q(x)  \, + \, \cdots \, 
 = \, x \, + \, \epsilon \cdot \, F(x) \, + \, \cdots \, $
and also $\, P \circ  H_{1\, + \, \epsilon}(x) \, = \, P(x \, + \epsilon  \cdot \, x)$
$  \, = \, P(x) \, + \, P'(x) \cdot \, \epsilon \cdot \,  x \, + \, \cdots \,  $
$\, = \, P(x) \, + \, \epsilon \cdot \, F(P(x)) \, + \,\cdots \, \,  $
yielding respectively 
\begin{eqnarray}
\label{covP}
\hspace{-0.95in}&& \quad \quad 
Q(x) \cdot \, P'(Q(x)) \, = \, \,   F(x) 
\quad \quad \hbox{and thus:} \quad \quad \quad 
x \cdot P'(x) \,\, = \, \,   \, \,\, F(P(x)). 
\end{eqnarray}
This last relation yields $\, Q(x) \cdot P'(Q(x)) \,\, = \, \,   \, \,\, F(x)$, 
which can also be written using $\, P \circ \, Q(x) \, = \, x$ 
(and thus $\, P'(Q(x)) \cdot \, Q'(x) \, = \, 1$): 
\begin{eqnarray}
\label{QF}
\hspace{-0.95in}&& \quad \quad \quad  \quad \quad  \quad \quad  \quad \quad 
 Q'(x) \cdot F(x)  \,\, = \, \,   \, \,\, Q(x).
\end{eqnarray}

\vskip .1cm

Inserting (\ref{R1pluseps}) in the rank-two condition (\ref{mad})
one immediately finds (at the first order in $\, \epsilon$) that 
$ \, F(x)$ is a {\em holonomic function},
 solution of a second order linear differential operator 
$\,\Omega^{*}$ which can be seen to be the 
{\em adjoint} of the  second order operator
$\, \Omega$ defined by (\ref{Omega}):
\begin{eqnarray}
\label{Fholo}
\hspace{-0.95in}&& \quad \, \quad  \quad \quad 
\Omega^{*} \, \, \, = \, \, \, \, \, 
D_x^2 \,\,  - \, A_R(x) \cdot \, D_x \,\,  - \,  A'_R(x)
\, \, \, = \, \, \, \, \, 
D_x \cdot \Bigl(D_x \, - \,  A_R(x)\Bigr).
\end{eqnarray}
  
\subsection{New results: $\, Q(x)$ and $\, P(x)$ as differentially algebraic functions \\} 
\label{resnewPQ}

\vskip .1cm

The two functions $\, Q(x)$ and its composition inverse 
$\, P(x) = \, \, Q^{-1}(x)$ are 
{\em differentially algebraic functions}~\cite{Selected,IsTheFull} 
as can be seen in~\cite{Hindawi}. The function $\, Q(x)$ 
is solution of the differentially algebraic equation:
\begin{eqnarray}
\label{logderiv1first}
\hspace{-0.97in}&& \quad 
A_R(x) \cdot \, G(x) \cdot \,  G'(x)
 \,\,  - \, \, A'_R(x)  \cdot \, G(x)^2
\,  \,+2 \, G'(x)^2 
 \,\,  - \, G(x) \cdot \,  G"(x)
 \, \,  =  \,\,  \, 0,
\end{eqnarray}
where $\, G(x)$ is the log-derivative\footnote[2]{With an extra log-derivative
step equation (\ref{logderiv1first}) can be written in an even simpler 
form. Introducing $\, H(x) \, = \, \, G'(x)/G(x)$, equation (\ref{logderiv1first}) becomes 
$\, A_R'(x) \, -A_R(x)\cdot \, H(x) \, + \, H'(x) \, - \, H(x)^2 \, = \, \, 0$.} 
of $\, Q(x)$, i.e. $\, G(x) \, = \, \, Q'(x)/Q(x)$. While equation (\ref{QF}) 
means that $\, F(x) \, = \, \, 1/G(x)$,
equation (\ref{logderiv1first}) is immediately obtained by 
imposing $\, F(x) \, = \, \, 1/G(x)$ to be a 
solution of $\, \Omega^{*}$.

\vskip .2cm

One remarks that this non-linear differential equation 
corresponds to a {\em homogeneous} quadratic
 equation in $\, G(x)$ and its derivatives. In terms 
of $\, Q(x)$ this equation corresponds to
a homogeneous cubic equation in $\, Q(x)$ 
and its derivatives:
\begin{eqnarray}
\label{QDAfirst}
\hspace{-0.95in}&& \quad \quad   
A_R(x) \cdot \, \Bigl( Q'(x)^2 \,
 -Q(x) \cdot  \, Q''(x)\Bigr)
 \cdot  \, Q'(x) \, \, 
+ \, A'_R(x) \cdot \, Q(x) \cdot \, Q'(x)^2
 \\
\hspace{-0.95in}&& \quad  \quad  \quad  \quad 
+Q''(x) \cdot  \, Q'(x)^2 \,
 +Q(x) \cdot \, Q'''(x) \cdot  \, Q'(x)
\,\,  -2 \,  Q(x) \cdot \, Q''(x)^2
 \,  \,= \, \, \, 0. 
\nonumber
\end{eqnarray}

\vskip .2cm
  
The function $\, P(x)$, being the composition inverse 
of a differentially algebraic function, is solution of the 
{\em differentially algebraic}~\cite{Selected,IsTheFull} 
equation\footnote[5]{Equation (\ref{simplerelation61first}) 
can be obtained using the Fa\`a di Bruno formulas for the 
higher derivatives of inverse functions.}:
\begin{eqnarray}
\label{simplerelation61first}
\hspace{-0.95in}&&  \, \quad \quad   \,
A_R(P(x)) \cdot \,  P'(x)^2 \cdot \, 
\Bigl( x \cdot \, P''(x) \, + P'(x)\Bigr)
\,\,\,  + \, \, x \cdot \, A_R'(P(x)) 
\cdot \, P'(x)^4 \, \, 
\nonumber \\ 
\hspace{-0.95in}&& \quad \quad \quad  \quad   \quad 
+ x \cdot \,  P"(x)^2  \,  \, \,
-x \cdot \, P'(x) \cdot \,  P'''(x)
\,  \,\, - P'(x) \cdot \, P"(x)
 \,\, \, = \, \, \, \, 0.
\end{eqnarray}
For instance, for the hypergeometric function (\ref{vid}), one verifies 
straightforwardly that $\, P(x) \, = \, \, sn^4(x, \, (-1)^{1/2})$,
given in~\cite{Hindawi}, verifies (\ref{simplerelation61first})
with $\, A_R(x)$ given by the first 
rational function in (\ref{respecAR}).

\vskip .1cm
  
\subsection{Assuming that $\, \Omega$ is globally nilpotent} 
\label{assuminglob}

The rank-two condition (\ref{mad}) turns out to identify exactly with 
the first Casale condition (\ref{cas2}), the only difference 
being that $\, A_R(x)$
is not meromorphic as in Casale's condition (\ref{cas2}), but a 
{\em rational function}: in lattice statistical mechanics and 
enumerative combinatorics, the differential operators are 
linear differential operators with polynomial coefficients. In fact,
the operators emerging in lattice models are not only Fuchsian,
but {\em globally nilpotent} operators~\cite{bo-bo-ha-ma-we-ze-09},
or $\, G$-operators~\cite{Andre}, thus their wronskians are the 
$\, N$-th root of a rational function~\cite{bo-bo-ha-ma-we-ze-09}. This 
naturally leads us to examine the case where $\, \Omega$ is taken to 
be globally nilpotent. Given $\, \Omega$ globally nilpotent, there 
exists an algebraic function $\, u(x)$ ($N$-th root 
of a rational function) such that $\, A_R(x)$
is the log-derivative of $\, u(x)$. Consequently 
$\, \Omega$ and $\, \Omega^{*}$,
which read respectively 
$\,\Omega \, = \, \, $
$ u(x)^{-1}  \cdot  \, D_x   \cdot  \,  u(x)  \cdot  \, D_x$
and $\, \Omega^{*} \, = \, \,  $
$ D_x \cdot  \, u(x)  \cdot  \, D_x   \cdot  \, u(x)^{-1}$, 
are  related by the simple conjugation:
\begin{eqnarray}
\label{simplerelati}
\hspace{-0.95in}&& \quad \quad  \quad \quad \quad \quad \quad \quad 
\,\, \Omega^{*} \cdot \,  u(x)
  \,\, \, = \,   \,  \,\,  u(x) \cdot \, \Omega. 
\end{eqnarray}
Thus, $ \, F(x)$ and $\, Y(x)$ are related through the simple 
equation:
\begin{eqnarray}
\label{simplerelation}
\hspace{-0.95in}&& \quad \quad \quad \, \, 
\quad \quad \quad \quad \quad
 u(x) \cdot \,  Y(x) \, \, = \, \, \, \, F(x).
\end{eqnarray}

\vskip .1cm

The fact that the holonomic function $\, Y(x)$ is solution of $\, \Omega$, 
amounts to writing that the log-derivative of  $\, Y'(x)$ is 
equal to $\, -A_R(x)$. If $\, \Omega$ is globally nilpotent then 
$\, -A_R(x)$ is the log-derivative of the reciprocal $\, 1/u(x)$, and
the logarithm of  $\, Y'(x)$ is equal to the logarithm of $\, 1/u(x)$, 
up to a constant of integration $\, \ln(\alpha)$, and thus:
\begin{eqnarray}
\label{deduces}
\hspace{-0.95in}&& \, \, \,  \quad  \, \,
 \alpha \cdot \, {{d Y(x)} \over { d x}} 
 \, \, = \, \, \, \, {{1} \over {u(x)}} 
\quad \quad \quad  \, \, \hbox{or:} \quad \quad \quad  \quad  \, 
\alpha \cdot \, {{ Y'(x)} \over {Y(x)}} \, \, = \, \, \, \, 
{{1} \over {u(x) \cdot \, Y(x)}}.
\end{eqnarray}

\vskip .1cm

Recalling the fact that the rank-two condition (\ref{mad}) gives (\ref{QF}), 
namely that the log-derivative of $\, Q(x)$ is equal to $\, 1/F(x)$, one 
deduces by combining (\ref{simplerelation}) with (\ref{deduces}):
\begin{eqnarray}
\label{combining}
\hspace{-0.95in}&&  \, \, \, \, \, \quad 
{{Q'(x) }  \over {Q(x)}}  \, \, = \, \, \, \, {{1} \over {F(x)}}  
\, \, = \, \, \, \, \alpha \cdot \, {{ Y'(x)} \over {Y(x)}}
 \quad \quad  \, \, \, \, \hbox{i.e.} \quad \quad \quad   
Q(x) \, \, = \, \, \, \lambda \cdot \, Y(x)^{\alpha}.
\end{eqnarray}
Note that, without any loss of generality, one can restrict $\, \lambda$
to $\, \lambda \, = \, \, 1$.

\vskip .1cm

$\, F(x)$ is solution of $\, \Omega^{*}$ as a consequence of the  
rank-two condition (\ref{mad}). This second order linear differential 
equation can be integrated into 
$\, F'(x) \, -A_R(x) \cdot \, F(x) \, = \, \, u(x) \cdot \, Y'(x)$, 
and taking into account (\ref{deduces}) this gives: 
\begin{eqnarray}
\label{integrated}
\hspace{-0.95in}&&  \, \, \quad \quad  
\quad \quad  \quad \quad  \quad \quad 
 F'(x) \, \,\, -A_R(x) \cdot \, F(x) 
\, \,\, = \, \, \, \, \, {{1} \over {\alpha}}.
\end{eqnarray}

\vskip .1cm

For the new results (see sections (\ref{moreHeun}) and (\ref{more2F1higher}) below), 
corresponding to a rank-two condition (\ref{mad}) like the hypergeometric examples
seen in the beginning of this section, the holonomic function $\,  Y(x)$ is 
of the form (\ref{form}). Thus the constant $\, \alpha$ is actually 
equal to a {\em positive integer} $\, N$ (see the case where 
$\, N \, = \, 3$  in \ref{more2F1examples} for a worked example). Further 
one deduces from (\ref{combining}) 
that $\, Q(x)$ is {\em always a holonomic function}:
$\, Q(x) \, = \, \, \lambda \cdot \, Y(x)^N$,  for instance,
for the hypergeometric functions
 (\ref{vid}),  (\ref{YM3first}),  (\ref{YM6first}),  (\ref{zero}), 
(\ref{more1})  and (\ref{QQQ8first}), we have $\, Q(x)\, = \, \, Y(x)^N$
with $\, N= \, 4, \, 3, \, 6, \, 2, \, 4, \, 6$ respectively.
                                                             
\vskip .1cm

Without assuming (\ref{form}), the constant $\, \alpha$ is not necessarily 
a positive integer, thus $\, Q(x)$ has no reason to be 
holonomic: it is just {\em differentially algebraic} (see (\ref{QDAfirst})). 
The log-derivatives of $ \, Q(x)$ and $\, Y(x)$ being
equal up to a multiplicative factor $\, \alpha$ (see (\ref{combining})), 
one deduces from the fact that (\ref{logderiv1first}) is a 
{\em homogeneous} (quadratic) condition in $\, G(x)$ and its derivatives,
that  $ \, Q(x)$ and $\, Y(x)$ verify necessarily the {\em same}
differentially algebraic condition (\ref{QDAfirst}).
                                                         
\vskip .1cm

With this global nilpotence assumption, the differentially algebraic 
function $\, P(x)$ is, in fact, solution of much simpler non-linear 
ODEs. From $\,\, u(x) \cdot \, Y(x) \, = \, F(x)\,$ one gets
 using (\ref{covP}):
\begin{eqnarray}
\label{muchsimpler}
\hspace{-0.95in}&&  \, \,  \, \, \quad   \quad  \quad \quad 
u(P(x))^{\alpha} \cdot \, Y(P(x))^{\alpha} \,\,\, = \, \,\, F(P(x))^{\alpha}
 \, \,\, =  \, \, \,\Bigl(x \cdot \, P'(x)\Bigr)^{\alpha}. 
\end{eqnarray}
Using $\,\, Q(x) = \,  \lambda \cdot \, Y(x)^{\alpha}$, 
and $\, \,Q(P(x)) \, = \, x$, one deduces:
\begin{eqnarray}
\label{muchsimpler}
\hspace{-0.95in}&&  \, \, \quad \, \, \,   \quad  \quad \quad  \quad \quad 
x  \cdot \, u(P(x))^{\alpha} 
 \, \,\, =  \, \, \, \,  \lambda \cdot \, 
\Bigl(x \cdot \, P'(x)\Bigr)^{\alpha}. 
\end{eqnarray}

\vskip .1cm
  
\section{More rational transformations:  an identity on a Heun function } 
\label{moreHeun}

In this section we write an identity similar to the $\, _2F_1$ 
hypergeometric identities  (\ref{vid}), (\ref{YM3first}),
(\ref{YM6first}), but, this time, on a {\em Heun function}, that is 
a holonomic function with {\em four} singularities instead of the
well-known three singularities $\, 0$, $\, 1$, $\, \infty$
of the hypergeometric functions.

\vskip .1cm
  
Let us consider the rational transformation\footnote[5]{Emerging 
as a symmetry of the complete elliptic integrals of the third kind
in the anisotropic Ising model (see~\cite{Barry}).}
\begin{eqnarray}
\label{next}
\hspace{-0.95in}&&  \, \quad \quad \quad  
\quad  \quad \quad \quad
x \quad \longrightarrow \quad \quad  
4 \cdot \,\frac{x \cdot \, (1-x) 
\cdot \, (1 \, -k^2 \, x)}{(1 \, -k^2 \, x^2)^2}, 
\end{eqnarray}
where one recognizes the transformation\footnote[2]{The general 
case $\,\,\theta \, \rightarrow \, \, p \, \theta \,\,$ 
is laid out in \ref{MiscellHeun}.}  
$\, \theta \, \rightarrow \, \, 2 \, \theta$  on the 
square of the elliptic sine $\, x \, = \, \, sn(\theta, \, k)^2$: 
\begin{eqnarray}
\label{doubling}
\hspace{-0.95in}&&     \quad \quad        \quad    
 sn(\theta, \, k)^2 \, \, 
\quad  \longrightarrow  \, \, \,  \quad   \quad  
sn(2 \, \theta, \, k)^2  \, \, \, =
\nonumber \\
\hspace{-0.95in}&&    \quad   \quad    \quad       \qquad  
\, \, \, \, = \, \, \, \,
 4 \cdot  \, {{  sn(\theta, \, k)^2 \cdot \, 
(1\, - sn(\theta, \, k)^2) \cdot \, 
(1\, - \, k^2 \cdot \, sn(\theta, \, k)^2) 
} \over {(1  \, -k^2 \cdot \, sn(\theta, \, k)^4 )^2 }}.
\end{eqnarray}

Denoting $\, M \, = \, \, 1/k^2$,  the transformation 
(\ref{next}) yields: 
\begin{eqnarray}
\label{Aadoubling}
\hspace{-0.95in}&&   \quad \quad \quad \quad  \quad \quad   
R(x) \, \, = \, \, \, 4 \cdot \,\frac{x \cdot \, (1-x) 
\cdot \, (1 \, -\, x/M)}{(1 \, - \, x^2/M)^2}.
\end{eqnarray}
For a given $\, M$, the transformations 
$\, \theta \, \rightarrow \, \, p \, \theta \, $ 
give rational transformations $\,x  \, \rightarrow \,  R_p(x)$ 
on the square of the elliptic sine, 
$\, x \, = \, \, sn(\theta, \, k)^2$, 
which are sketched for the first primes $\, p \, \, $
 in \ref{MiscellHeun}. The series expansions of these rational
transformations read 
$\,  R_p(x) \, = \, \, p^2 \cdot \, x \, + \, \cdots \, $ 
With these rational functions $\, R_p(x)$
we have the following identity on a Heun function\footnote[9]{The Heun 
function is the Heun {\em general} function, HeunG function in Maple, 
not a confluent Heun function.}:
\begin{eqnarray}
\label{Fdoublingidefirst}
\hspace{-0.95in}&&   \quad \quad \quad   \quad \quad   
 R_p(x) \cdot \, 
Heun\Bigl(M, \, {{M\, +1} \over {4}}, \,  {{1} \over {2}}, 
\, 1, \, {{3} \over {2}}, \,{{1} \over {2}}, \, R_p(x) \Bigr)^2
\nonumber \\
\hspace{-0.95in}&&   \quad \quad \quad  \quad \quad \quad \quad \quad   
 \, \, = \, \, \, 
p^2 \cdot \, x \cdot \, 
Heun\Bigl(M, \, {{M\, +1} \over {4}}, \,  {{1} \over {2}}, 
\, 1, \, {{3} \over {2}}, \,{{1} \over {2}}, \, x\Bigr)^2. 
\end{eqnarray}
Using the formalism introduced 
in section (\ref{recalls}), we write
\begin{eqnarray}
\label{AadoublingAa}
\hspace{-0.95in}&&  
A_R(x) \, = \, \, {{ u'(x)} \over {u(x)}} \, = \, \,
 -{{1} \over {2 \, (M\, -x)}}
 \, +{{2 \, x \, -1} \over { 2 \, x \, (x-1)}}
\,\, \, = \,\, \,  \,
{{1} \over {2}} \cdot \, 
{{3\, x^2 \, -2\, (M+1)\, x\,  +M } \over {
x \cdot \,  (1-x) \cdot \,  (M \, -x) }}, 
\nonumber \\
\hspace{-0.95in}&&      \,  \quad \quad  \quad  \quad  
\hbox{where:} \quad    \quad \quad \quad  \,  \,  \, 
u(x) \, \, = \, \, \,  \Bigl( x \cdot \,  (1-x) \cdot \,  (1 \, -x/M) \Bigr)^{1/2}.
\end{eqnarray}
The Liouvillian solution of the 
operator $\,\, \Omega \, = \, \, (D_x \, +A_R(x)) \cdot \, D_x\,$ 
corresponds to the 
{\em incomplete elliptic integral of the first kind} 
(introducing $\, u \, =\, \, sin^2(\theta)$ and  
$\, x \, =\, \, sin^2(\phi)$):
\begin{eqnarray}
\label{incomplete}
\hspace{-0.95in}&&  
F(\phi, \, m) \,\,  = \, \, \, 
\int_{0}^{\phi} \, {{ d \theta } \over { (1\, - \, m \cdot \, sin^2(\theta))^{1/2}}}
\,\,  = \, \, \, {{1} \over {2}} \cdot \, 
 \int_{0}^{x} \,  
{{ d u} \over {
 u^{1/2} \cdot \, (1\, -u)^{1/2} \cdot \, (1\, - \, m \cdot \, u)^{1/2}}}.
\nonumber 
\end{eqnarray}
This corresponds to a Heun function,
 or equivalently to the {\em inverse Jacobi sine \footnote[5]{The Jacobi sine function
arises  from the inversion of the incomplete elliptic integral of the first kind.}}:
\begin{eqnarray}
\label{otherwords}
\hspace{-0.95in}&&  
x^{1/2} \cdot \, Heun\Bigl(M, \, {{M\, +1} \over {4}}, \,  {{1} \over {2}}, 
\, 1, \, {{3} \over {2}}, \,{{1} \over {2}}, \, x\Bigr)
 \,  = \, \,   
InverseJacobiSN\Bigl(x^{1/2}, \, {{1} \over {M^{1/2}}}\Bigr).
\end{eqnarray}
The Heun solution of $\, \Omega$ reads with 
$\, x \, =\, \, sin^2(\phi)$:
\begin{eqnarray}
\label{Fdoubling}
\hspace{-0.95in}&&     \quad       \quad    
Y(x) \, = \, \,\,  
x^{1/2} \cdot \, 
Heun\Bigl(M, \, {{M\, +1} \over {4}}, \,  {{1} \over {2}}, 
\, 1, \, {{3} \over {2}}, \,{{1} \over {2}}, \, x\Bigr)
\, = \,  \,\,  \, {{1} \over {M^{1/2}}} 
\cdot \, F\Bigl(\phi, \, {{1} \over {M}}\Bigr)
\nonumber \\
\hspace{-0.95in}&&   \quad \quad \quad   \quad    \quad      \quad \quad    
 \,\,  = \, \, \, {{1} \over {2}} \cdot \, 
\int_0^{x} \, {{ d u} \over { u^{1/2} \cdot \, (1\, -u)^{1/2} \cdot \, (M\, -u)^{1/2}}}. 
\end{eqnarray}
The Heun identity (\ref{Fdoublingidefirst}) amounts to writing 
a covariance on this Heun function given by:
\begin{eqnarray}
\label{Ydoublingidentity}
\hspace{-0.95in}&& \quad \quad \quad \quad    \quad \quad  \quad  \quad 
Y\Bigl(R_p(x)\Bigr) \, \,  \, = \, \, \,   \, 
p \, \cdot \,  Y(x).
\end{eqnarray}

\vskip .1cm 

The adjoint operator 
$\,\, \Omega^{*} \, = \, \, D_x \cdot (D_x \, -A_R(x))\,$
has  the following Heun function solution: 
\begin{eqnarray}
\label{Fdoubling}
\hspace{-0.95in}&&      
F(x) \, \, = \, \, \,  \,  
x \cdot \, (1 \, -x)^{1/2} \cdot \,
 \Bigl(1 \, -{{x} \over {M}} \Bigr)^{1/2} \cdot \, 
Heun\Bigl(M, \, {{M\, +1} \over {4}}, \,
  {{1} \over {2}}, \, 1, \, {{3} \over {2}}, \,{{1} \over {2}}, \, x\Bigr). 
\end{eqnarray}

All the rational transformations $\,  R_p(x)$ verify 
a rank-two condition (\ref{mad}) with 
$\, A_R(x)$ given by (\ref{AadoublingAa}). 
More generally, the one-parameter series solution of 
the rank-two condition  (\ref{mad}) are, again, 
{\em commuting} series: 
\begin{eqnarray}
\label{Heunonepara}
\hspace{-0.95in}&&  
R(a, \, x) \, \, =  \, \,  \, \,
 a \cdot \, x \, \,\,\, + a \cdot (a-1) \cdot \, S_a(x) 
\quad \quad \quad \quad \quad \hbox{where:} 
\\
\hspace{-0.95in}&& 
 S_a(x) \, \, =  \, \,  \, \,-\,{\frac { (M\, +1) }{ 3 \, M}} \cdot \, x^2 \, \, \, 
 \, + \, \, {\frac { (2\, \cdot \, (M^{2}+1) 
 \cdot \, (a \, -4)  \, + \, (13\,a \, -7) \cdot \,  M) }{ 
45 \cdot \, M^{2}}} \cdot \, x^3 \, \, 
\nonumber \\ 
\hspace{-0.95in}&& \,  \, \,   
- \, {{(M\, +1)} \over { 315 \cdot \, M^3 }} \cdot  
\Bigl((M^{2}+1)  \cdot \, (a \, -4)  \cdot \, (a \, -9)
  \, + \, \,  (29\,a^2 \,-62\,a \,-6) \cdot \, M\Bigr) 
 \cdot \, x^4 
\, \,   + \, \,\cdots 
\nonumber 
\end{eqnarray}
with $\,\,\, R(a_1, \, R(a_2, \, x)) \, = \, \, $
$R(a_2, \, R(a_1, \, x)) \, = \, \, R( a_1\,a_2, \, \, x)$.
The one-parameter series (\ref{Heunonepara}) reduces to the 
series expansion of the rational functions $\,  R_p(x)$  for $\, a \, = \, p^2$ 
{\em for every integer} $\, p$. One thus sees that the rank-two
condition (\ref{mad}) with $\, A_R(x)$ given by (\ref{AadoublingAa}), 
{\em encapsulates an infinite number of commuting rational transformations} 
$\, R_p(x)$.

Finally, as far as the Koenig-Siegel   
linearization~\cite{Siegel,Siegel2,Almost,Milnor}  
of the one-parameter series is concerned, one has 
$ \,\, Q(x) \, \, = \, \, Y(x)^2\,$ and: 
\begin{eqnarray}
\label{Pdoubling}
\hspace{-0.95in}&&   \quad \quad  \quad \quad \quad    \quad    \quad   \quad     
P(x) \, = \, \, \,   sn\Bigl(x^{1/2}, \, {{1} \over {M^{1/2}}}\Bigr)^2.
\end{eqnarray}
One easily verifies that this exact expression (\ref{Pdoubling}) 
in terms of the elliptic sine is solution of the 
differentially algebraic equation (\ref{simplerelation61first})    
with $\, A_R(x)$ given by (\ref{AadoublingAa}). 

\vskip .1cm
 
One can verify (though it is not totally straightforward) that the rational
function (\ref{Aadoubling}), and more generally the $\, R_p(x)$, 
have the decomposition 
\begin{eqnarray}
\label{decompo}
\hspace{-0.95in}&&   
\, \, \,\,  \,  4 \cdot \,\frac{x \cdot \, (1 \, -x) 
\cdot \, (1 \, -\, x/M)}{(1 \, - \, x^2/M)^2} 
\, \, = \,\, \,  P(4 \cdot  Q(x)),
\quad \, \, \,  \, \, 
R_p(x) \, = \,\, \,  P(p^2 \cdot Q(x)),
\end{eqnarray}
with $\, P(x)$ and $\, Q(x)$ given respectively by (\ref{Pdoubling}) 
and  $ \,\, Q(x) \, \, = \, \, Y(x)^2$.

\vskip .2cm
 
\subsection{$\, _2F_1$ hypergeometric functions deduced from the Heun example} 
\label{more2F1Heun}

\vskip .2cm
 
We know from~\cite{maier-05,Belyi3} for example, that selected 
Heun functions can reduce to pullbacked $\, _2F_1$ 
hypergeometric functions. This is also the case
for the Heun function  (\ref{Fdoubling}) 
in section (\ref{moreHeun})  for selected 
values of $\, M$. For $\, M\, = \, \, 2$ we have: 
\begin{eqnarray}
\label{M2}
\hspace{-0.95in}&&  \quad  \quad  \quad  \quad   \quad   
Heun\Bigl(M, \, {{M\, +1} \over {4}}, \,  {{1} \over {2}}, 
\, 1, \, {{3} \over {2}}, \,{{1} \over {2}}, \, x\Bigr)
\nonumber \\
\hspace{-0.95in}&&  \quad  \quad  
 \quad  \quad  \quad  \quad   \quad  \quad  \quad 
 \, \, = \, \, \, (1\, -x)^{-1/4} \cdot \,
 _2F_1\Bigl([{{1} \over {4}}, \, {{3} \over {4}}], \, [{{5} \over {4}}],
 \, {{-x^2} \over { 4 \, \cdot \, (1 \, -x)}}\Bigr),
\end{eqnarray}
for $\, M\, = \, -1$:
\begin{eqnarray}
\label{Mm1}
\hspace{-0.95in}&&  \quad  \quad  \quad    \quad  \quad    
Heun\Bigl(M, \, {{M\, +1} \over {4}}, \,  {{1} \over {2}}, 
\, 1, \, {{3} \over {2}}, \,{{1} \over {2}}, \, x\Bigr) 
\nonumber \\
\hspace{-0.95in}&&  \quad  \quad  \quad \quad 
 \quad  \quad \quad  \quad      \quad  
\, \, = \, \, \, (1\, -x^2)^{-1/4} \cdot \,
 _2F_1\Bigl([{{1} \over {4}}, \, {{3} \over {4}}], \, [{{5} \over {4}}], 
\, {{-x^2} \over { 1 \, -x^2}}\Bigr),
\end{eqnarray}
and for $\, M\, = \, 1/2$:
\begin{eqnarray}
\label{M1over2}
\hspace{-0.95in}&&  \quad  \quad  \quad     \quad  \quad     
Heun\Bigl(M, \, {{M\, +1} \over {4}}, \,  {{1} \over {2}}, 
\, 1, \, {{3} \over {2}}, \,{{1} \over {2}}, \, x\Bigr)
\nonumber \\
\hspace{-0.95in}&&  \quad  \quad  \quad
 \quad  \quad  \quad     \quad  \quad   \quad   
\, \, = \, \, \, (1\, -2\, x)^{-1/4} \cdot \,
 _2F_1\Bigl([{{1} \over {4}}, \, {{3} \over {4}}], \, [{{5} \over {4}}],
 \, {{-x^2} \over { 1 \, -2\, x}}\Bigr).
\end{eqnarray}
Besides, the three previous values of $\, M \, = \, 1/k^2$
such that the Heun function 
(or the inverse Jacobi sine, InverseJacobiSN in Maple) reduces 
to pullbacked hypergeometric functions, correspond to a 
 {\em complex multiplication value} of 
the $\, j$-function~\cite{j},
namely~\cite{Heegner} $\, j \, = \, \, (12)^3 \, = \, 1728$:
\begin{eqnarray}
\label{jfunc}
\hspace{-0.95in}&&   \quad \quad
j \, \, = \, \, \, 
256 \cdot \,{\frac { ({M}^{2}-M+1)^{3}}{ M^{2}  \cdot \, (M \, -1)^{2}}}, 
\quad \quad \,   \,  \, 
j \, =\, 1728 \, 
\quad \longleftrightarrow \quad
 M \, = \, \,  \, 2, \,\, {{1} \over {2}}, \,\, -1.
\end{eqnarray}
The other complex multiplication values (Heegner numbers 
see~\cite{Heegner}) do not seem to correspond to a reduction of 
the Heun function to pullbacked hypergeometric functions. 

\vskip .2cm

Recalling (\ref{M2}), (\ref{Mm1}), (\ref{M1over2}), 
and specifying the Heun identity (\ref{Fdoublingidefirst}), 
or (\ref{Ydoublingidentity}),
for $\, M\, = \, 2$, $\, M\, = \, -1$, and $\, M\, = \, 1/2 \, $
respectively, one gets three identities on the 
hypergeometric function  $\, _2F_1([1/4,3/4],[5/4],x)$. These three
identities are in fact consequences of the simple identity:
\begin{eqnarray}
\label{NEWident}
\hspace{-0.95in}&&  \, Y(x) \, = \, \, \, \, x^{1/4} \cdot \,
 _2F_1\Bigl([{{1} \over {4}}, \, {{3} \over {4}}], \, [{{5} \over {4}}], \, x\Bigr)
\, = \, \,  \, 
 {{1} \over {2}} \cdot \,    {\cal P}(x)^{1/4} \cdot \,
 _2F_1\Bigl([{{1} \over {4}}, \, {{3} \over {4}}], \, [{{5} \over {4}}], \,  {\cal P}(x) \Bigr), 
\end{eqnarray}
where 
\begin{eqnarray}
\label{pullNEWident}
\hspace{-0.95in}&&  \quad  \quad  \quad  \quad  \quad  \quad  \quad  \quad  
{\cal P}(x) \, \, = \, \, \, 
16 \cdot \, \,{\frac {x \cdot \, (1 \, -x) }{ (1 \,+4\,x\,-4\,{x}^{2})^{2}}},
\end{eqnarray}
together with the ``transmutation'' relations 
\begin{eqnarray}
\label{transmut}
\hspace{-0.95in}&&  \quad  \quad  \quad  \quad  \quad   \, \, 
{\cal P}(p_k(x)) \, \, = \, \, \, p_k({\cal P}_k(x)), 
\quad  \quad  \quad   \quad  
k \, \, = \, \, 1, \, 2, \, 3, 
\end{eqnarray}
where the pullbacks $\, {\cal P}(p_k(z))$ are transformation 
(\ref{Aadoubling}) for respectively  $\, M\, = \, 2$, 
$\, M\, = \, -1$, and $\, M\, = \, 1/2$
\begin{eqnarray}
\label{transmutwhere}
\hspace{-0.95in}&&  \quad  \quad  \quad    
{\cal P}_1(x) \, \, = \, \, \, 
8 \cdot \, 
{\frac { x  \cdot \, (1\, - x)  \cdot \, (2\, -x) }{ ({x}^{2}-2)^{2}}}, 
 \quad   \quad   \quad  
{\cal P}_2(x) \, \, = \, \, \, 
4 \cdot \,{\frac {x \cdot \, (1\, -x^2)  }{ (1 + \,{x}^{2})^{2}}}, 
\nonumber \\ 
\hspace{-0.95in}&&  \quad  \quad  \quad  \quad \quad  \quad  
 \quad  
{\cal P}_3(x) \, \, = \, \, \, 
4 \cdot \,{\frac {x \cdot \, (1\, -x) 
 \cdot \, (1 \, -2\,x) }{ (1 \, - 2\,{x}^{2})^{2}}}, 
\end{eqnarray}
and where $\, p_k(x)$ are the pullbacks emerging in the $\, _2F_1$ 
representations (\ref{M2}), (\ref{Mm1}), (\ref{M1over2}) of the 
Heun function:
\begin{eqnarray}
\label{transmutwhere2}
\hspace{-0.95in}&&  \,  \quad    
p_1(x) \, \, = \, \, \, 
- {{1} \over {4}} \cdot \,{\frac {{x}^{2}}{1 \, -x}},
 \quad   \quad  
p_2(x) \, \, = \, \, \, 
-{\frac {{x}^{2}}{1 \, -{x}^{2}}}, 
\quad   \quad  
p_3(x) \, \, = \, \, \,
 -{\frac {{x}^{2}}{1 \, -2\,x}}. 
\end{eqnarray}

The hypergeometric function $\, Y(x)$ given by  (\ref{NEWident}),
is solution of the order-two linear differential operator 
$\,\, \Omega \, = \, \, (D_x \, + \, A_R(x)) \cdot \, D_x \,\,$ 
where 
\begin{eqnarray}
\label{Y2F11over4Az}
\hspace{-0.95in}&& \quad \, \quad \quad \quad \quad \quad  \quad  
A_R(x) \, \, = \, \, \, 
{{3} \over {4}} \cdot \, {\frac {1 \, -2\,x}{x \cdot \, (1 \, -x) }}, 
\end{eqnarray}
verifies the rank-two condition (\ref{mad}):
\begin{eqnarray}
\label{Y2F11over4Rota}
\hspace{-0.95in}&& \quad \quad  \quad \quad \quad  \quad  
A_R({\cal P}(x)) \cdot \, {\cal P}'(x)^2 
\,\,  = \,\,  \,  \,\,  A_R(x) \cdot \,  {\cal P}'(x)  \, \, \,  
 + \, {\cal P}''(x). 
\end{eqnarray}
One notes that the hypergeometric functions (\ref{more1}) and 
(\ref{NEWident}) are associated with the same $\, A_R(x)$ 
given by (\ref{Y2F11over4Az}):  one can easily show 
that these two hypergeometric
functions are  equal. Therefore  (\ref{NEWident}) shares the 
same rank-two  condition (\ref{mad}) with  (\ref{Y2F11over4Az}), 
a condition that is also verified for the rational 
transformation (\ref{more1pull}) together 
with the pullback (\ref{pullNEWident}), with (\ref{more1pull})
and (\ref{pullNEWident}) {\em commuting}. The hypergeometric 
function (\ref{NEWident}) verifies an identity with the pullback 
(\ref{more1pull}), namely $\, Q(R(x)) \, =  \, \, -4 \cdot \, Q(x)$ 
with $\, R(x)$ given by (\ref{more1pull}), where 
$\, Q(x) \, = \, \, Y(x)^4$.

\vskip .2cm 

{\bf Remark:} For these selected values of $\, M$, 
one could be surprised that the function $\, Q(x)$ in the case of the 
Heun function is such that $\, Q(x) \, = \, Y(x)^2$, when the $\, Q(x)$
for the hypergeometric function (\ref{NEWident}) closely related to this 
Heun function (see identities (\ref{M2}), (\ref{Mm1}), (\ref{M1over2})) 
is such that $\, Q(x) \, = \, Y(x)^4$. This difference comes from the 
pullbacks (\ref{transmutwhere2}): the pullbacked hypergeometric 
functions (\ref{M2}), (\ref{Mm1}), (\ref{M1over2}) also 
correspond to $\, Q(x) \, = \, Y(x)^2$.

\vskip .2cm 

\subsection{A comment on the non globally bounded character of the Heun function } 
\label{moreHeuncomment}

Heun functions with generic parameters
are generally not reducible to $\, _2F_1$ hypergeometric functions 
with one or several pullbacks\footnote[5]{This corresponds 
to the emergence of a modular form represented 
as a  $\, _2F_1$ hypergeometric functions with two 
possible pullbacks~\cite{Christol}: 
the series expansion can be recast into a series with 
{\em integer coefficients}~\cite{Kratten}.}.

 Unlike $\, _2F_1$ functions the corresponding linear 
differential Heun operators are generally not globally nilpotent,
and the series of Heun functions are 
not globally bounded. While, for Heun functions, the reducibility 
to pullbacked $\, _2F_1$ hypergeometric functions, 
the  global nilpotence, and the  global boundedness implicate 
each other in general, this is not true 
when the corresponding linear differential operator {\em factors}.
Note that the series (\ref{Fdoubling}) as well as the 
series $\, Q(x)\, = \, \, Y(x)^N$ for the various hypergeometric 
functions ((\ref{vid}),  (\ref{YM3first}),  (\ref{YM6first}), 
... with $\, N= \, 4, \, 3, \, 6$, ...)) are 
{\em not globally bounded}\footnote[2]{A globally bounded
series is a series that can be recast into a series 
with {\em integer coefficients}~\cite{Christol}.}.

\vskip .1cm

In this light, the fact that the series (\ref{Fdoubling}) 
as well as the series $\, Q(x)\, = \, \, Y(x)^N$
are not globally bounded, does not seem to be in agreement with 
the previous modular form emergence and the previous remarkable 
identities (\ref{Ydoublingidentity}), or  
$\,\, Q(R(x)) \, = \, \, 4 \cdot \, Q(x)$. The series 
\begin{eqnarray}
\label{Gdoubling}
\hspace{-0.95in}&&   \quad \quad \quad   \quad    \quad  \quad       
G(x) \,\, = \, \, \, \,
 Heun\Bigl(M, \, {{M\, +1} \over {4}}, \,  {{1} \over {2}}, 
\, 1, \, {{3} \over {2}}, \,{{1} \over {2}}, \, \, 4 \, M \, x\Bigr), 
\end{eqnarray}
might not be globally bounded, yet it is ``almost globally bounded'': 
the denominator of the coefficients of $\, x^n$ are of the form 
$\, 2 \, n \, +1$. Therefore one finds that the 
closely related series 
\begin{eqnarray}
\label{Udoubling}
\hspace{-0.96in}&&   \, \, \,     
 {\tilde G}(x) \, \, = \, \, \,  \,  
2 \, \, x \cdot \, G'(x) \,\, + \, \, G(x)
\nonumber \\ 
\hspace{-0.96in}&&   \,\,\,\, = \,
1\, 
+ 2 \, (M+1) \cdot \, x
\,+ 2 \, (3\,{M}^{2} +2\,M +3) \cdot \, {x}^{2} 
\,+4\, (M+1) \, (5\,{M}^{2} -2\,M +5)  \cdot \, {x}^{3}
\nonumber \\ 
\hspace{-0.96in}&&   \quad   \quad   \quad \quad   
\,+ \, 2 \cdot \, (35\,{M}^{4} +20\,{M}^{3} +18\,{M}^{2} +20\,M +35) 
 \cdot \, {x}^{4}
\,\, \, \, + \,\, \,\cdots 
\end{eqnarray}
is actually globally bounded for any rational number 
value of $\, M$: the coefficient of $\, x^n$ is a 
polynomial in $\, M$ {\em with integer coefficients} 
of degree $\, n$. ${\tilde G}(x)$ is solution of 
an order-one linear differential operator
and  is an algebraic 
function: 
$\,  {\tilde G}(x) \,\, = \, \,\, \,$
$ (1 \, -4 \, x)^{-1/2} \, \cdot \,   (1 \, -4 M \, x)^{-1/2}$. 
Thus the series (\ref{Udoubling}) is globally bounded 
for any rational number $\, M$.  

\vskip .1cm 

{\bf Remark:} To be globally bounded~\cite{Christol} is a property 
that is preserved by operator homomorphisms:  the transformation 
by a linear differential operator of 
a globally bounded  series is also globally bounded, however, 
{\em it is not preserved by integration}. 

\vskip .1cm 
\vskip .1cm

\section{$\, _2F_1$ hypergeometric function: a higher genus case } 
\label{more2F1higher}

The $\, _2F_1$ hypergeometric examples (\ref{vid}), 
(\ref{YM3first}), (\ref{YM6first}), 
and (\ref{zero}) are associated with 
{\em elliptic or rational} (see (\ref{zero}))
 {\em curves}. It is tempting to imagine the  rank-two conditions (\ref{mad})
to be {\em only} associated with hypergeometric functions connected to
{\em elliptic curves}, and with pullbacks given by 
{\em rational functions}\footnote[1]{Casale showed in~\cite{Casale} that the only 
{\em rational} functions from $\mathbb{P}_1$ to $\mathbb{P}_1$ with a non-trivial 
$\, {\cal D}$-envelope are Chebyshev polynomials 
and {\em Latt\`es transformations}. Latt\`es transformations 
are rational transformations associated with elliptic 
curves (see for instance~\cite{Eremenko2}).}.
This is not the case though, as we shall see in the next {\em genus-two} 
hypergeometric example with {\em algebraic} function pullbacks.  

\vskip .1cm
 
Let us consider the hypergeometric function 
\begin{eqnarray}
\label{Ygenus}
\hspace{-0.95in}&& \, \, \, \,   \,\, \,  
 Y(x) \,\, = \, \, \, \,
x^{1/6} \cdot \, 
_2F_1\Bigl([{{1} \over {6}}, \, {{1} \over {3}}], \, [{{7} \over {6}}], \,  x  \Bigr)
 \,\, = \, \, \,\,
 {{1} \over {6}} \cdot \, \int_0^{x} \, (1\, -t)^{-1/3} \cdot \, t^{-5/6}  \cdot \, dt, 
\end{eqnarray}
 solution of the (factorized) order-two operator
$\, \Omega  \, = $
$\, \, (D_x \, + \, A_R(x)) \cdot \, D_x$ 
where:
\begin{eqnarray}
\label{aAgenus}
\hspace{-0.95in}&& \,    
A_R(x) \, = \,  \, \, 
 {{1}  \over { 6}} \,{\frac {7\,x \, -5}{x \cdot \, (x  \, -1) }}
\, = \,  \, \, {{u'(x)} \over {u(x)}} 
\quad  \,  \,  \,    \,  \,  \hbox{where:} \quad \quad
u(x) \, \, = \, \, \, (1\, -x)^{1/3} \cdot \, x^{5/6}, 
\end{eqnarray}
and one gets  $\, 6 \cdot \, Y'(x) \, = \, \, 1/u(x)$. Introducing 
$\, u \, = \, \, 6 \cdot \, Y'(x)$, one can 
canonically associate 
to (\ref{aAgenus}) the algebraic curve 
\begin{eqnarray}
\label{curvegenus}
\hspace{-0.95in}&& \quad \quad  \quad \quad \quad \quad \quad
u^6 \, \,\, - \,  (1\, -x)^2 \cdot \, x^5
 \, \,\, = \, \,\, \, 0, 
\end{eqnarray}
which is a  {\em genus-two algebraic curve}. 
We are seeking  an identity on this 
hypergeometric function (\ref{Ygenus}) of the form:
\begin{eqnarray}
\label{Ygenusrewrit}
\hspace{-0.95in}&& \quad \quad  \quad \quad \quad 
{\cal A}(x) \cdot \,
 _2F_1\Bigl([{{1} \over {6}}, \, {{1} \over {3}}], \, [{{7} \over {6}}], \,  x  \Bigr)
\, \, = \, \, \,  \,
 _2F_1\Bigl([{{1} \over {6}}, \, {{1} \over {3}}], \, [{{7} \over {6}}], \,  y(x)  \Bigr). 
\end{eqnarray}
Introducing the order-two linear differential operators 
annihilating respectively the LHS and RHS of (\ref{Ygenusrewrit}),
the identification of the wronskians of these two operators 
gives the algebraic function $\, {\cal A}(x)$ in terms 
of the pullback $\, y(x)$:
\begin{eqnarray}
\label{YgenusrewritcalA}
\hspace{-0.95in}&& \quad \quad  \quad \quad \quad \quad  \quad 
{\cal A}(x)\, \,\,  = \, \, \, 
 \Bigl({{ -27 \cdot \, x} \over { y(x)}}\Bigr)^{1/6}.
\end{eqnarray}
The pullback $\, y(x)$ must be some symmetry (isogeny) of the 
genus-two algebraic curve (\ref{curvegenus}). At first sight, this 
{\em seems to exclude rational function} pullbacks similar to 
the ones previously introduced. In fact, remarkably, there 
exists a simple identity on this
(higher genus) hypergeometric function: 
\begin{eqnarray}
\label{identityparam2}
\hspace{-0.95in}&& \quad \quad  \,  \,   \,  
 _2F_1\Bigl([{{1} \over {6}}, \, {{1} \over {3}}], \, [{{7} \over {6}}], \,\, \, 
 -27 \cdot \, \,
{\frac {v \cdot  \, (1-v)  \cdot \, (1+v)^{4}}{ 
(1 \, +3\,v)  \cdot \, ( 1 \, -3\,v)^{4}}}  \Bigr)^6
\nonumber \\
\hspace{-0.95in}&& \quad \quad   \quad  \quad  
\, \, \, \,     = \, \, \,  \,  \,  
{\frac { (1 \, +3\,v)^2 \cdot \,  (1 \, -3\,v)^4 }{
(1 \, -v)^2 \cdot \,(1+v)^{4} }} \cdot \, 
 _2F_1\Bigl([{{1} \over {6}}, \, {{1} \over {3}}], \, [{{7} \over {6}}], \, \, \, 
 {\frac {v \cdot \, (1 \, +3\,v) }{1 \, -v}} \Bigr)^6.
\end{eqnarray}
The two pullbacks in this remarkable identity (\ref{identityparam2})
yield the simple rational parametrization
\begin{eqnarray}
\label{param}
\hspace{-0.95in}&& \quad \quad  \quad  \quad 
x \, \, = \, \,  \, {\frac {v \cdot \, ( 1 \, +3\,v) }{1 \, -v}}, 
\quad \quad \quad
y \, \, = \, \,  \, 
-27 \cdot \,{\frac {v \cdot  \, (1 \, -v) 
 \cdot \, (1 \, +v)^{4}}{ (1 \, +3\,v)  \cdot \, ( 1 \, -3\,v)^{4}}},
\end{eqnarray}
which parametrizes the following
 {\em genus-zero (i.e. rational) curve}\footnote[9]{One 
should not confuse these two algebraic curves: the genus-two
curve (\ref{curvegenus}) is associated with integrant of the 
hypergeometric integral (\ref{Ygenus}), when 
the rational curve (\ref{curverationalgenus}) is associated with 
the pullback in the hypergeometric identity (\ref{Ygenusrewrit}).
}: 
\begin{eqnarray}
\label{curverationalgenus}
\hspace{-0.95in}&& \quad   \quad
-27\, \cdot \, x \cdot \, (x \, -1)^{4}  \cdot \, ({y}^{2} \, +1) \, \, 
 \nonumber \\
\hspace{-0.95in}&& \quad   \quad  \quad \quad \,
 - \, ({x}^{6} -12\,{x}^{5} 
+807\,{x}^{4} +2504\,{x}^{3} +807\,{x}^{2} -12\,x +1)
 \cdot \,  y 
\, \,  \, = \, \,  \, \, 0. 
\end{eqnarray}
The algebraic function $\, y \, = \, y(x)$, defined by the  
genus-zero  curve (\ref{curverationalgenus}),
is an example of a 
pullback\footnote[5]{This is a consequence of identity (\ref{identityparam2}).} 
$\, y(x)$ occurring in 
the $\, _2F_1$ hypergeometric identity  (\ref{Ygenusrewrit}).
This (multivalued) algebraic function $\, y \, = \, y(x)$
has the following series expansions:
\begin{eqnarray}
\label{seriescurverationalgenus}
\hspace{-0.95in}&& \quad   \quad 
y_1 \, \, \, = \, \, \, 
-27\,x \,\,  \, -216\,{x}^{2}\, \, 
-648\,{x}^{3}\, \,\,  -1944\,{x}^{4}\,\,  
-648\,{x}^{5}\,\,  -27864\,{x}^{6}\, +203256\,{x}^{7}
\nonumber  \\
\hspace{-0.95in}&& \quad   \quad  \quad    \quad \quad\, 
-2123928\,{x}^{8}\,\,  +21844728\,{x}^{9}\,\,  -233611992\,{x}^{10}
\, \, \, + \, \, \, \cdots 
 \\
\hspace{-0.95in}&& \quad   \quad 
\label{y2genus} 
y_2 \, \,  \, = \, \, \, \, -{{1} \over { 27 \, x}}
  \,\, +{\frac {8}{27}}\,\,\,  -{\frac {40\,x}{27}}\,\, 
+{\frac {200\,{x}^{2}}{27}}\,\, 
 -{\frac {1192\,{x}^{3}}{27}}\,\,  
+{\frac {8456\,{x}^{4}}{27}}\, \,  -{\frac {68264\,{x}^{5}}{27}}
\nonumber \\
\hspace{-0.95in}&& \quad   \quad  \quad \quad  \quad \,\, 
+{\frac {604360\,{x}^{6}}{27}} \,\, \,  -{\frac {5722664\,{x}^{7}}{27}}
 \, \,\,  +{\frac {6332872\,{x}^{8}}{3}} \, 
\, \,\, + \, \, \cdots 
\end{eqnarray}
Note that the rational curve (\ref{curverationalgenus}) has 
the obvious symmetry $\, y \, \leftrightarrow \, \, 1/y$ (as well 
as the $\, x \, \leftrightarrow \, \, 1/x$ symmetry, consequence 
of the palindromic form of (\ref{curverationalgenus})), therefore 
the series (\ref{y2genus}) is
the reciprocal of 
(\ref{seriescurverationalgenus}): $\, y_2 \, = \, 1/y_1$.
Clearly $\, x$ and $\, y$ are not on the same footing.
The  composition inverse of the previous series 
gives the series
\begin{eqnarray}
\label{seriesreversiongenus}
\hspace{-0.95in}&&  \quad   
 {{ - \, y} \over { 27}} \, \, \,  -{\frac {8\,{y}^{2}}{729}} \, \,   \,
-{\frac {104\,{y}^{3}}{19683}} \,  \, \, 
-{\frac {1672\,{y}^{4}}{531441}} \,  \, \, 
 -{\frac {30248\,{y}^{5}}{14348907}} \,  \, \, 
-{\frac {196568\,{y}^{6}}{129140163}} \,\, \, 
  \, +  \,  \,  \,  \cdots 
 \\
\hspace{-0.95in}&&  \quad 
\label{reverse2}
 {{-27} \over {y}} \, \, +8 \,  \,+{\frac {40\,y}{27}} \, \,
+{\frac {520\,{y}^{2}}{729}} \, \,
+{\frac {8552\,{y}^{3}}{19683}} \,
 \,+{\frac {158344\,{y}^{4}}{531441}} \, \,
+{\frac {3151144\,{y}^{5}}{14348907}} \,\,
 \, + \, \,  \cdots,   
\end{eqnarray}
the second series being the reciprocal of the first 
one\footnote[9]{Note that the rational curve (\ref{curverationalgenus})
provides additional Puiseux series.}. 

\vskip .1cm 

Furthermore the two series
(\ref{seriescurverationalgenus}) and (\ref{y2genus})
verify\footnote[1]{These two series are related by
 $\, y \, \leftrightarrow \, \, 1/y$. Note that 
$\, y \, \leftrightarrow \, \, 1/y$ is not a symmetry 
of (\ref{madgenus}) in general.
} the rank-two condition (\ref{mad}) with $\, A_R(x)$ given by 
(\ref{aAgenus}): 
\begin{eqnarray}
\label{madgenus}
\hspace{-0.95in}&& \quad \quad  \quad  \quad  \quad \quad \quad 
A_R(y(x)) \cdot \, y'(x)^2 
\,\,     = \,\,  \,  \,\,
   A_R(x) \cdot \,  y'(x) \, \, \, + \, y''(x). 
\end{eqnarray}
Do note that the series, corresponding to the 
 composition inverse of 
these two series (\ref{seriescurverationalgenus}) 
and (\ref{y2genus}) (namely (\ref{seriesreversiongenus})
 and  (\ref{reverse2}) where one changes $\, y$ into $ \, x$),
also verify the rank-two  condition (\ref{mad}) with $\, A_R(x)$ given by 
(\ref{aAgenus}). For example, introducing  $\, Q(x) \, =\, \, Y(x)^6$,
 one finds that  $\, Q(y(x))  \,  \, =\, \,  \,     -27 \cdot \, Q(x)$
for $\, y(x)$ the algebraic function corresponding to series 
(\ref{seriescurverationalgenus}). The
composition inverse  of series\footnote[2]{Namely series 
(\ref{seriesreversiongenus}) where one changes 
$\, y$ into $ \, x$.} (\ref{seriescurverationalgenus})
gives the (reversed) result: 
$\, Q(x) \, =\, \,  - 27 \cdot \,  Q(y(x))$.

\vskip .1cm
\vskip .1cm 

{\bf Remark:} The rank-two condition (\ref{madgenus}) 
with $\, A_R(x)$ given by (\ref{aAgenus}) has a one-parameter 
family of {\em commuting} solution series:
\begin{eqnarray}
\label{madseriesgenus}
\hspace{-0.95in}&& \quad \quad 
R(a, \, x) \, \, =  \, \,  \, \,
 a \cdot \, x \, \,\, + a \cdot (a-1) \cdot \, S_a(x) 
\quad \quad \ \quad \quad \hbox{where:} 
\\
\hspace{-0.95in}&& \quad  \quad 
 S_a(x) \, \, =  \, \,  \, \, -{{2} \over {7}} \cdot \, x^2 \, \, 
 +{{17\,a \, -87} \over {637}} \cdot \, x^3  \,\, 
 +{{2 \cdot \, (113\,a^2 \,-856\,a \,+3438)} \over {84721}} \cdot \, x^4
\nonumber \\ 
\hspace{-0.95in}&& \quad  \quad \quad \quad  \quad \quad 
-{{3674\,a^3 +121194\,a^2 -552261\,a +2095059 } \over {38548055}} \cdot \, x^5
\, \,\,\,  + \, \, \, \cdots 
\end{eqnarray}
with $\,\,\, R(a_1, \, R(a_2, \, x)) \, = \, \, $
$R(a_2, \, R(a_1, \, x)) \, = \, \, R( a_1\,a_2, \, \, x)$,
where (\ref{madseriesgenus}) reduces to  the algebraic series 
(\ref{seriescurverationalgenus}) and  (\ref{seriesreversiongenus}) 
for $\, a \, = \, -27$ and  $\, a \, = \, -1/27$ 
respectively. Consequently the occurrence of a {\em higher genus} 
curve like (\ref{curvegenus})
is not an obstruction to the existence of a family of 
one-parameter {\em  abelian} series.

\vskip .2cm 
 
\section{Schwarzian condition on an algebraic 
transformation: $\, _2F_1$ representation of a modular form} 
\label{Schwarz}

The typical situation emerging in physics with 
{\em modular forms}~\cite{IsingCalabi,IsingCalabi2,Diagselect} 
is that some ``selected'' hypergeometric 
function $\,\, _2F_1([\alpha, \, \beta], \, [\gamma], \, x)$ 
verifies an identity with 
{\em two different pullbacks}\footnote[5]{The modular
forms occuring in physics often correspond to cases
where the two different pullbacks $\, p_1(x)$ and $\, p_2(x)$ 
are rational functions, but they can also be 
algebraic functions~\cite{Christol,Morain}.
} related 
by an {\em algebraic} curve, the {\em modular equation} curve 
$\,\, M(p_1(x), \,p_2(x)) \, = \, \, \,0$:
\begin{eqnarray}
\label{modularform}
\hspace{-0.95in}&& \quad \quad  \quad \quad \quad \quad 
 _2F_1\Bigl([\alpha, \, \beta], \, [\gamma], \,  p_1(x)  \Bigr)
\, = \, \, \, \,\, {\tilde A}(x) \cdot \,
 _2F_1\Bigl([\alpha, \, \beta], \, [\gamma], \, p_2(x)  \Bigr), 
\end{eqnarray}
where $\, {\tilde A}(x)$ is an algebraic function. 

This representation of {\em modular forms} in terms of
hypergeometric functions with {\em many pullbacks}, is well 
described in Maier's papers~\cite{SuperMaier,BelyiMaier}. It
is different from the ``mainstream'' mathematical definition of
modular forms as (complex) analytic functions on the upper half-plane 
satisfying functional equations with respect to the group action 
of the modular group. {\em However, this hypergeometric 
representation is the one we do need in 
physics}~\cite{IsingCalabi,IsingCalabi2}. The reason why 
this hypergeometric function representation of modular forms
exists is a consequence of a not very well-known equality 
between the Eisenstein~\cite{Sebbar} series $\, E_4$  
(of weight four under the modular group), and a 
hypergeometric function of the (weight zero) modular 
$\,j$-invariant~\cite{Heegner,Canada} 
(see Theorem 3 page 226 in~\cite{Stiller}, see also page 216 of~\cite{Shen}): 
\begin{eqnarray}
\label{paradox}
\hspace{-0.95in}&&  \quad  \quad 
E_4(\tau) \,  \, \, = \, \,  \, \,
1 \, + \, 240 \, \sum_{n=1}^{\infty} \, n^3\cdot \, {{q(\tau)^n} \over  {1-q(\tau)^n}}
\,  \, \, = \, \,  \, \,
 _2F_1\Bigl([{{1} \over {12}}, \, {{5} \over {12}}],
 \, [1], {{1728} \over {j(\tau)}}\Bigr)^4.
\end{eqnarray}
In terms of $\, k$ the modulus of the elliptic functions,  the  $\, E_4$ 
Eisenstein series (\ref{paradox}) can also be written as:
\begin{eqnarray}
\label{paradoxE4}
\hspace{-0.95in}&&  
 _2F_1\Bigl([{{1} \over {12}},  {{5} \over {12}}], [1], \,
{{27} \over {4}} \, 
{\frac { {k}^{4} \cdot \, (1 \, -{k}^{2})^{2}}{ ({k}^{4}-{k}^{2}+1)^{3}}}\Bigr)^4 
\, = \,\,  (1-k^2+k^4) \cdot \,  _2F_1\Bigl([{{1} \over {2}},  {{1} \over {2}}],
  [1], \, k^2\Bigr)^4.
\end{eqnarray}
Another relation between hypergeometric functions and modular forms
corresponds to the representation of the Eisenstein series $\, E_6$ in
terms of the hypergeometric functions\footnote[9]{One easily verifies
that the expressions (\ref{paradoxE4}) and (\ref{paradoxE6})
 for respectively $\, E_4$ and $\, E_6$, are 
such that $\, (E_4^3 \, -E_6^2)/E_4^3 \, $ is actually the well-known 
expression of the Hauptmodul $\, 1728/j$ given as a rational function 
of the modulus $\, k$ (see (\ref{jjprime}) and 
(\ref{Haupt})).
}  (\ref{paradoxE4}) (see page 216 of~\cite{Shen}):
\begin{eqnarray}
\label{paradoxE6}
\hspace{-0.95in}&&  \quad  \quad   
  E_6 \,  \, = \,\, \, \, 
 (1+k^2)\cdot \, (1 \, -2\,k^2) \cdot \, \Bigl(1 \, -{{k^2} \over {2}}\Bigr) 
\cdot \,  _2F_1\Bigl([{{1} \over {2}}, \, {{1} \over {2}}],\,  [1], \, k^2\Bigr)^6
\\
\label{paradoxE6bis}
\hspace{-0.95in}&&  \quad  \quad  \quad  \quad  
 \, \, \,\,  = \,\, \, \, 
(1+k^2) \cdot \, (1 \, -2\,k^2)  \cdot \,  \Bigl(1 \, -{{k^2} \over {2}}\Bigr) 
\nonumber \\
\hspace{-0.95in}&&  \quad  \quad  \quad  \quad   \quad  
 \, \, \,\,  \,\, \, \, 
\times (1-k^2+k^4)^{-3/2} \cdot \, 
 _2F_1\Bigl([{{1} \over {12}},  {{5} \over {12}}], [1], \,
{{27} \over {4}} \, 
{\frac { {k}^{4} \cdot \, (1 \, -{k}^{2})^{2}}{ ({k}^{4}-{k}^{2}+1)^{3}}}\Bigr)^6.
\end{eqnarray}
One can rewrite a remarkable hypergeometric identity 
like (\ref{modularform}) in the form
\begin{eqnarray}
\label{modularform2}
\hspace{-0.95in}&& \quad \quad  \quad \quad \quad \quad 
  {\cal A}(x) \cdot \, 
_2F_1\Bigl([\alpha, \, \beta], \, [\gamma], \,  x  \Bigr)
\, = \, \, \, \,\,  
 _2F_1\Bigl([\alpha, \, \beta], \, [\gamma], \, y(x)  \Bigr), 
\end{eqnarray}
where $\, {\cal A}(x)$ is an algebraic function
and where $\, y(x)$ is an algebraic function
corresponding to the previous modular curve 
$\,\, M(x, \, y(x)) \, = \,\, \, 0$.

The Gauss hypergeometric function 
$\,\, _2F_1([\alpha, \, \beta], \, [\gamma], \, x) \, $
is solution of the second order linear 
differential operator\footnote[1]{Note that $\, A(x)$ is 
the log-derivative of 
$\, \, u(x) \, \,= \,\, \, x^{\gamma} \cdot \, (1 \, -x)^{\alpha+\beta+1-\gamma}$.}: 
\begin{eqnarray}
\label{Gaussdiff}
\hspace{-0.95in}&& \quad \quad  \quad \quad 
 \Omega \,\,  = \, \, \, \,  
D_x^2    \,  \, + \, A(x) \cdot \,  D_x  \, \, + \, B(x), 
\quad \quad \quad  \quad \quad  \hbox{where:} 
\nonumber \\
\hspace{-0.95in}&& \quad    
A(x) \,\,  = \, \, \, 
{{ (\alpha +\beta+1) \cdot  \, x \,  \, -\gamma} \over { x \cdot \, (x\, -1)}}
  \, \, = \, \,  \,  {{u'(x)} \over { u(x)}}, 
\quad  \quad \quad 
B(x) \,\,  = \, \, \,  {{\alpha  \, \beta } \over {x \cdot \, (x\, -1) }}.
\end{eqnarray}
We would like now, to identify the two order-two linear differential 
operators of the LHS and RHS of identity (\ref{modularform}). 
A  straightforward calculation 
enables us to find the algebraic function $\, {\cal A}(x)$
in terms of the algebraic function pullback $\, y(x)$
in (\ref{modularform2}):
\begin{eqnarray}
\label{modularform3}
\hspace{-0.95in}&& \quad   \quad \quad  \quad \quad  \quad \quad
{\cal A}(x)  \, \, = \, \,  \,  
 \Bigl( {{u(x)} \over { u(y(x)) }}
 \cdot \,  y'(x) \Bigr)^{-1/2}. 
\end{eqnarray}
Expression (\ref{modularform3}) for  $\, {\cal A}(x)$
is such that the two order-two linear differential operators 
(of a similar form as (\ref{Gaussdiff}))
have the same $\, D_x$ coefficient. The identification
of these two operators thus corresponds 
(beyond (\ref{modularform3})) to just one condition that can 
be rewritten (after some algebra ...) in the following 
Schwarzian form:
\begin{eqnarray}
\label{condition1}
\hspace{-0.95in}&& \quad   \quad \quad  \quad \quad  \quad 
 W(x) 
\, \, \,  \,-W(y(x)) \cdot  \, y'(x)^2
\, \, \,  \,+ \,  \{ y(x), \, x\} 
\, \,\, \, = \,\, \, \,  \, 0, 
\end{eqnarray}
or: 
\begin{eqnarray}
\label{condition}
\hspace{-0.95in}&& \quad   \quad \quad  \quad \quad  \quad 
 {{W(x)} \over {y'(x)}}  
\, \, \,  \,-W(y(x)) \cdot  \, y'(x)
\, \, \,  \,+ \, {{ \{ y(x), \, x\} } \over {y'(x)}}
\, \,\, \, = \,\, \, \,  \, 0, 
\end{eqnarray}
where  
\begin{eqnarray}
\label{wherecond}
\hspace{-0.95in}&& \quad   \quad \quad  \quad \quad  \quad 
W(x)  \, \, = \, \,  \, \, \,  A'(x) \, \, \,  + \, \, 
   {{A(x)^2} \over {2 }} \, \,  \, \,  -2 \cdot \, B(x),
\end{eqnarray}
and where $\,\{ y(x), \, x\}$ denotes the 
{\em Schwarzian derivative}~\cite{What}:
\begin{eqnarray}
\label{Schwa}
\hspace{-0.95in}&&  \, \,  \,    \, 
\{ y(x), \, x\}    \, \, = \, \,  \,  \,  \,
{{y'''(x) } \over{ y'(x)}} 
 \,  \,  - \, \, {{3} \over {2}}
 \cdot \, \Bigl({{y''(x)} \over{y'(x)}}\Bigr)^2
 \, \, = \, \,  \,  \,  \,
{{ d } \over { dx  }} \Bigl( {{y''(x) } \over{ y'(x)}}   \Bigr) 
\, \, - {{1} \over {2}} \cdot \, \Bigl( {{y''(x) } \over{ y'(x)}}   \Bigr)^2. 
\end{eqnarray}

In the identity (\ref{modularform}), characteristic of modular forms, 
the two pullbacks $\, p_1(x)$ and $\, p_2(x)$ are clearly on the 
{\em same footing}, while identity (\ref{modularform2}) breaks 
this fundamental symmetry, seeing $\, y$ as a function of 
$\, x$. We can perform the same calculations 
seeing the variable $\, x$ as a function of $\, y$ 
in (\ref{modularform2}). 
Despite the simplicity of condition (\ref{condition})
 it is not clear whether $\, x$ and $\, y$ are on the same footing
in condition (\ref{condition}). This is actually the case, 
since if one considers  $\, x$ as a function of $\, y$, we 
have the well-known classical 
result that the Schwarzian derivative of  $\, x$ with 
respect to $\, y$ is simply related to (\ref{Schwa}), 
the Schwarzian derivative of  $\, y$ with respect to $\, x$:
\begin{eqnarray}
\label{Schwarew}
\hspace{-0.95in}&& \quad  \quad   \quad  \quad  \quad  \quad 
\{ y(x), \, x\}    \, \, \, = \, \,  \,  \,
 -y'(x)^2 \cdot \,  \{ x(y), \, y\}. 
\end{eqnarray}
In other words, if one introduces the following Schwarzian bracket
\begin{eqnarray}
\label{Schwabis}
\hspace{-0.95in}&& \quad    \quad  \quad  \quad  \quad 
[ y, \, x ]    \, \, = \, \,  \,  \,  \,
{{ \{ y(x), \, x\} } \over {y'(x)}}  \, \, = \, \,  \,  \,  \,
{{y'''(x) } \over{ y'(x)^2 }} 
 \,  \,  \, - \, \, {{3} \over {2}} \cdot \, {{y''(x)^2} \over{y'(x)^3}}, 
\end{eqnarray}
it is antisymmetric: $\, [ y, \, x ]    \,  = \, \,   -\,   [ x, \, y ]$.
With this appropriate notation,
$\, x$ and $\, y$ {\em can be seen on the same footing}.
With this in mind we can now rewrite  condition (\ref{condition}) 
in a balanced way:
\begin{eqnarray}
\label{conditionNEW}
\hspace{-0.95in}&& \quad   \quad \quad  \quad 
 2 \cdot \, W(x) \cdot \, {{ d x} \over { dy}}   
\, \, 
\, \, \,+ \, [ y, \, x] 
\, \,\, \, = \,\, \, \,  \,
2 \cdot \, W(y) \cdot  
\,  {{ d y} \over { dx}}  \,  \,  \, \, +  \,  [ x, \, y].
\end{eqnarray}
If one denotes by $\, L(x, \, y)$ the LHS of (\ref{conditionNEW})
\begin{eqnarray}
\label{L}
\hspace{-0.95in}&& \quad  \quad  \quad   \quad  \quad \quad \quad 
 L(x, \, y)\,\, \, = \,\, \, \,  \, 
 {{ 2 \cdot \, W(x) \, + \, \, \{ y, \, x\} } \over {y'}}, 
\end{eqnarray}
the Schwarzian condition (\ref{condition}), or (\ref{conditionNEW}),
reads $\, \,  L(x, \, y) \, = \, \, L(y, \, x)$.

Being the result of the covariance (\ref{modularform2}), a
Schwarzian identity like (\ref{conditionNEW}) {\em has to be compatible} 
with the composition of functions. For instance, 
from (\ref{modularform2}) one immediately deduces:
\begin{eqnarray}
\label{modularform22}
\hspace{-0.95in}&& \quad \quad  \quad \quad  \quad 
 _2F_1\Bigl([\alpha, \, \beta], \, [\gamma], \, y(y(x))  \Bigr)
\, = \, \, \, \,\,  
  {\cal A}(y(x)) \cdot \,
 _2F_1\Bigl([\alpha, \, \beta], \, [\gamma], \,  y(x)  \Bigr)
\nonumber \\
\hspace{-0.95in}&& \quad \quad 
 \quad \quad  \quad  \quad \quad \quad  \quad \quad 
\, = \, \, \, \,\,  
 {\cal A}(y(x)) \cdot \, {\cal A}(x) \cdot \,
 _2F_1\Bigl([\alpha, \, \beta], \, [\gamma], \, x  \Bigr). 
\end{eqnarray}
One thus expects condition (\ref{condition}) to be compatible with 
the composition of function (similarly to the previous compatibility
of the rank-two condition (\ref{mad}) with the iteration 
of $\, x \, \rightarrow \, R(x)$): this is actually the case.
Recalling the (well-known)  chain rule for the Schwarzian 
derivative of the composition of functions 
\begin{eqnarray}
\label{chainrule}
\hspace{-0.95in}&& \quad \quad  \quad \quad  \quad \quad 
\{ z(y(x)), \, x\} \, \,  \,= \, \, \,  \, \,
\{ z(y), \, y \} \cdot y'(x)^2 
\, \,\, + \, \, \{ y(x), \, x\}, 
\end{eqnarray}
it is straightforward to show directly (without referring to
the covariance (\ref{modularform2})) that condition (\ref{condition}) 
is actually compatible with the composition of functions
 (see \ref{Schwarzcomp}
for a demonstration).

\vskip .1cm

The Schwarzian derivative is the perfect tool~\cite{What}  
to describe the {\em composition of functions} and the 
{\em reversibility} of an iteration (the previously mentioned 
$\, x \, \longleftrightarrow \, y$
symmetry): it is not a surprise to see 
the emergence of a Schwarzian derivative in the description
of the modular forms~\cite{Schwarzian2,Schwarzian,Chakra} 
corresponding to identities like (\ref{modularform2}). We 
are going to see, for a given (selected ...) hypergeometric 
function $\,\, _2F_1([\alpha, \, \beta], \, [\gamma], \, x)$,  
that the condition (\ref{condition}) ``encapsulates'' 
all the isogenies corresponding to all the modular equations 
associated to transformations on the ratio of periods 
$\, \tau \, \rightarrow \, N \cdot \, \tau$ 
(resp. $\, \tau \, \rightarrow \, \, \tau/N$),
for various values of the integer $\, N$ corresponding 
to the different modular equations. 

\vskip .2cm
 
\subsection{Schwarzian condition and the simplest example of modular forms: a series viewpoint} 
\label{Schwarzsimplest}

Let us focus on an example of a modular form that emerged
many times in the analysis of $\, n$-fold integrals of the 
square Ising model~\cite{broglie,bo-ha-ma-ze-07b,Heegner,Christol}. Let 
us recall the simplest example of a modular form and of 
a modular equation curve 
\begin{eqnarray}
\label{modularform2explicit}
\hspace{-0.95in}&& \quad \quad  \quad \quad \quad  
  {\cal A}(x) \cdot \,
 _2F_1\Bigl([{{1} \over {12}}, \, {{5} \over {12}}], \, [1], \,  x  \Bigr)
\, = \, \, \, \,\,  
 _2F_1\Bigl([{{1} \over {12}}, \, {{5} \over {12}}], \, [1], \, y  \Bigr), 
\end{eqnarray}
where $\, {\cal A}(x)$ is an algebraic function
and where $\, y \, = \, \, y(x)$ is an algebraic function
corresponding to the modular equation (\ref{modularcurve}).
The algebraic function $\, y \, = \, \, y(x)$ is a
multivalued function, but we can single out the series 
expansion\footnote[1]{This series (\ref{seriesmodularcurve})
has a radius of convergence $\, 1$, even if the discriminant 
of the modular equation (\ref{modularcurve}) which vanishes
at $\, x \, = \, 1$, vanishes for values inside the 
unit radius of convergence, for instance at 
$\, x \, = \, -64/125$.}:
\begin{eqnarray}
\label{seriesmodularcurve}
\hspace{-0.95in}&& \quad  \quad  \quad 
 y \, \, \, = \, \, \,  \,  \,
{\frac {1}{1728}} \cdot \, {x}^{2} \,  \, \,
+{\frac {31}{62208}} \cdot \, {x}^{3}\,
\, +{\frac {1337}{3359232}} \cdot \,{x}^{4} \,   \,\,
+{\frac {349115}{1088391168}} \cdot \,{x}^{5}
\nonumber \\ 
\hspace{-0.95in}&& \quad \quad  \quad  \quad  \quad 
+{\frac {20662501}{78364164096}} \cdot \, {x}^{6} \,\,  \, 
+{\frac {1870139801}{8463329722368}} \cdot \,{x}^{7} \, \,
 \, + \, \, \cdots 
\end{eqnarray}
One verifies easily that the Schwarzian 
condition (\ref{condition}) is verified with: 
\begin{eqnarray}
\label{whereconda}
\hspace{-0.95in}&& 
W(x)  \,  = \,   
 -{\frac {32\,{x}^{2}-41\,x+36}{72 \cdot \, {x}^{2} \cdot \, (x \, -1)^{2}}},
\, \,   
A(x) \,  = \,  
{{ 3 \cdot  \, x \,  \, -2} \over { 2  \, x \cdot \, (x\, -1)}}, 
\, \,  
B(x) \,  = \, 
  {{ 5} \over { 144 \, x \cdot \, (x\, -1) }}.
\end{eqnarray}

\subsubsection{Other algebraic transformations:  other modular equations \\} 
\label{other}

\vskip .1cm 
\vskip .1cm 
\vskip .1cm 

The {\em modular equations} $\, {\cal M}_N(x, \, y) \, = \, \, 0$,  
corresponding to the transformation 
$\, \tau \, \rightarrow \, N \cdot \, \tau$,
or $\, \tau \, \rightarrow \,  \, \tau/N$,  define algebraic 
transformations (isogenies) $\, x \, \, \rightarrow \, \, y$  
for the identity (\ref{modularform2explicit})
with $\, {\cal A}(x)$ given by an algebraic function.

Let us consider another important modular equation. 
The modular equation of order three corresponding to 
$\, \, \tau \, \rightarrow \, 3 \cdot \, \tau$,
or $\, \tau \, \rightarrow \,  \, \tau/3$, 
reads\footnote[5]{Legendre already knew (1824) this 
order three modular equation in the form 
$\, (k \lambda)^{1/2} +   (k'  \lambda')^{1/2} 
 =  1$, where $\, k$ and $\, k'$, and $\,\lambda$, $\,\lambda'$
are pairs of complementary moduli $\, k^2+k'^2=1$, 
$\, \lambda^2+\lambda'^2=1$, and Jacobi derived that 
modular equation~\cite{Jacobi,Nova}.}: 
\begin{eqnarray}
\label{orderthree}
\hspace{-0.95in}&&  \quad  \quad  \quad  \quad 
{k}^{4} \,\, +12\,{k}^{3}\lambda \, \, \, +6\,{k}^{2}{\lambda}^{2} \,\, 
+12\,k{\lambda}^{3}\,\, +{\lambda}^{4} \,  \,  \,  \,
  -16\,{k}^{3}{\lambda}^{3} \,  -16\,k\lambda
\,\, \,  \,  = \, \,\,  \,   0.
\end{eqnarray}
Recalling that 
\begin{eqnarray}
\label{orderthreexy}
\hspace{-0.95in}&&  \,  
x \,\, = \, \,\, {{27} \over {4}} \cdot \, 
{\frac { {k}^{4} \cdot \, (1 \, -{k}^{2})^{2}}{ ({k}^{4}-{k}^{2}+1)^{3}}}
\,\, = \, \,\, {{1728} \over {j(k)}}, \quad \, 
y \, \,  = \, \,\,{{27} \over {4}} \cdot \,  {\frac {{\lambda}^{4} \cdot \,
 \left( 1-{\lambda}^{2} \right)^{2}}{({\lambda}^{4}-{\lambda}^{2}+1)^{3}}}
\,\, = \, \,\, {{1728} \over {j(\lambda)}}, 
\end{eqnarray}
the modular equation 
(\ref{orderthree}) becomes the modular curve: 
\begin{eqnarray}
\label{orderthreemod}
\hspace{-0.95in}&&  \quad  
262144000000000 \cdot \, {x}^{3}{y}^{3} \cdot \, (x+y) \,  \, 
+4096000000 \cdot \, {x}^{2}{y}^{2} 
\cdot \, (27\,{x}^{2}-45946\,xy+27\,{y}^{2})\, 
\nonumber \\
\hspace{-0.95in}&& \quad  \quad  \quad 
 +15552000  \cdot \, xy \cdot \, 
 (x+y) \cdot \,  ({x}^{2}+241433\,xy+{y}^{2})
\nonumber \\
\hspace{-0.95in}&& \quad \,  \quad  \quad 
 +729\,{x}^{4} \, -779997924\,{x}^{3}y  \,  \, 
+1886592284694\,{x}^{2}{y}^{2}  \,  \, -779997924\,x{y}^{3} \, 
+729\,{y}^{4} 
\nonumber \\
\hspace{-0.95in}&& \quad \, \quad  \quad 
  +2811677184 \cdot \,xy \cdot \, (x+y) \, 
\, -2176782336 \cdot \,x\, y \, \, \, = \, \, \, \, 0.
\end{eqnarray}
which gives the series expansion:
\begin{eqnarray}
\label{orderthreey}
\hspace{-0.95in}&& \quad 
 y \,\, = \, \,\,\,  {\frac {{x}^{3}}{2985984}}
\,  \,\, +{\frac {31 \,x^4 }{71663616}} \,  \,
+{\frac {36221\, x^5  }{82556485632}} \, 
\,\,  +{\frac {29537101 \, {x}^{6}}{71328803586048}} 
\,\, \,   \, + \, \,\, \cdots 
\end{eqnarray}

\vskip .2cm 

One can easily get the the polynomial
with integer coefficients $\, {\cal M}_4(x, \, y)$, in the modular equation 
$\, {\cal M}_4(x, \, y) \, = \, 0$ corresponding to the transformation 
$\, \tau \, \rightarrow \, 4 \cdot \, \tau$,
or $\, \tau \, \rightarrow \,  \, \tau/4$, as follows:
if one denotes by $\,  {\cal M}_2(x, \, y)$ the LHS of the 
modular equation (\ref{modularcurve}),  
the polynomial $\, {\cal M}_4(x, \, y)$
is straightforwardly obtained by calculating the resultant 
of $\, {\cal M}_2(x, \, z)$ and $\, {\cal M}_2(z, \, y)$ 
in $\, z$, which factorizes in the form\footnote[8]{The 
exact expression of $\, {\cal M}_4(x, \, y)$ is a bit too large 
to be given here.} $\, (x\, -y)^2 \cdot \,  {\cal M}_4(x, \, y)$. 
The modular equation $\,\,  {\cal M}_4(x, \, y) \, = \, \, 0\, $
defines several algebraic series corresponding to the different 
branches\footnote[1]{These series can be obtained using the 
command ``algeqtoseries'' in the ``gfun'' package of Maple. 
} of the (multivalued) algebraic function transformation 
$\, x \, \rightarrow \, \, y$. We find Puiseux series 
and two analytic series at $\, x=\,0$ given by 
\begin{eqnarray}
\label{back}
\hspace{-0.95in}&&  \quad  \quad 
 y \, \, = \, \, \, \, {\frac {{x}^{4}}{5159780352}} \, \, 
+{\frac {31\,{x}^{5}}{92876046336}} \, \, 
+{\frac {43909\,{x}^{6}}{106993205379072}} \, 
\,\,\, + \, \, \cdots 
\end{eqnarray}
which is clearly similar to the previous series 
(\ref{seriesmodularcurve}) and (\ref{orderthreey}), 
but also an {\em involutive}\footnote[2]{The series (\ref{othersolution}) is the 
{\em only involutive series}  of the form $\,\, - x \, + \, \, \cdots \, $
 {\em which verifies the Schwarzian condition} (\ref{condition}).
} series of radius of convergence $\, 1$, of the (quite unexpected) 
simple  form  $\, \,  -x \, + \, \cdots \, $ namely:
\begin{eqnarray}
\label{othersolution}
\hspace{-0.95in}&& 
y \, \, = \, \, \,  \, 
-x \, \,   \, \,   -{\frac {31\,{x}^{2}}{36}}\,  \, 
-{\frac {961 }{1296}} \cdot \, x^3 \,  \, 
-{\frac {203713 }{314928}}\cdot \, x^4 \, \,  
-{\frac {4318517 }{7558272}} \cdot \, x^5\, \, 
 -{\frac {832777775}{1632586752}}\cdot \, x^6\,
\nonumber \\
\hspace{-0.95in}&& \,  \,   \,   \,  
-{\frac {729205556393 }{1586874322944}} \cdot \,{x}^{7}
-{\frac {2978790628903 }{7140934453248}} \cdot \,{x}^{8}  
-{\frac {43549893886943 }{114254951251968}} \cdot \,{x}^{9}  
\, \,  + \,   \dots 
\end{eqnarray}

One easily verifies that all these series  (\ref{seriesmodularcurve}),
 (\ref{orderthreey}), (\ref{back}), (\ref{othersolution}) 
(as well as the other Puiseux series) are solutions 
of the Schwarzian condition 
(\ref{condition}), and that the series (\ref{seriesmodularcurve}),
(\ref{orderthreey}), (\ref{back})  {\em commute} when composed, while  (\ref{back}) 
and (\ref{othersolution}) {\em do not} ! This 
is a consequence of the fact that they correspond to the various commuting
isogenies $\, \tau \, \rightarrow \, N \cdot \, \tau \, $ 
(resp. $ \tau \, \rightarrow \, \, \tau/N$). 

\subsubsection{A one-parameter solution series of  the Schwarzian condition\\} 
\label{oneparam}

\vskip .1cm 
\vskip .1cm 
\vskip .1cm 
\vskip .1cm 

Let us first seek solution-series of the 
Schwarzian condition (\ref{condition}) of the form 
$\, \, e \cdot \, x \,\, + \, \cdots\, \, $ 
with $\, W(x)$ given by (\ref{whereconda}).
One finds that the Schwarzian condition 
(\ref{condition}) has a {\em one-parameter family} of 
solution-series as well of the form 
$\, \, e \cdot \, x \,  \, + \, \cdots\, \,\, $
namely\footnote[5]{The  one-parameter series (\ref{seriesmodcurve1a})
is {\em completely defined} by the fact that it is a series of the form 
$\, e \cdot  \, x \, + \, \, \cdots \, $ commuting with 
the algebraic series (\ref{othersolution})
and the hypergeometric series (\ref{seriesmodcurve1aeps}),
{\em  without referring to the Schwarzian condition} 
(\ref{condition}).}:
\begin{eqnarray}
\label{seriesmodcurve1a}
\hspace{-0.95in}&& \, \, \quad \,       \quad   \quad       
y(e, \, x) \, \, \, = \, \, \,  \, 
e \cdot \, x   \, \,\, \, + \, e  \cdot \, (e-1) \cdot \, S_e(x),
 \quad \quad \quad     \quad  \quad \hbox{where:} 
\\
\hspace{-0.95in}&& \, \,  \quad   \quad \quad   \quad   \quad \,  \,       
 S_e(x)\, \, = \, \, \, \, 
-{\frac {31}{72}}  \cdot \,  {x}^{2} \, 
\, \, \, +{\frac { (9907\,e -20845) }{82944}} \cdot \,  {x}^{3}
\nonumber \\
\hspace{-0.95in}&& \, \,  \,   \,   \, \, 
\quad \quad    \quad   \quad  \quad    \quad      \quad  
\,\,  -{\frac {
 (4386286\,{e}^{2}-20490191\,e +27274051)
 }{161243136}}\cdot \,  {x}^{4}
\,\,  \,  \, \,+ \, \,\, \cdots 
\end{eqnarray}
The series (\ref{seriesmodcurve1a}) is a one-parameter 
family of {\em commuting series}:
\begin{eqnarray}
\label{seriesmodcurve1axx}
y(e, \, y(\tilde{e}, \, x)) 
\, \, \, = \, \, \,  \,
y(\tilde{e}, \, y(e, \, x)) 
 \, \, \, = \, \, \,  \,y(e \, \tilde{e}, \, x), 
\end{eqnarray}
and in the $\, e \, \rightarrow \, \, 1$ limit 
of the one-parameter family (\ref{seriesmodcurve1a}),
one has:
\begin{eqnarray}
\label{seriesmodcurve1aeps}
\hspace{-0.95in}&& \quad  \quad 
y(e, \, x) \, \, \, = \, \, \,  \, 
x \,  \, \,  + \, \,  \, \epsilon \cdot \, F(x)\,  \,  \, 
+ \, \,\epsilon^2 \cdot \, G(x)  \,  \,\, \,+ \, \, \cdots
\qquad  \quad \,  \quad  \, \,  \hbox{where:}
\\
\label{holo1eps}
\hspace{-0.95in}&& 
F(x)\, \, = \, \,  x \cdot \, (1\, -x)^{1/2} \cdot \,
 _2F_1\Bigl([{{1} \over{12}}, \, {{5} \over{12}}], \, [1], \, x\Bigr)^2, 
 \quad   \,\, 
 G(x) \,  = \, \,\, 
 {{1} \over {2}}  \cdot \, F(x) \cdot \, (F'(x) \, -1).
\nonumber 
\end{eqnarray}

\subsubsection{Other one-parameter solution series of  the Schwarzian condition\\} 
\label{oneparam}

\vskip .1cm 

\vskip .1cm 
\vskip .1cm 

Clearly the analytic series (\ref{seriesmodularcurve}),
 (\ref{orderthreey}), (\ref{back})  
corresponding to the various isogenies 
$\, \tau \, \rightarrow \, N \cdot \, \tau$,
are not series of the form 
$\, \, e \cdot \, x \, + \, \cdots\, \, $, instead they are 
solution-series of the 
Schwarzian condition (\ref{condition}) of the form 
$\, \, a \cdot \, x^N \, + \, \cdots\, \, $

In order to generalize the solution-series (\ref{seriesmodularcurve}),
we will first seek solution-series of the 
Schwarzian condition (\ref{condition}) of the form 
$\, \, a \cdot \, x^2 \, + \, \cdots\, \, $
A straightforward calculation gives a one-parameter 
family of solution-series of (\ref{condition}) of the form 
$\, \, a \cdot \, x^2 \, + \, \cdots\, \, $:
\begin{eqnarray}
\label{seriesmodcurvea}
\hspace{-0.96in}&&  
 y_2 \, \, = \, \, \,
 a \cdot  \,{x}^{2} \, 
+{\frac {31 \cdot \, a{x}^{3}}{36}}
 \,-{\frac {a \cdot \,
 \left( 5952\,a-9511 \right) }{13824}}\cdot \, {x}^{4}
\,-{\frac {a \cdot \, 
\left( 14945472\,a-11180329 \right) }{20155392}} \cdot \, {x}^{5}
\nonumber \\ 
\hspace{-0.96in}&& \quad \quad  \,  
 \,+{\frac {a \cdot \, 
\left( 88746430464\,{a}^{2}-677409785856\,a+338926406215 \right)
}{743008370688}} \cdot \,  {x}^{6} 
\, \, \,\, + \, \, \cdots 
\end{eqnarray}
which actually reduces to (\ref{seriesmodularcurve}) for
$\, a \, = \, \, 1/1728$.
Similarly, one also finds a one-parameter 
family of solution-series of (\ref{condition}) of the form 
$\, \, b \cdot \, x^3 \, + \, \cdots\, \, $:
\begin{eqnarray}
\label{seriesmodcurve3a}
\hspace{-0.95in}&& \quad  \quad  \,
y_3 \, \, \, = \, \, \,  \, \,
b \cdot \, {x}^{3} \, \, 
+{\frac {31\,b }{24}} \cdot \, {x}^{4}\, \, 
+{\frac {36221\,b }{27648}} \cdot \, {x}^{5}  \, \, \, 
-{\frac {b  \cdot \, \left( 23141376\,b-66458485 \right)
}{53747712}} \cdot \,  {x}^{6}
\nonumber \\ 
\hspace{-0.95in}&& \quad  \quad \quad  \quad  \quad   \, 
\, -{\frac {b   \cdot \, (183649959936\,b-187769367601) 
 }{165112971264}}  \cdot \, {x}^{7}
\, \, \,  \, + \, \, \cdots 
\end{eqnarray}
which reduces to  (\ref{orderthreemod}) for 
$\, b \, = \, \, 1/2985984 \, = \, 1/1728^2$,
and  another one-parameter 
family of solution-series of (\ref{condition}) of the form 
$\, \, c \cdot \, x^4 \, + \, \cdots\, \, $:
\begin{eqnarray}
\label{corresponding}
\hspace{-0.95in}&& \quad \quad \quad \quad 
y_4 \, \, = \, \, \, \,\,  c \cdot \, {x}^{4} \,  \, \, 
+{\frac {31\,c }{18}}\cdot \,{x}^{5}
 \,\,  \, +{\frac {43909\,c }{20736}} \cdot \,{x}^{6} \, \, \, 
+{\frac {46242779\,c }{20155392}} \cdot \, {x}^{7}
\nonumber \\
\hspace{-0.95in}&& \quad \quad \quad \quad \quad \quad \,  \, \, \, \, 
+{\frac {c  \cdot \, (869687301215 -159953190912\,c) }{371504185344}}
 \cdot \, {x}^{8}\, 
\, \,\,  \, +  \, \, \, \cdots 
\end{eqnarray}
which reduces to  (\ref{back}) for 
$\, c \, = \, 1/5159780352  \, = \, \,1/1728^3$.
The series (\ref{seriesmodcurvea}), (\ref{seriesmodcurve3a}), 
(\ref{corresponding}), {\em do not commute}. The 
composition of the one-parameter series (\ref{seriesmodcurve3a})
with the one-parameter series (\ref{seriesmodcurvea}) 
gives the series\footnote[2]{If one seeks for the solution series 
of the Schwarzian condition (\ref{condition}) of the form 
$\, d \cdot \, x^6 \, +  \, \, \cdots \, \,$
one recovers the one-parameter family (\ref{curious2}).}:
\begin{eqnarray}
\label{curious2}
\hspace{-0.95in}&& \quad   \quad   \,\,  
y_2(y_3(x)) \, \, = \, \, \,  \, \, d \cdot \, {x}^{6}
\,  \,\,  
+{\frac {31\,d \cdot \, {x}^{7}}{12}} \,  \, 
\,+{\frac {59285\,d }{13824}} \cdot \, {x}^{8} \,  \, 
\,+{\frac {19676177\,d }{3359232}} \cdot \, {x}^{9}
\nonumber \\ 
\hspace{-0.95in}&& \quad  \,\,  \quad \quad    \quad    \quad   
\,+{\frac {197722802303\,d }{27518828544}} \cdot \, {x}^{10} \,  \, 
\,\, +{\frac {8173747929317\,d }{990677827584}}\cdot \, {x}^{11}
\,\,\, \,\,    + \,\,\, \cdots 
\end{eqnarray}
where $\, d \, \, = \, \, a \cdot \, b^2$. The composition of 
 the one-parameter series (\ref{seriesmodcurvea}) 
with the one-parameter series (\ref{seriesmodcurve3a})
gives a similar result where, now, 
$\, d \, \, = \, \, b \cdot \, a^3$.
These two one-parameter series commute when 
$\, a \cdot \, b^2 \, \, = \, \, b \cdot \, a^3$, i.e. 
$\, b \, = \, a^2$, and the modular equation series 
corresponds to $\,\, b \, = \, a^2\,$ with $\, a \, = \, \, 1/1728$.

The composition of the one-parameter series (\ref{seriesmodcurvea})
with the one-parameter series (\ref{seriesmodcurve1a}) 
gives the series  (\ref{seriesmodcurvea}) for $\, a \, e$ and 
$\, a \, e^2 \, $  respectively: 
\begin{eqnarray}
\label{curious5}
\hspace{-0.95in}&& \quad   \quad   \quad   \quad 
y_1(e, \, y_2(a, \, x)) \, \, = \, \, y_2(a \, e, \, x), 
\qquad y_2(a, \,y_1(e, \, x))  \, \, = \, \, y_2(a \, e^2, \, x). 
\end{eqnarray}
In other words if one introduces the modular equation series $ \, Y_2(x)$
given by (\ref{seriesmodularcurve}), corresponding to 
$ \, y_2(a, \, x)$ for $\, a \, = \, \, 1/1728$, 
the  one-parameter series $\, y_2(a, \, x)$ given by (\ref{seriesmodcurvea}), 
can be obtained as $\,\, y_1(1728 \,a, \, Y_2(x))\,$ or as 
$\, \, Y_2(y_1((1728 \,a)^{1/2}, \, x))$. 

\vskip .2cm 

Therefore, all the one-parameter families (\ref{seriesmodcurvea}),
(\ref{seriesmodcurve3a}), (\ref{corresponding}), are nothing 
but the {\em isogeny-series} (\ref{seriesmodularcurve}), 
(\ref{orderthreey}), (\ref{back}) {\em transformed by the one-parameter series} 
(\ref{seriesmodcurve1a}).

\vskip .2cm 

\subsection{The equivalent of $\, P(z)$ and $ \, Q(z)$ for the Schwarzian condition: the mirror maps \\} 
\label{Schwarzmirror}

\vskip .2cm 

Let us recall the concept of 
{\em mirror map}~\cite{IsingCalabi,IsingCalabi2,Candelas,Doran,Doran2,LianYau} 
relating the reciprocal of the $\,j$-function and the nome, 
with the well-known series with integer coefficients:
\begin{eqnarray}
\label{mirror}
\hspace{-0.95in}&& \quad 
\tilde{X}(q) \, \, = \, \, \, \, q \,\, \, -744\,{q}^{2} 
\,\, +356652\,{q}^{3} \,\, -140361152\,{q}^{4} \,\, +49336682190\,{q}^{5}
 \nonumber \\ 
\hspace{-0.95in}&& \quad \quad \, 
-16114625669088\,{q}^{6}  \,
\, +4999042477430456\,{q}^{7} \,  \,-1492669384085015040\,{q}^{8} 
\nonumber \\ 
\hspace{-0.95in}&& \quad \quad \, \, 
+432762759484818142437\,{q}^{9} \,  \, \, + \, \, \cdots
\end{eqnarray}
and\footnote[1]{In Maple the series (\ref{mirror}) 
can be obtained substituting $\, L=EllipticModulus(q^{1/2})^2$,
 in $ 1/j \,  = \,  \,$
$ \,L^2 \cdot \,(L-1)^2/(L^2-L+1)^3/256$. See https://oeis.org/A066395 
for the series (\ref{mirror})
and https://oeis.org/A091406 for the series (\ref{mirror2}).} 
its composition  inverse:
\begin{eqnarray}
\label{mirror2}
\hspace{-0.95in}&&
\tilde{Q}(x) \, \, = \, \, \,  \,
x \,  \, \, +744\,{x}^{2} \, \, +750420\,{x}^{3} 
\, \, +872769632\,{x}^{4} \,  \,
+1102652742882\,{x}^{5} 
\nonumber \\ 
\hspace{-0.95in}&& \quad \,  \, +1470561136292880\,{x}^{6}
 \, +2037518752496883080\,{x}^{7} \,
+2904264865530359889600\,{x}^{8} 
\nonumber \\ 
\hspace{-0.95in}&& \quad \, \, 
+4231393254051181981976079\,{x}^{9}\, \, \, + \, \, \cdots 
\end{eqnarray}
These series correspond to $\, x$ being the reciprocal of the
$\, j$-function: $\, 1/j$ . In this paper, as a consequence of the (modular 
form) hypergeometric identities (\ref{modularform2explicit}) 
(see  (\ref{Haupt}), (\ref{modularcurve}) and also (\ref{paradox})), 
we need $\, x$ to be identified 
with the {\em Hauptmodul} $\, 1728/j$. Consequently we introduce 
$\, X(q) \, = \, 1728 \cdot \, \tilde{X}(q)$
and $\, Q(x)  \, = \, \tilde{Q}(x/1728)$. With these appropriate
changes of variables one finds that the series (\ref{seriesmodcurve1a}) 
is nothing but  $\, \, \, \, X( e \cdot \, Q(x))$. 

Thus an interpretation of the one-parameter series 
(\ref{seriesmodcurve1a}) through the prism of the mirror map, is
that the  one-parameter series 
amounts to the multiplication of the nome of 
elliptic functions~\cite{Heegner} by an arbitrary complex
number $\, e$:   $\, q \, \, \longrightarrow \, \,  e \cdot \, q$.
The isogenies correspond to $\, \, q \, \longrightarrow \, \, q^N$
(resp. $\,\,  q \, \longrightarrow \, \, q^{1/N}$) for an integer $\, N$
and the one parameter families we have encountered 
(namely (\ref{seriesmodcurvea}),
(\ref{seriesmodcurve3a}))
correspond to the composition of  
$\,\,  q \, \, \longrightarrow \, \,  e \cdot \, q\, $
and  $\,\,  q \, \longrightarrow \, \, q^N$ 
(resp. $\,\,  q \, \longrightarrow \, \, q^{1/N}$),
namely   $\,\, q \, \, \longrightarrow \, \,  e \cdot \, q^N$ 
(resp. $\,\,  q \, \longrightarrow \, \, e \cdot \,  q^{1/N}$).

The series $\, X(q) \, = \, \,  1728 \cdot \, \tilde{X}(q)$
(with $\tilde{X}(q)$ given by (\ref{mirror})) is solution of 
the Schwarzian equation
\begin{eqnarray}
\label{Harnad11}
\hspace{-0.95in}&&  \quad \, 
\{X(q), \, q \} \, \,\,  -{{1} \over {2 \,  q ^2}} \,\,  \,  \,  
 + \, {{1} \over {72}} \,  \cdot \,
 {{ 32 \, X(q)^2 \, -41 \, X(q) \, +36} \over {
 X(q)^2 \cdot \, (1\, - X(q))^2 }} \cdot \, 
 \Bigl( {{ d X(q)} \over {d q}}  \Bigr)^2
 \, \, = \,\,  \, 0. 
\end{eqnarray}
which is nothing but:
\begin{eqnarray}
\label{Harnad111}
\hspace{-0.95in}&& \quad  \quad  \quad \quad  \quad 
\{X(q), \, q \} \, \, \,\,   -{{1} \over { 2 \, q ^2}} \, \, \,    \,
 - \, W(X(q))
\cdot \,  \Bigl( {{ d X(q)} \over {d q}}  \Bigr)^2
\, \, \, = \,\,\, \, 0. 
\end{eqnarray}
The series  $\, Q(x)  \, = \, \tilde{Q}(x/1728)$
(with $\,\tilde{Q}(x)$ given by (\ref{mirror2})) is solution 
of the Schwarzian equation
\begin{eqnarray}
\label{Harnad213}
\hspace{-0.95in}&& 
 -\, \{Q(x), \, x \} \, \, \,
 - {{1} \over {2  \cdot \, Q(x)^2 }} 
 \cdot  \Bigl({{ d Q(x)} \over {d x}} \Bigr)^2
\, + \, {{1} \over {72}}  \cdot \,
\Bigl( {\frac {32\, x^{2} \, -41 \, x +36}{
 x^{2} \cdot \, (1-\, x)^{2}}} \Bigr)
\, \, =  \, \, \, \,  \, 0,
\end{eqnarray}
equivalently written as:
\begin{eqnarray}
\label{Harnad214}
\hspace{-0.95in}&& \quad  \quad  \quad  \quad  \quad 
 \, \{Q(x), \, x \} \, \, \, \, 
 + {{1} \over {2  \cdot \, Q(x)^2 }} 
 \cdot  \Bigl({{ d Q(x)} \over {d x}} \Bigr)^2
\,  \, +  \, W(x) \, \,  \, =  \, \, \, \, \, 0.
\end{eqnarray}
The two mirror map series (\ref{mirror}), (\ref{mirror2}) 
thus correspond to differentially 
algebraic~\cite{Selected,IsTheFull} functions,
and are solutions of simple Schwarzian equations like in 
(\ref{condition1}).

These differentially algebraic mirror maps transformations 
$\, Q(x)$ and $\, X(q)$ are the well-suited changes of variables
such that the transformation $\, x \, \longrightarrow \, \, y(x)$ 
 verifying the Schwarzian equation (\ref{condition1}) become  
simple transformations, ``simple'' meaning
transformations like 
$\, q \,  \longrightarrow \, \, S(q) \, = \, \, e \cdot \, q^N$
(or $\, S(q) \, = \, \, e \cdot \, q^{1/N}$)
in the nome $\, q$ of elliptic functions~\cite{Heegner}. Generalizing 
the Koenig-Siegel 
linearization~\cite{Siegel,Siegel2,Almost,Milnor},  
we thus decompose $\, y(x)$ as  
$\,y(x) \, = \, \, X(S(Q(x)))$.
 
The Schwarzian conditions (\ref{Harnad111}), (\ref{Harnad214}) are 
essentially the well-known Schwarzian equation discovered 
by Jacobi~\cite{Jacobi,Nova} on the $\, j$-function (see for instance 
equation (1.26) in~\cite{HarnadHalphen}). The compatibility of the
Schwarzian equations (\ref{Harnad111}), (\ref{Harnad214})
on the mirror maps with the Schwarzian condition (\ref{condition1})
on $\, y(x)$ emerging from a more general Malgrange's pseudo-group 
perspective~\cite{Casale,Casale2,Casale3,Casale4}, is 
shown in \ref{Schwarzmirror}. The fact that the {\em same function} $\, W(x)$ 
occurs in the Schwarzian conditions (\ref{Harnad111}), (\ref{Harnad214}) 
on the mirror maps, and on the Schwarzian condition (\ref{condition1}), 
is crucial for this demonstration and is not a mere coincidence.

\subsection{The general case: $\, _2F_1([\alpha, \, \beta], \, [\gamma],x)$ hypergeometric function.} 
\label{general}

\subsubsection{The $\, _2F_1([1/6,1/3],[1],x)$ hypergeometric function. \\} 
\label{16131}

\vskip .1cm 
\vskip .1cm 
\vskip .1cm
 
We have analyzed in some detail  in section (\ref{Schwarzsimplest})
the modular form example (\ref{modularform2explicit}). For other values 
of the $\, [[\alpha, \, \beta], \, [\gamma]]\, $ parameters of 
the $\, _2F_1$ (see (\ref{modularform2}), (\ref{Gaussdiff}))
one can easily  find series expansions of the solution 
$\, y(x)$ of the Schwarzian condition. A set of values like 
$\, [[1/2, \, 1/2], \, [1]]$,
 $\, [[1/4, \, 1/4], \, [1]]$,
 $\, [[1/3, \, 1/3], \, [1]]$, $\, [[1/3, \, 2/3], \, [1]]$  
or  $\, [[1/3, \, 1/6], \, [1]]$ 
(see for instance~\cite{Christol,SuperMaier} and Ramanujan's cubic theory
of alternative bases~\cite{Canada}) which are known to yield  
modular form hypergeometric identities like (\ref{modularform2explicit})
with algebraic pullbacks $\, y(x)$ associated with modular equations. 
For these values of the  $\, [[\alpha, \, \beta], \, [\gamma]]\, $  
parameters one finds a set of one-parameter series totally similar to 
what is described in section (\ref{Schwarzsimplest}). The example of the
 $\, _2F_1([1/6,1/3],[1],x)$ hypergeometric function is sketched in \ref{16131app}.

\subsubsection{The general case: $\, _2F_1([\alpha, \, \beta], \, [\gamma],x) \, $ hypergeometric function. \\} 
\label{general2}

\vskip .1cm 
\vskip .1cm
\vskip .1cm
 
Let us now consider arbitrary parameters of the Gauss hypergeometric 
function $ \, [[\alpha, \, \beta], \, [\gamma]] \,$ that are not 
in the previous selected set, and are different from the 
cases given in sections (\ref{recalls}), (\ref{more2F1Heun}), 
and (\ref{more2F1higher}) corresponding to the rank-two condition.

\vskip .1cm
 
A simple calculation shows that one always finds a series of the form 
$\, e \cdot \, x \, + \, \cdots \, $ (like (\ref{seriesmodcurve1a}) 
or (\ref{361})), solution of the Schwarzian condition,  
{\em but it is only for} $\, \gamma \, = \, 1$ that series  
of the form $\, a \cdot \, x^2 \, + \, \cdots$,
$\, b \cdot \, x^3 \, + \, \cdots$, etc ... (like 
(\ref{seriesmodcurvea}) or (\ref{seriesmodcurve3a})) can be
solutions of the Schwarzian condition.

When $\, \gamma \, = \, 1 \,$ one gets the following series  
of the form $\, a \cdot \, x^2 \, + \, \cdots \, \,$ 
solution of the Schwarzian condition
\begin{eqnarray}
\label{cequalone}
\hspace{-0.96in}&&   \quad
y_2(u, \, x) \,\, = \, \,\, \,\,  a \cdot \, {x}^{2}
 \, \,\,\,  
-2 \, a \cdot \, (2\,\alpha \beta \, -\alpha -\beta) \cdot \,  {x}^{3}
\,  \,\,+ \, \, {{a} \over {2}} \cdot \, C_4 \cdot   {x}^{4} \,\, \, 
   + \,\,  \cdots  \quad \quad \hbox{with:} \
\nonumber  \\
\hspace{-0.96in}&& \,  \, \,     
 C_4 \, = \, \,\, \,
 2\,\,(2\,\alpha \beta \, -\alpha -\beta) \cdot a \, \,
+ (\alpha \beta \,  -1)  (\alpha \beta \, -\alpha -\beta)  \, 
+\, \,  5\,\, (2\,\alpha \beta\, -\alpha -\beta)^{2}, 
\end{eqnarray}
and one also gets the following series  
of the form $\,\, b \cdot \, x^3 \, + \, \cdots \, \,\,$ 
solution of the Schwarzian condition
\begin{eqnarray}
\label{cequalone}
\hspace{-0.95in}&&   \, \,   \quad  \quad 
y_3(v, \, x) \, \, = \, \,\, \,\,  b \cdot \, {x}^{3}
 \, \,\,\,  -3 \, b \cdot \, (2\,\alpha \beta\, -\alpha -\beta) \cdot \,  {x}^{4} \,\,
 \\
\hspace{-0.95in}&&  \quad  \quad  \quad  \quad  \,  \, \,     \,  \, \,     
\, \, +\,{{3 \, b} \over {4}} \cdot \, 
 \Bigl( (\alpha \beta \, -1) \cdot \,   (\alpha \beta\, -\alpha -\beta)  \, 
+7 \,  \,   (2\,\alpha \beta\, -\alpha -\beta)^{2} \Bigr) \cdot   {x}^{5} 
 \, \, \, \, + \,\, \cdots \nonumber 
\end{eqnarray}
together with the one-parameter family of commuting series  
of the form $\, e \cdot \, x \, +  \cdots \, $
\begin{eqnarray}
\label{cequalone}
\hspace{-0.95in}&&   
y_1(e, \, x) \, = \, \, e \cdot \, x
 \,\, 
+ e \cdot \,  \, (e\, -1) \cdot \, 
(2\,\alpha \beta\, -\alpha -\beta) \cdot \,  {x}^{2} 
\,  \,\, +\,{{ e \cdot \, (e-1)} \over {4}} 
\cdot  \, C_3 \cdot   {x}^{3} \, \, 
   + \,\,  \cdots 
\nonumber  \\
\hspace{-0.95in}&&  \hbox{with:}  \quad  \quad  
C_3 \, = \, \, 
(\alpha \beta   -1)  (\alpha \beta -\alpha -\beta)  \cdot \, (e\, +1) 
\, + (2\,\alpha \beta -\alpha -\beta)^{2}  \cdot \, (5\, e -3).
\end{eqnarray}
Again one has the equalities 
\begin{eqnarray}
\label{suchsuch}
\hspace{-0.95in}&&   \quad   \quad  \quad 
 y_1(e, \,y_2(a, \, x)) \, = \, \, y_2(a\, e, \, x), 
\quad   \quad  \quad 
 y_2(a, \,y_1(e, \, x)) \, = \, \, y_2(a \, e^2, \, x),
\nonumber \\
\hspace{-0.95in}&&   \quad   \quad  \quad 
 y_1(e, \,y_3(b, \, x)) \, = \, \, y_3(b \, e, \, x), 
\quad   \quad  \quad 
 y_3(b, \,y_1(e, \, x)) \, = \, \, y_3(b \, e^3, \, x),
\end{eqnarray}
and, again, the two series $\, y_2(a, \, x)$ and $\, y_3(b, \, x)$ 
commute for $\, b \, = \, a^2$. As far as series analysis is concerned we 
have {\em exactly the same structure} (\ref{suchsuch}) as the one 
previously described (see (\ref{Schwarzsimplest}) and (\ref{16131})) 
where {\em modular correspondences}~\cite{Goro} take place.
However, it is not clear if such one-parameter series
can reduce to algebraic functions for some selected 
values of the parameter $\, a, \, b, \cdots \, $ In other words, 
are these series modular correspondences, or are they just 
``similar'' to modular correspondences ?
{\em The question of the reduction of these Schwarzian conditions to 
modular correspondences remains an open question}.

\vskip .1cm

When $ \, \gamma \, \ne \, 1$ the situation is drastically 
different\footnote[2]{Recall that {\em globally bounded} $\, _nF_{n-1}$ 
series of ``weight zero''~\cite{Heckman} (no ``down'' parameter 
is equal to $1$ or to an integrer, 
i.e. in the case of globally bounded $\, _2F_1$
series,  $\, \gamma \, \, $ is 
different from an integer), are {\em algebraic functions}.}: one does
not have solution of the Schwarzian equation of the form 
$\, a \cdot \, x^2 \, + \, \cdots \,\, $ 
or $\, b \cdot \, x^3 \, + \, \cdots \,\, $ etc ...
One only has a one-parameter family of commuting series:
\begin{eqnarray}
\label{357}
\hspace{-0.95in}&& \, \, \,  \,      
y(e, \, x) \, \, \, = \, \, \,  \, 
e \cdot \, x   \, \,\, \,
 - \, e  \cdot \, (e\, -1) \cdot \,
 {{ \gamma^2 \, -(\alpha+\beta+1)\cdot \, \gamma \, +2\, \alpha \, \beta } \over { 
\gamma \cdot \, (\gamma\,-2) }} \cdot \, x^2 
\, \,  \, \,  +\, \, \cdots 
\end{eqnarray}
Again, it is not clear to see if such a one-parameter series
can reduce to algebraic functions for some selected 
values of the parameter $\, e$.

\vskip .1cm
\vskip .1cm
 
\section{Rank-two condition on the rational transformations as  a subcase of the Schwarzian condition} 
\label{subcase}

\vskip .1cm 

\subsection{Preliminary result: factorization 
of the  order-two  linear differential operator}
\label{prelim}

When 
\begin{eqnarray}
\label{Bx}
\hspace{-0.95in}&& \quad \quad  \quad \quad \quad \quad 
B(x) \, \, = \, \, \, \, {{C(x)} \over {4}} \cdot \, (2\,A(x) \, -C(x))
 \, \,\, + \, \, {{1} \over {2}} \cdot \, C'(x), 
\end{eqnarray}
the second order linear differential operator 
\begin{eqnarray}
\label{Omeg}
\hspace{-0.95in}&& \quad \quad  \quad \quad  \quad \quad 
 \Omega \,\,  = \, \, \, \,  \,  
D_x^2    \,\,  \,  + \, A(x) \cdot \,  D_x \,  \, + \, B(x), 
\end{eqnarray}
factorizes as follows:
\begin{eqnarray}
\label{OmegFact}
\hspace{-0.95in}&& \quad \quad  \quad \quad  \quad \quad 
 \Omega \,\,  = \, \, \, \,  \,  
\Bigl(D_x    \, + \, A(x) \, -{{C(x)} \over {2}}\Bigr)
 \cdot  \, \Bigl(D_x \, +\, {{C(x)} \over {2}}\Bigr).
\end{eqnarray}
Let us assume that $\, C(x)$ is a log-derivative:
\begin{eqnarray}
\label{Assume}
\hspace{-0.95in}&& \quad \quad  \quad \quad  \quad \quad 
 C(x)  \,\,  = \, \, \, \,  \, 
 2 \cdot \, {{ d \ln(\rho(x))} \over {dx}}, 
\end{eqnarray}
one immediately finds that a conjugation of 
(\ref{OmegFact}) factors as follows:
\begin{eqnarray}
\label{Immediat}
\hspace{-0.95in}&& \quad \quad  \quad \quad  \quad \quad 
\rho(x)  \cdot \, \Omega  \cdot \, {{1} \over {\rho(x)}}
\,\,\,  = \, \, \, \,  \,  
\Bigl(D_x    \, + \, A(x) \, - C(x)\Bigr)  \cdot  \, D_x.
\end{eqnarray}
Therefore the $\, A_R(x)$ in the rank-two condition 
(\ref{mad}) is not the $\, A(x)$ in (\ref{Omeg}) but 
$\, A_R(x) \,\, = \,\, A(x) \, - C(x) \,\, $ 
where $\, B(x)$ is of the form (\ref{Bx}).

The  rank-two  condition reads:
\begin{eqnarray}
\label{rotaA}
\hspace{-0.95in}&& \quad \quad \quad 
 y''(x) \,\,   \,   = \,\,  \,  
(A(y(x)) \, - \, C(y(x))) \cdot  \, y'(x)^2 \, \,  
 - \,  (A(x) \, - \, C(x)) \cdot \,  y'(x), 
\end{eqnarray}
to be compared with the Schwarzian condition
\begin{eqnarray}
\label{condition1n}
\hspace{-0.95in}&& \quad   \quad \quad  \quad \quad  \quad \quad  
 W(x) 
\, \, \,  \,-W(y(x)) \cdot  \, y'(x)^2
\, \, \,  \,+ \,  \{ y(x), \, x\} 
\, \,\, \, = \,\, \, \,  \, 0, 
\end{eqnarray}
where:
\begin{eqnarray}
\label{wherecondn}
\hspace{-0.95in}&& \quad   \quad \quad  \quad \quad  \quad \quad  
W(x)  \, \, = \, \,  \, \, \,  A'(x) \, \, \,  + \, \, 
   {{A(x)^2} \over {2 }} \, \,  \, \,  -2 \cdot \, B(x).
\end{eqnarray}
{\bf Remark:} For a general Gauss hypergeometric function 
$\, _2F_1([\alpha, \, \beta], \, [\gamma], \,  x)$,
$\, A(x)$ and $\, B(x)$ are given by (\ref{Gaussdiff}). The 
factorization condition (\ref{Bx}) can be satisfied 
only for selected values of the $[[\alpha, \, \beta], \, [\gamma]]$ 
parameters\footnote[1]{For these conditions on the parameters the function $\, C(x)$
read respectively $\, C(x)=\, 2\,\alpha/x$, $\, C(x)=\, 2\,\beta/x$,  
$\, C(x)=\,  2\,(\beta\,x -\gamma+1)/x/(x-1)$,  
$\, C(x)=\, 2\,(\alpha\,x-\gamma+1)/x/(x-1)$, 
$\, C(x)=\, 2\,\alpha/(x-1)$, $\, C(x)=\,2\, \gamma\, \beta/x/(x-1)$,
 $\, C(x)=0$, $\, C(x)=0$.}: 
$\, \gamma= \, \alpha\, +1$,  $\, \gamma= \, \beta\, +1$, $\, \gamma= \,1$, 
$\, \beta= \,1$, $\gamma \, = \, \beta$, 
$ \, \gamma \, = \, \alpha$, $ \,\alpha=0 \,$ and $\, \beta=0$.

\vskip .1cm 

\subsection{Condition on the rational transformation as a subcase of the Schwarzian condition} 
\label{subcase}

Let us assume that the rank-two condition  (\ref{rotaA}) 
is satisfied, then we can use 
it to express the second derivative $\, y''(x)$
in terms of  $\, y(x)$ and the first derivative the  $\, y'(x)$.
One finds that the Schwarzian condition (\ref{condition1n}) 
is automatically 
verified provided $\, A(x)$, $\, B(x)$, $\, C(x)$ are related
though the condition  (\ref{Bx}) which amounts to
a factorization condition for the second order linear differential
operator (\ref{Omeg}).
The  $\, A_R(x)$ in the rank-two condition (see (\ref{mad})):
\begin{eqnarray}
\label{rota}
\hspace{-0.95in}&& \quad \quad  \quad  \quad  \quad  
  y''(x) \,\,     = \,\,  \,  
A_R(y(x)) \cdot  \, y'(x)^2  \, \,     - \,  A_R(x) \cdot \,  y'(x), 
\end{eqnarray}
is nothing but  $\, A_R(x) \, = \, \, A(x) \, - \, C(x)$,
or after rearranging $\, A(x) \, = \, \, A_R(x) \, +C(x)$. Now 
substituting (\ref{Bx}) in (\ref{wherecondn}) one gets:
\begin{eqnarray}
\label{wherecondn2before}
\hspace{-0.95in}&& \quad   \quad \quad  \quad \quad 
W(x) \,\, =  \,\, \,\, A'_{R}(x)\, \, \, + \frac{A(x)^2}{2} 
\,\, - C(x) \, A(x) \,\, \, - \frac{C^2(x)}{2},
\end{eqnarray}
with the last three terms being equivalent to $ \, A_{R}(x)^2$. Thus one finds 
that $\, W(x)$ is {\em only a function of} $\, A_R(x)$:
\begin{eqnarray}
\label{wherecondn2}
\hspace{-0.95in}&& \quad   \quad \quad  \quad \quad  \quad \quad  \quad 
W(x)  \, \, = \, \,\,  \, \,   A'_R(x) 
 \, \, \,  + \, \,   {{A_R(x)^2} \over {2 }}.
\end{eqnarray}
With this expression (\ref{wherecondn2}) of $\, W(x)$
the Schwarzian condition reads:
\begin{eqnarray}
\label{condition1nbis}
\hspace{-0.95in}&& \quad  \quad  \quad \quad  \quad \quad  
 W(x) 
\, \, \,  \,-W(y(x)) \cdot  \, y'(x)^2
\, \, \,  \,+ \,  \{ y(x), \, x\} 
\, \,\, \, = \,\, \, \,  \, 0, 
\end{eqnarray}
In order to see the compatibility of the rank-two condition (\ref{rota}) 
with the Schwarzian condition (\ref{condition1nbis}) when  the 
function $\, W(x)$ is given by (\ref{wherecondn2}),  
let us rewrite the rank-two condition (\ref{rota}) as
\begin{eqnarray}
\label{Rotarewr}
\hspace{-0.95in}&& \quad \quad  \quad    \quad \quad  \quad \quad  \quad 
{{y''(x)} \over {y'(x)}}
 \, \, = \, \,  \, \, A_R(y(x))  \cdot \, y'(x)  \,\,\, -A_R(x). 
\end{eqnarray}
Using (\ref{Rotarewr}), one can rewrite the Schwarzian 
derivative as
\begin{eqnarray}
\label{Schwarwrbefore}
\hspace{-0.95in}&& \quad   \quad   \quad   \quad  \quad   \quad    
\{y(x), \, x\}  \, \, = \, \,  \, \, 
{{d } \over { dx}} \Bigl({{y''(x)} \over {y'(x)}}\Bigr)
 \,\, -{{1} \over {2}} \cdot \, \Bigl({{y''(x)} \over {y'(x)}}\Bigr)^2
\end{eqnarray}
as $\,\,  W(y(x)) \cdot  \, y'(x)^2 \,\,\, -W(x)\, + \, \Delta$,  where $\, W(x)$ 
is given by (\ref{wherecondn2}),  
and where $\, \Delta$ is given by:  
\begin{eqnarray}
\label{Schwarwr}
\hspace{-0.95in}&& \,  \, 
\Delta \, \,  \, = \, \, \, 
 A_R(y(x))  \cdot \, y''(x) \, \,  - \, A_R(y(x))^2  \cdot \, y'(x)^2   \, \, 
  \, + \, A_R(x) \cdot \, A_R(y(x)) \cdot \, y'(x).
\end{eqnarray}
Note that  $\, \Delta$ is clearly zero when the rank-two condition 
is fulfilled.  This shows that the Schwarzian condition (\ref{condition1nbis}) 
when  the function $\, W(x)$ is given by (\ref{wherecondn2}), actually reduces to
the rank-two condition (\ref{rota}), as expected.

\vskip .1cm 

\vskip .1cm 

{\bf Remark:} The Heun function case of section (\ref{moreHeun})
was a case where the rank-two  condition was verified with
 $\, A_R(x)$ given by (\ref{AadoublingAa}). One also verifies that
the rational transformation  (\ref{Aadoubling}), and more generally 
the rational transformations $\, R_p(x)$ 
(pullbacks on the Heun function, see (\ref{Fdoublingidefirst})), 
are solutions of a Schwarzian equation (\ref{condition1nbis}) 
with $\, W(x)$ deduced from (\ref{wherecondn2}) with $\, A_R(x)$ 
given by (\ref{AadoublingAa}), 
namely:
\begin{eqnarray}
\label{AadoublingbisW}
\hspace{-0.95in}&&  
W(x) \, \, = \, \,  \, -{{3} \over {8 \cdot \, (x-M)^2}} 
\,  \, 
-{{1} \over {4}} \cdot \, {\frac {2\,x  \, -1}{ (M \, -x) \cdot \, x \cdot \, (x-1) }}
\, \, 
-{{1} \over {8}} \cdot \, {\frac {4\,{x}^{2}-4\,x+3}{ x^{2} \cdot \, (x-1)^{2}}}.
\end{eqnarray}

\vskip .1cm
 
In the previous case where the rank-two  condition can be seen as a subcase of the 
Schwarzian condition  (\ref{condition1nbis}) on $\, y(x)$, it is tempting to imagine, 
in a Koenig-Siegel linearization perspective, that the differentially algebraic 
function $\, Q(x)$ (see (\ref{QDAfirst})) also verifies a Schwarzian condition
similar to the Schwarzian condition  (\ref{Harnad214}) 
on $\, Q(x)$ now seen as a mirror map and we show in \ref{subcase} 
that this is actually the case.

\vskip .1cm

\section{Schwarzian condition for generalized hypergeometric functions} 
\label{Schwarzgeneralized}

\vskip .1cm
 
\subsection{Schwarzian condition and $\, _3F_2$ hypergeometric identities} 
\label{Schwarz3F2}

Generalizing the modular form identity considered in section (\ref{int}),
let us seek a $\, _3F_2$ hypergeometric identity of the form 
\begin{eqnarray}
\label{modularform3F2}
\hspace{-0.95in}&& \quad \quad  \quad \quad \quad 
  {\cal A}(x) \cdot \, _3F_2\Bigl([a, \, b,  \,  c], \, [d, \, e], \,  x  \Bigr)
\, = \, \, \, \,\,  
 _3F_2\Bigl([a, \, b,  \,  c], \, [d, \, e], \, y(x)  \Bigr), 
\end{eqnarray}
where $\, {\cal A}(x)$ is an algebraic function.
Similarly to what has been performed in section (\ref{int}), 
we consider the two order-three linear differential operators 
associated respectively to the LHS and RHS of 
(\ref{modularform3F2}).

A  straightforward calculation 
enables us to find (from the equality of the wronskians of these 
two operators) the algebraic function $\, {\cal A}(x)$
in terms of the algebraic function pullback $\, y(x)$
in (\ref{modularform3F2}):
\begin{eqnarray}
\label{modularform3F2calA}
\hspace{-0.95in}&& \quad   \quad  \quad \quad \quad \quad
{\cal A}(x)  \, \, = \, \,  \,  
 \Bigl( {{ y(x)^{\eta} \cdot \, 
(1 \, -y(x))^{\nu} } \over {
 x^{\eta} \cdot \, (1 \, -x)^{\nu} }} \Bigr) 
\cdot \,    \Bigl( {{d y(x)} \over {dx }} \Bigr)^{-1}, 
\nonumber \\
\hspace{-0.95in}&&     \quad \quad \quad \quad \quad
\quad 
\eta \, \, = \, \,  \,  {{d+e+1} \over {3}}, 
 \quad \quad \quad \quad
 \nu \, \, = \, \,  \,  {{a+b+c+2-d-e} \over { 3}},  
\end{eqnarray}
The identification 
of the $\, D_x$ coefficients of these two linear differential 
operators, gives (beyond (\ref{modularform3F2calA})) 
a first condition that can be rewritten 
 in the following Schwarzian form:
\begin{eqnarray}
\label{condition3F2}
\hspace{-0.95in}&& \quad  \quad \quad \quad \quad \quad  \quad 
 W(x) 
\, \, \,  \,-W(y(x)) \cdot  \, y'(x)^2
\, \, \,  \,+ \,  \{ y(x), \, x\} 
\, \,\, \, = \,\, \, \,  \, 0, 
\end{eqnarray}
where $\, W(x)$ reads:
\begin{eqnarray}
\label{condition3F2W}
\hspace{-0.95in}&& \quad    \quad \quad  \quad  \quad 
 W(x) \,\, \, = \,\, \, \,
 {{1} \over {6}} \cdot \, {{P_W(x)}\over { x^2 \cdot \, (1\, -x)^2}}, 
\quad \quad \quad \quad \quad  \quad \hbox{where:} 
\\
\hspace{-0.95in}&&    
P_W(x)\,\, \, = \,\, \, 
({a}^{2} \, +{b}^{2} \, +{c}^{2} \, - a b \, - \, a c \, \, -bc  \, -3) \cdot \,  {x}^{2}
\nonumber \\
\hspace{-0.95in}&&    
 + \, ( 3\,  (a  b \, +\,  a c \, +\,bc  \, +\,de \, +1)
-2\,  (a  d \, +\, a e \, \, +\,bd \, +\,be \,+\,cd \,+\,ce) \,  \, +a+b+c) \cdot \, x
\nonumber \\
\hspace{-0.95in}&& \quad   \quad   
\, \, +{d}^{2}\, +{e}^{2} \, -de  \, -d \, -e \, -2. 
\end{eqnarray}
The identification 
of the coefficients with no $\, D_x$ of these two linear differential 
operators gives a second condition where the fourth derivative of 
$\, y(x)$ takes place. The analysis of this set of conditions 
corresponds to tedious but straightforward differential algebra 
calculations which are performed in \ref{Reduc3F2}. 

One finds that all the 
conditions on the parameters $\, a, \, b, \, c, \, d, \, e$ 
of the $\, _3F_2$ hypergeometric function associated with 
$\, Q(x)  \, = \, \, \, 0$,
correspond to cases where the order-three operator 
is the symmetric square of a second order operator having 
$\, _2F_1$ solutions. In other words this situation correspond to
 the {\em Clausen identity}, the $\, _3F_2$ hypergeometric function
reducing to the square of a $\, _2F_1$ hypergeometric function:
\begin{eqnarray}
\label{Clausen}
\hspace{-0.95in}&& \quad \quad \quad \quad \quad \quad \quad 
 _3F_2\Bigl([2\,a, \, a\, +b, \, 2\, b], \, 
[a\, +b\, +{{1} \over {2}}, \, 2\,a \, +2 \, b], \, x  \Bigr) 
\nonumber \\
\hspace{-0.95in}&& \quad \quad \quad \quad  
\quad \quad \quad \quad \quad \quad \, \, \, 
\, = \, \, \, 
 _2F_1\Bigl([a, \, b], \, [a\, +b\, +{{1} \over {2}}], \, y(x)  \Bigr)^2.
\end{eqnarray}
In that Clausen identity case, the Schwarzian condition (\ref{condition3F2}) 
we found for the $\, _3F_2$ is nothing but the Schwarzian condition 
on the underlying $\, _2F_1$. 

\vskip .1cm

\subsubsection{The intriguing $\, _3F_2([1/9,4/9,5/9],[1/3,1],x)$ case \\} 
\label{intring}

\vskip .1cm
\vskip .1cm
\vskip .1cm

Beyond the trivial transformation $\, y(x) \, = \, \, x$
one hopes to find a condition (\ref{modularform3F2})
where the pullback $\, y \, = \, \, y(x)$ is an algebraic function. 

For the intriguing hypergeometric function 
$\, _3F_2([1/9,4/9,5/9],[1/3,1],x)$,  known to be a globally 
bounded\footnote[5]{The series 
$ \, _3F_2([1/9,4/9,5/9],[1/3,1], \, 3^5 \, x) \, $ 
is a series with {\em integer coefficients}~\cite{Christol}.} 
series~\cite{Christol}, one does not know if it is the 
{\em diagonal of a rational function},
or not. It is natural to apply the previous conditions 
to see if we could have an identity like (\ref{modularform3F2})  generalizing 
the identities one gets for modular forms. The occurrence of a series with
{\em integer coefficients} is a strong argument for a 
``modular form interpretation'' of this intriguing $\, _3F_2$ 
hypergeometric function. Therefore, it is tempting 
to imagine that a remarkable identity like (\ref{modularform3F2}) exists
for this $\, _3F_2$ hypergeometric function.

The corresponding order-three 
operator has a differential Galois group that is an 
extension\footnote[1]{See the Boucher-Weil criterion~\cite{Boucher}. The 
symmetric square and exterior square of a normalized order-three 
operator has no rational solutions. One sees also clearly that 
this order-three operator is not homomorphic to its adjoint.} of 
$\, SL(3, \, \mathbb{C})$. Therefore,
this operator cannot be homomorphic to the symmetric square 
of an order-two operator: {\em an identity of the Clausen type
is thus excluded for this $\, _3F_2$ hypergeometric function}. The previous 
calculations showing that an identity like (\ref{modularform3F2})
exists only when the $\, _3F_2$ hypergeometric function reduces to 
square of $\, _2F_1$ hypergeometric functions discards an identity 
like (\ref{modularform3F2}) for $\, _3F_2([1/9,4/9,5/9],[1/3,1],x)$.
This is easily seen: for 
this hypergeometric function the ``invariant'' 
$\,{\cal I}(x) \, = \, \, {\cal I}y(x)$ (see \ref{Reduc3F2}), and  
the rational function $\, W(x)$ in the Schwarzian condition 
read respectively
\begin{eqnarray}
\label{invariant3F2}
\hspace{-0.95in}&& 
{\cal I}(x) = \, \, 
 {{p_8^3} \over {(140\,{x}^{3} +81\,{x}^{2}+2403\,x-864)^8 }}, 
\quad 
W(x)  = \, \, 
-{\frac {230\,{x}^{2}-261\,x+207}{486\,{x}^{2} \left( x-1 \right) ^{2}}}, 
\end{eqnarray}
where:
\begin{eqnarray}
\label{invariant3F2more}
\hspace{-0.95in}&& \quad    \,  
p_8 \, \, = \, \, \, 254800\,{x}^{8}\,  +7247520\,{x}^{7}\,
+223006266\,{x}^{6} \,\, -533339127\,{x}^{5}\,\, -62800191\,{x}^{4}
\nonumber \\
\hspace{-0.95in}&& \quad \quad  \quad  \quad \quad  
 +1082145339\,{x}^{3}\,\, -244855791\,{x}^{2}\, \, -290993472\,x \,\, +26873856.
\end{eqnarray}
Reinjecting the invariance condition $\,{\cal I}(x) \, = \, \, {\cal I}y(x)$  
with (\ref{invariant3F2})
in the Schwarzian condition (\ref{condition3F2}), one finds that there 
is no (algebraic) solution $\, y(x)$ except the trivial solution 
$\,\, y(x) \, = \, \, x$.

\vskip .1cm   
 
\subsection{Schwarzian condition and other generalized hypergeometric functions} 
\label{Schwarz4F3}

\vskip .1cm

In \ref{Schwarz4F32F2} we seek an identity of the form 
(\ref{modularform3F2}) but where the $\, _3F_2$ hypergeometric function 
is replaced by a $\, _4F_3$ hypergeometric function known to correspond 
to a {\em Calabi-Yau ODE}~\cite{IsingCalabi,IsingCalabi2}, 
or a  hypergeometric function with {\em irregular} singularities 
namely a simple $\, _2F_2$ hypergeometric function. One finds, unfortunately, 
that the only solution, for these two examples sketched respectively 
in \ref{Schwarz4F3} and \ref{Schwarz2F2}, is the trivial solution
$\, y(x) \, = \, \, x$. Keeping in mind the non trivial results previously 
obtained on a Heun function, or on a $\, _2F_1$ hypergeometric function 
associated with a higher genus curve, these two negative results should
rather be seen as an incentive to find more non trivial examples of these
extremely rich and deep Schwarzian equations.   

\vskip .1cm
\vskip .1cm 
\vskip .1cm
 
\section{Conclusion} 
\label{Conclusion}

In this paper we focus essentially on identities relating the same 
hypergeometric function with two different algebraic pullback 
transformations related by modular equations. This corresponds
to the modular forms that emerged so many 
times in physics~\cite{IsingCalabi,IsingCalabi2,Christol}: these  
algebraic transformations can be seen as  simple illustrations 
of exact representations of the renormalization group~\cite{Hindawi}.  
Malgrange's pseudo-group approach aims at
generalizing differential Galois theory to non-linear differential 
equations. In his analysis of Malgrange's pseudo-group  Casale found two 
non-linear differential equations (\ref{cas2}) and (\ref{Casale}) 
yet these two conditions were presented separately with no explicit link.
In a previous paper~\cite{Hindawi}, where we  gave simple 
examples of exact representations of the renormalization group, 
associated with selected linear differential operators covariant by 
rational pullbacks, we found simple exact examples 
of Casale's condition (\ref{cas2}). Building on this work we revisited 
these previous examples and provided non-trivial new examples 
associated with a Heun function and a $ \, _2F_1$ hypergeometric function
associated with higher genus curves. Then we instantiated, for the first 
time, Casale's second condition (\ref{Casale}) with the examples 
given in section (\ref{Schwarz}).  Furthermore we found that
Casale's condition (\ref{cas2}) can be seen as a subcase of the
Schwarzian condition (\ref{Casale}), corresponding to a factorization 
of a linear differential operator $\, \Omega$.
Seemingly, this Schwarzian condition  (\ref{Casale}) is seen 
to ``encapsulate'' in one differentially algebraic (Schwarzian) 
equation, all the {\em modular forms} 
and {\em modular equations} of the theory of elliptic curves. 
The Schwarzian condition (\ref{Casale}) can thus be seen as some 
quite fascinating ``pandora box'', which encapsulates an infinite number of 
highly remarkable modular equations, and a whole ``universe'' of  
{\em Belyi-maps}\footnote[2]{Belyi-maps~\cite{Belyi3,Belyi,Belyi2,Belyi4,Belyi5}
 are central to Grothendieck's program 
of ``dessins d'enfants''.}.  Furthermore we found, only when 
$\, \gamma\, = \, 1$, that one-parameter series starting with quadratic, 
cubic, or higher order terms satisfy the rank-three condition. The 
question of a modular correspondence interpretation of these series 
is an open question.

Recalling the two previous higher-genus and Heun 
examples, it is important to underline that these conditions (\ref{cas2})
and (\ref{Casale}) are  actually richer than just elliptic curves, and 
go beyond  ``simple'' restriction to $\, _2F_1$ hypergeometric functions.

This paper provides a simple and pedagogical illustration of 
such exact non-linear symmmetries in physics (exact representations of the 
renormalization group transformations like the Landen transformation for 
the square Ising model, ...) and is a strong incentive to discover more 
differentially algebraic equations involving fundamental symmetries, 
developping more differentially algebraic analysis 
in physics~\cite{Selected,IsTheFull}, beyond obvious candidates 
like the full susceptibility of the 
square-lattice Ising model~\cite{IsTheFull,Automat}.

\vskip .1cm

\vskip .4cm 

\vskip .4cm 

{\bf Acknowledgments:} 
We would like to thank
 S. Boukraa, G. Casale, S. Hassani, E. Paul, C. Penson 
and J-A. Weil for very fruitful 
discussions. We would like to thank the anonymous referee 
for his very careful reading of our manuscript and his
valuable suggestions and corrections.
 This work has been performed 
without any ERC, ANR, PES or MAE financial support. 

\vskip .5cm
\vskip .5cm

\appendix

\section{$\, _2F_1$ hypergeometric example: $\, N\, = \, 3$ } 
\label{more2F1examples}

\vskip .2cm 

Recalling Vidunas paper~\cite{Vidunas} one introduces the 
following hypergeometric function:
\begin{eqnarray}
\label{YM3}
\hspace{-0.95in}&& \quad \quad \quad  \quad  \quad  \quad  
Y(x) \, \, = \, \, \, 
x^{1/3} \cdot \, 
_2F_1\Bigl([{{1} \over {3}}, \, {{2} \over {3}}], \, [{{4} \over {3}}], \, x\Bigr), 
\end{eqnarray}
for which one has the following exact expressions 
for $\, A_R(x)$, $\, u(x)$ and $\, R(x)$:
\begin{eqnarray}
\label{Aa3}
\hspace{-0.96in}&&  \quad  \, 
A_R(x) \, \, = \, \, \, 
{{2} \over {3 }} \cdot \, 
 {{2 \, x \, -1 } \over { x \cdot \, (x \, -1)}}\, \, = \, \, \, {{u'(x)} \over {u(x)}}, 
\,  \, \, \quad \,  \, \hbox{where:} \quad \quad \, \, \, 
u(x) \,  \, = \, \, \, x^{2/3} \cdot \, (1-x)^{2/3}, 
\nonumber \\
\hspace{-0.96in}&&  \quad \quad  \quad \quad \quad \quad \quad 
R(x) \,\, = \, \, \,
 {{x \cdot \, (x  \, -2)^3} \over { (1\, -2 \, x)^3}}.
\end{eqnarray}
One verifies that 
$\, Q(x)\, = \, \, Y(x)^3$:
\begin{eqnarray}
\hspace{-0.95in}&& 
{{d  Q(x)} \over { dx}}/Q(x)
 \,\,  = \, \, \,  3 \cdot \, {{d  Y(x)} \over { dx}}/Y(x)
\,\,  = \, \, \,  {{1} \over {F(x)}}, 
\quad \hbox{where:} \quad F(x) \,\,  = \, \, \, 
u(x) \cdot Y(x). 
\end{eqnarray}
One has the identity:
\begin{eqnarray}
\label{Q8}
\hspace{-0.95in}&& \quad \quad \quad \quad  \quad \quad 
Q(R(x)) \,\, = \, \,\,\,  -8   \cdot \,  Q(x) \, = \, \, \,  \, 
-8 \cdot \, x \, \cdot \,
 _2F_1\Bigl([{{1} \over {3}}, \, {{2} \over {3}}], \, [{{4} \over {3}}], \, x\Bigr)^3
\nonumber \\
\hspace{-0.95in}&& \quad \quad \quad \quad  \quad \quad \quad \quad \,  \, 
\,\,  = \, \, \,  \, 
{{x \cdot \, (x  \, -2)^3} \over { (1\, -2 \, z)^3}} \cdot \, 
_2F_1\Bigl([{{1} \over {3}}, \, {{2} \over {3}}], \, [{{4} \over {3}}], \, 
{{x \cdot \, (x  \, -2)^3} \over { (1\, -2 \, x)^3}}\Bigr)^3. 
\nonumber
\end{eqnarray}
The rational function\footnote[1]{Note a typo in~\cite{Vidunas}: the
$\, R(x)$ in equation (64) of~\cite{Vidunas} is $\, -R(x)$. }:
\begin{eqnarray}
\label{other}
\hspace{-0.95in}&& \quad \quad 
 \quad  \quad \quad  \quad  \quad \quad 
\tilde{R}(x) \, = \, \, 
{{ 27 \, x \cdot \, (1\, -x) \, (1 \, -x \, +x^2)^3} \over {
(1\, +3\, x -6\, x^2\,  +\, x^3)^3 }}, 
\end{eqnarray}
commutes with $\, R(x)$ given by (\ref{Aa3}). Also note that 
$\, R(x)$ given by (\ref{Aa3}) commutes with the two known symmetries
of the hypergeometric function, namely 
$\,\, R(x) \, = \, \, 1 \, -x \,$  and  $\, R(x) \, = \, \, 1/x$.
These last two transformations yield the involution 
$\, R(x) \, = \, -x/(1\, -x)\, $ which commutes with the two previous 
rational transformations (\ref{Aa3}), (\ref{other}), 
and corresponds to  $\, Q(-x/(1\, -x)) \, = \, Q(x)$.

The composition of  $\, R(x) \, = \, -x/(1\, -x)\,\,  $
with (\ref{Aa3}) and  (\ref{other}) gives respectively:
\begin{eqnarray}
\label{other2}
\hspace{-0.95in}&& \quad \quad  \quad  \quad \quad 
 {{x \cdot \, (2\, -x)} \over {(1\, -x) \cdot \, (1\, +x)^3 }}, 
\quad  \quad \quad 
{{ -27 \, x \cdot \, (1\, -x) \, (1 \, -x \, +x^2)^3} \over {
(1\, -6\, x +3 \, x^2\,  +\, x^3)^3 }}. 
\end{eqnarray}
Note that $\,\, R(x) \,= \, 1/x\,$ and  $\,\, R(x) \,= \, 1\, -x\,$
also verify the rank-two  condition.

As we can see, the one-parameter family of solution of 
\begin{eqnarray}
\label{mad2one}
\hspace{-0.95in}&& \quad \quad  \quad \quad 
\Bigl({{ d R(a, \, x)} \over {dx}}\Bigr)^2 \cdot A(R(a, \,x)) 
\,\,     = \,\,  \,  \,\,
    {{ d R(a, \,x)} \over {dx}} \cdot A(x) \, \, \,  
 + {{ d^2 R(a, \,x)} \over {dx^2}}, 
\end{eqnarray}
namely the differentially algebraic series 
\begin{eqnarray}
\label{mad2oneseries}
\hspace{-0.95in}&& \,   
R(a, \, x) \,\,     = \,\,  \,  \,\,\,
a \cdot \, x \, \, \,\,\,
 - {{1} \over {2}} \,a \cdot \, (a-1) \cdot \, {x}^{2} \, \, \, \, 
+ {{1} \over {28}}\,a \cdot \, (a-1) \cdot  \, (5\,a-9) \cdot \, {x}^{3}
 \\
\hspace{-0.95in}&& \quad  
\,\, \, \, -{\frac {a \cdot \, (a-1)  
\, (3\,{a}^{2}-12\,a+13) }{56}} \, \cdot {x}^{4}
 \,\, \,  + \, \, \cdots  \, \,  \,
+ \,  a \cdot \, (a-1) \cdot {{P_{18}(a)} \over {D_{20}}}   \cdot \, x^{20}
 \, \,\, +  \,\,  \cdots
 \nonumber
\end{eqnarray}
corresponds to {\em movable singularities}. For (an infinite number of) 
selected values of the 
parameter $\, a$, this series becomes a rational function, for instance 
(\ref{Aa3}) for $\, a \, = \, -8$, (\ref{other}) for $\, a \, = \, 27$, 
(\ref{other2}) for  $\, a \, = \, 8$ and $\, a \, = \, -27$. For a generic 
parameter $\, a$ the series is much more complex, it is not globally bounded.
For instance, $\, P_{18}(a)$ in (\ref{mad2oneseries})
is a polynomial with integer coefficients of degree $\, 18$ in $\, a$, 
and the denominator $\, D_{20} \, = \, \, 1277610230161807653119590400$ 
is an integer that factors in many primes: $\, \, D_{20} \, = \, \,$
$2^{17} \cdot \, 5^2  \cdot \, 7^9 \cdot \, 13^4 \cdot \, 19^3 \,
 \cdot \, 31  \cdot \, 37 \cdot \, 43 $. 

One verifies easily on this series that the two differentially 
algebraic series $\, R(a, x)$ and $\, R(b, x)$ commute and that
\begin{eqnarray}
\label{mad2oneseries}
\hspace{-0.95in}&& \quad \quad  \quad \quad \quad \quad 
R(a, \, R(b, \, x)) \,\,     = \,\,  \,  \,\,
R(b, \, R(a, \, x)) \,\,     = \,\,  \,  \,\, 
R(a\, b, \, x).
\end{eqnarray}
Note that the $\, a \, \rightarrow \, \, 1$
limit of the one-parameter series (\ref{mad2oneseries})  
gives as expected
\begin{eqnarray}
\label{mad2oneseries}
\hspace{-0.95in}&& \quad \quad  \quad \quad \quad \quad 
R(1\, + \, \epsilon \cdot \, x) 
\,\,     = \,\,  \,  \,\, \,
x \,\, \,+ \, \,  \epsilon \cdot \, F(x)  \,  \,\, + \, \, \cdots 
\end{eqnarray}
where:
\begin{eqnarray}
\label{FM3}
\hspace{-0.95in}&& 
F(x) \, \, = \, \, \, 
x \cdot \, (1 \, -x)^{2/3} \cdot \, 
_2F_1\Bigl([{{1} \over {3}}, \, {{2} \over {3}}], \, [{{4} \over {3}}], \, x\Bigr)
 \, \, \, = \, \, \,\,
 x  \,\, \,\, -{\frac {x^2}{2}} \, \,  \,
-{\frac {x^3}{7}}  \,  
\, \, + \,\, \cdots 
\end{eqnarray}

\vskip .1cm 
\vskip .1cm 

\section{$\, _2F_1$ hypergeometric functions deduced from Goursat and Darboux identity} 
\label{more2F1GoursatDarboux}

\vskip .2cm 

\subsection{$\, _2F_1$ hypergeometric functions deduced from the quadratic identity \\} 
\label{more2F1quadra}

Using the quadratic identity
\begin{eqnarray}
\hspace{-0.95in}&& \quad \,  \, \, 
_2F_1\Bigl([\alpha, \, \beta], \, [ {{\alpha+\beta+1} \over {2}}], \, x\Bigr)
\, \, = \, \, \, \, 
_2F_1\Bigl([{{\alpha} \over {2}}, \, {{\beta} \over {2}}], \, [ {{\alpha+\beta+1} \over {2}}], \, 
4 \, x \, (1\, -x) \Bigr), 
\end{eqnarray}
one deduces:
\begin{eqnarray}
\hspace{-0.95in}&& \quad \quad \quad \quad  \, \, 
_2F_1\Bigl([{{1} \over {2}}, \, 1], \, [ {{5} \over {4}}], \, x\Bigr)
\, \, = \, \, \,  \, 
_2F_1\Bigl([{{1} \over {4}}, \, {{1} \over {2}}], \, [ {{5} \over {4}}], \, 
4 \, x \, (1\, -x) \Bigr).
\end{eqnarray}
The previously described relations on $\, _2F_1([1/4,\, 1/2], [5/4], x)$,
together with the rational function $\, R(x) $
$\, = \,\, -4 \,x/(1\, -x)^2$, yields the new identity
\begin{eqnarray}
\label{newidR1R2R3}
\hspace{-0.95in}&& \quad \quad  \, 
(1 \, -2 \, x) \cdot \, 
_2F_1\Bigl([{{1} \over {2}}, \, 1], \, [ {{5} \over {4}}], \, x\Bigr)
\, \, \,= \, \, \,\,
 _2F_1\Bigl([{{1} \over {2}}, \, 1], \, [ {{5} \over {4}}], \, 
- \, 4\,{\frac {x \cdot \, (1 \, -x) }{ (1 \, -2\,x)^2}}\Bigr), 
\end{eqnarray}
where we have used the relation 
 $ \, \, R_3(R_1(x)) \,  =  \, \,  R_2(R_3(x))\,  \, $ 
with:
\begin{eqnarray}
\label{transmutation}
\hspace{-0.95in}&&
R_1(x) \, \, =  \, \, \, 
- \, 4\,{\frac {x \cdot \, (1 \, -x) }{ (1 \, -2\,x)^2 }},
\quad 
R_2(x) \,\,  =  \, \,\,  {{ - \, 4 \, x} \over { (1\, -\, x)^2}}, 
\quad 
R_3(x) \,\,  =  \, \,\,  4 \, x \cdot \, (1\, -x). 
\end{eqnarray}

Introducing
\begin{eqnarray}
\label{newYY}
\hspace{-0.95in}&& \quad \quad \quad \quad \quad 
Y(x) \,  \,\, = \, \,  \, \,
 x^{1/4} \cdot \, (1 \, -x)^{1/4} \cdot \, 
 _2F_1\Bigl([{{1} \over {2}}, \, 1], \, [ {{5} \over {4}}], \, x\Bigr),
\end{eqnarray}
one sees that it is solution of 
$\,\,\, \Omega \, = \, (D_x \, + \, A_R(x)) \cdot\, D_x\,\, $ with:
\begin{eqnarray}
\label{newOmega}
\hspace{-0.95in}&& \quad  \quad  \, \,  
A_R(x) \, \, = \, \, \, 
 {{3} \over { 4}} \cdot \,{\frac {2\,x-1}{x \left( x-1 \right) }}
\, \, = \, \, \, {{u'(x)} \over { u(x)}},  
\quad  \quad \, \,  
u(x) \, \, = \, \, \, x^{3/4} \cdot \, (1 \, -x)^{3/4}. 
\end{eqnarray}
The rank-two  condition is verified with $\, A_R(x)$ 
given by (\ref{newOmega})
and $\, R(x)$ given by $\, R_1(x)$ in (\ref{transmutation}).

\vskip .2cm 

\subsection{$\, _2F_1$ hypergeometric functions deduced from the Goursat  identity \\} 
\label{more2F1Goursat}

Using the Goursat identity
\begin{eqnarray}
\hspace{-0.96in}&& 
_2F_1\Bigl([\alpha, \, \beta], \, [ 2\, \beta], \, x\Bigr)
\, \, = \, \, \, (1\, -\, x/2)^{-a} \cdot \, 
_2F_1\Bigl([{{\alpha} \over {2}}, \, {{\alpha+1} \over {2}}], \, [ \beta \, + \, {{1} \over {2}}], \, 
{{ \, x^2} \over {(2 \, -\, x)^2}}\Bigr).
\end{eqnarray}
for $\, \alpha \, = \, 1/3$, $\, \beta \, = \, 2/3$,
one gets: 
\begin{eqnarray}
\hspace{-0.95in}&& \quad \quad 
_2F_1\Bigl([{{1} \over {3}}, \, {{2} \over {3}}], \, [{{4} \over {3}}], \, x\Bigr)
\, \, = \, \, \,\, (1\, -\, x/2)^{-1/3} \cdot \, 
_2F_1\Bigl([{{1} \over {6}}, \, {{2} \over {3}}], \, [{{7} \over {6}}], \, 
{{ \, x^2} \over {(2 \, -\, x)^2}}\Bigr).
\end{eqnarray}
Combining this last identity with (\ref{Q8}) one gets:
\begin{eqnarray}
\hspace{-0.95in}&& 
\label{QQ8}
{{x \cdot \, (x  \, -2)^3} \over { (1\, -2 \, x)^3}} \cdot \, 
_2F_1\Bigl([{{1} \over {3}}, \, {{2} \over {3}}], \, [{{4} \over {3}}], \, 
{{x \cdot \, (x  \, -2)^3} \over { (1\, -2 \, x)^3}}\Bigr)^3 
\, = \, \, \, -8 \cdot \, x \, \cdot \,
 _2F_1\Bigl([{{1} \over {3}}, \, {{2} \over {3}}], \, [{{4} \over {3}}], \, x\Bigr)^3
\nonumber\\
\hspace{-0.95in}&& \quad  
  \, \, = \, \, \,  \, 
{{16 \, x} \over {x \, -2}} \cdot \, 
_2F_1\Bigl([{{1} \over {6}}, \, {{2} \over {3}}], \, [{{7} \over {6}}], \, 
{{ \, x^2} \over {(2 \, -\, x)^2}}\Bigr)^3
\\
\hspace{-0.95in}&& \quad  \, = \, \, \,  \, 
{{-2 \, x \cdot \, (x-2)^3} \over {x^4+10 \, x^3-12\, x^2+4\, x-2}} \cdot \, 
_2F_1\Bigl([{{1} \over {6}}, \, {{2} \over {3}}], \, [{{7} \over {6}}], \, 
{{ \, x^2 \cdot \, (x-2)^6} \over {(x^4+10 \, x^3-12\, x^2+4\, x-2)^2}} \Bigr)^3.
\nonumber
\end{eqnarray}
It yields the identity on this new hypergeometric function:
\begin{eqnarray}
\hspace{-0.95in}&& 
\label{QQQ8}
\quad \quad \quad \quad \, \, 
_2F_1\Bigl([{{1} \over {6}}, \, {{2} \over {3}}], \, [{{7} \over {6}}], \, 
{{64 \, x } \over{ (1\, +18 \, x \, -27 \, x^2)^2}}\Bigr) 
\nonumber \\
\hspace{-0.95in}&& \quad \quad   \quad \quad  \quad  \quad \quad  \quad \, \, 
\, = \, \, \,  \, (1+18\,x-27\,x^2)^{1/3} \cdot \, 
_2F_1\Bigl([{{1} \over {6}}, \, {{2} \over {3}}], \, [{{7} \over {6}}], \, x\Bigr).
\end{eqnarray}
We have used the relation 
 $ \, \, R_3(R_1(x)) \,  =  \, \,  R_2(R_3(x))\,  \, $ 
with:
\begin{eqnarray}
\label{transmutation}
\hspace{-0.96in}&&
R_1(x) \,  =  \, \,  
\,{\frac {x \cdot \, (x \, -2)^3 }{ (1 \, -2\,x)^3 }},
\, \,  \,\,    
R_2(x) \,  =  \, \, 
 {{ 64 \, x} \over { (1\, +18 \, x \, -27 \, x^2)^2}}, 
\, \,  \,  \,  
R_3(z) \,  =  \, \, 
 {{ x^2} \over {(2\, -x)^2 }}. 
\end{eqnarray}

\vskip .1cm
  
\section{Miscellaneous rational functions for the covariance of a Heun function}
\label{MiscellHeun}

Let us consider the well-known formula for the addition on elliptic sine:
\begin{eqnarray}
\label{additionellipticsinus}
\hspace{-0.95in}&& \quad  \quad  \quad \quad 
sn(u \,+ \,  v) \,\,\, = \, \, \, \, {{ 
sn(u) \, cn(v) \, dn(v) \,\,  + \, \, sn(v) \, cn(u) \, dn(u) 
} \over { 1 \, \, - k^2 \, sn(u)^2 \,  sn(v)^2
}}.
\end{eqnarray}
Introducing the variables $\, x \, = \, sn(u)^2$,  $\, y \, = \, sn(v)^2$ 
and  $\, z \, = \, sn(u \, +v)^2$, and $\, M \, = \, \, 1/k^2$, 
the previous addition formula 
(\ref{additionellipticsinus}) for the elliptic sine reads:
\begin{eqnarray}
\label{additionellipticsinusbis}
\hspace{-0.95in}&& \quad  \quad  \quad   \quad  
(M-xy)^{2}\cdot \, z^{2} \, \, \,  +2\,M \cdot \, \Bigl(2\,x y \cdot \, (M \, +1) \,
 - \, (x+y)  \cdot \, (xy \, +M) \Bigr) \cdot \, z 
\nonumber \\ 
\hspace{-0.95in}&& \quad  \quad \quad  \quad \quad  \quad \quad  \quad  \, \, 
+ \,(x-y)^{2} \cdot \, M^{2}
\, \, = \, \, \, 0.
\end{eqnarray}
Note that, since $\, y$ is the square of the elliptic sine, 
$\, y \, = \, sn(v)^2\, = \, sn(-v)^2$,
the ``master equation'' (\ref{additionellipticsinusbis}) 
{\em is also a representation of the difference on elliptic sine}: 
$\, x \, = \, sn(u)^2$,  $\, y \, = \, sn(v)^2$, $\, z \, = \,   sn(u \, -v)^2$.
Actually $\, x$, $\, y$ and $\, z$ are on the same footing in this
 ``master equation'' (\ref{additionellipticsinusbis}) that can be rewritten
in a symmetric way as an algebraic surface:
\begin{eqnarray}
\label{mastersym}
\hspace{-0.95in}&& \quad  \quad  \quad    \,    \quad   
x^2  \, y^2  \, z^2 \, \,  \,
-2\cdot \, M  \cdot \, (x \,+y \, +z) \cdot \, x  \, y  \, z \, \,  \,
 +4 \cdot \, M \cdot \, (M+1) \cdot \, x  \, y  \, z 
\nonumber \\ 
\hspace{-0.95in}&& \quad  \quad \quad  \, \,  \quad \quad  \quad 
\, \, +M^2 \cdot \, ((x \,+y \, +z)^2 \, \, -4\cdot \, (x \, y \, +x \, z \, +y \, z)) 
\, \, \,\, = \, \, \,\, \, 0.
\end{eqnarray}
For every fixed $\, z$ and $\, M$ (except $z \, = \, 0, \, 1, \, M, \, \infty$ 
and $\, M= \, 0, \, 1, \, \infty$), 
condition (\ref{mastersym}) reduces to an 
algebraic curve of {\em genus one}. The  algebraic surface (\ref{mastersym}) 
is thus foliated in elliptic curves\footnote[2]{In mathematics, an 
{\em elliptic surface} is a surface that has an elliptic fibration: almost 
all fibers are smooth curves of genus $\,1$.}. This algebraic surface is 
left invariant by an {\em infinite set of birational transformations} 
generated by the three involutions:
\begin{eqnarray}
\label{involution}
\hspace{-0.95in}&& \quad  \quad  \quad   \quad  \quad   \,\,
(x, \,\, y ,\, \, z) \quad \, \,\, \,\longrightarrow \, \quad \quad \,
 \Bigl( {{M^2\cdot \,(y-z)^2} \over {(M\,-y\,z)^2 \cdot \, x}}, \,\, y, \,\, z \Bigr),  
\end{eqnarray}
and the two other ones corresponding to the permutation of $\, x$, $\, y$ and $\, z$.

\vskip .1cm

{\bf Remark:} For fixed $\, z$ condition (\ref{mastersym}) is an elliptic 
curve (except $\, M=0, \, 1, \, \infty$). If one calculates its 
$\, j$-invariant\footnote[1]{In 
Maple use with(algcurves) and the command $\, j\_invariant$.} 
one gets the same result as (\ref{jfunc}) namely 
\begin{eqnarray}
\label{jfuncbis}
\hspace{-0.95in}&&   \quad \quad \quad \quad \quad \quad
j \, \, = \, \, \, 
256 \cdot \,{\frac { ({M}^{2}-M+1)^{3}}{ M^{2}  \cdot \, (M \, -1)^{2}}}. 
\end{eqnarray}
{\em which does not depend on} $\, z$. Of course one gets the same result for the 
elliptic curves corresponding to  condition (\ref{mastersym}) for fixed $\, x$ 
or fixed $\, y$.

The rational transformation (\ref{Aadoubling}) corresponding to 
$\, \theta \, \rightarrow \, \, 2 \, \theta \,$ is obtained by imposing 
$\,\, y \, = \, x \,$ in (\ref{additionellipticsinusbis}). For 
$\,\, y \, = \, x \,$ the relation (\ref{additionellipticsinusbis}) 
factorizes\footnote[9]{Other cases of factorizations are,
 up to permutations in $\, x$, $\, y$ and $\, z$: $\, y \, =$
$ \, 0, \, 1, \, M, \, \infty$, 
 $\, y \, = \, M/x$, $\, y \, = \, (M-x)/(1-x)$,
$\, y \, = \, M \cdot \,(1-x)/(M-x)$.} into:
\begin{eqnarray}
\label{additionellipticsinusbisfacto}
\hspace{-0.95in}&& \quad  \quad  \quad   \quad  
z \cdot \, \Bigl( (M-x^2)^2 \cdot \, z  \,\,
 -4\cdot \, M \cdot \, x \cdot \, (1-x) \cdot \, (M-x)   \Bigr)
 \,\, = \,\,\, \, 0. 
\end{eqnarray}
Discarding the trivial solution $\, z \, = \, \, 0$,  one gets:
\begin{eqnarray}
\label{additionellipticsinusbisfacto}
\hspace{-0.95in}&& \quad  \quad  \quad   \quad   \quad   \quad  
z \, \, = \, \, \,  
4 \cdot \, {{ x \cdot \, (1\, -x) \cdot \, (1-x/M)} \over { (1\, -x^2/M)^2}}, 
\end{eqnarray}
which is exactly (\ref{Aadoubling}).

Imposing in (\ref{additionellipticsinusbis}) $\, y$ to be equal 
to (\ref{Aadoubling}) one deduces the rational transformation 
 corresponding to $\,  \theta \, \rightarrow \, \, 3 \, \theta $, and
one can deduce from the ``master'' equation (\ref{additionellipticsinusbis})
all the rational transformations corresponding to 
$\, \theta \, \rightarrow \, \, p \, \theta $. When $\, p$ is a 
prime number different from $\, p\, = \, 2$, the 
corresponding rational transformations have a simple form. 

Introducing the square of the elliptic sine 
$\, x \, = \, \, sn(\theta, \, k)^2$,  the rational transformations 
corresponding to 
$\,  \,\theta \, \rightarrow \, \, p \, \theta  \,$
give for a given $\, M$: 
\begin{eqnarray}
\label{Aaprime}
\hspace{-0.95in}&&    \quad  \quad  \quad  \quad  \,\, 
R_p(x, \, M) \, \, = \, \, \, 
 x \cdot \,  \Bigl({{ P_p(x, \, M)} \over { Q_p(x, \, M)}} \Bigr)^2, 
\quad  \quad  \quad   \quad \quad  \hbox{where:} 
\\
\hspace{-0.95in}&&    \quad  \quad  \quad  \quad  \,\, 
Q_p(x, \, M) \, \, = \, \, \,  \,   \, 
x^{(p^2-1)/2} \cdot \, M^{(p^2-1)/4} \cdot \, 
P_p\Bigl({{1} \over {x}}, \, \, {{1} \over {M}}\Bigr), 
\nonumber 
\end{eqnarray}
where $ \,P_p(x, \, M)$ are polynomials in $\, x$ and $\, M$ 
of degree $\, (p^2-1)/2 \, $ in $\, x$ and of degree $\, (p^2-1)/4 \,$ 
in $\, M$. For instance,  $ \,P_3(x, \, M)$ reads:
\begin{eqnarray}
\label{P3zM}
\hspace{-0.95in}&&    \quad  \quad  \quad  \quad 
P_3(x, \, M) \, \, = \, \, \, \, \, 
{x}^{4} \, \,\,  -6\,M \cdot \, {x}^{2} \,\,  
+4 \cdot \,M \cdot \, (M \,+1) \cdot \, x \,\,  \, -3\,{M}^{2}. 
\end{eqnarray}
The polynomial $\, P_p(x, \, M)$ reads for $\, p \, = \, 5$:
\begin{eqnarray}
\label{ppfirstprime}
\hspace{-0.95in}&&   
 P_5(z, \, M) \, \, = \, \, \, {x}^{12} \, 
-50\,M\, {x}^{10}\, +140\,M \, (M+1) \cdot \, {x}^{9}\, 
-5\,M \, (32\,{M}^{2}+89\,M+32) \cdot \,  {x}^{8}\, 
\nonumber \\
\hspace{-0.95in}&& \quad  \quad
+16\,M \, (M+1)  \, (4\,{M}^{2}+31\,M+4) \cdot \,  {x}^{7}
\, \, \, 
-60\,{M}^{2} \, (4\,{M}^{2}+13\,M+4) \cdot \,  {x}^{6}
\nonumber \\
\hspace{-0.95in}&& \quad \quad \, 
+360\,{M}^{3} \, (M+1) \cdot \,  {x}^{5} \,  \,\, 
-105\,{M}^{4}\cdot \, {x}^{4}\, \, \, 
-80\,{M}^{4} \, (M+1) \cdot \,  {x}^{3}
\nonumber \\
\hspace{-0.95in}&& \quad \quad \, 
+2\,{M}^{4} \, (8\,{M}^{2}+47\,M+8) \cdot \,  {x}^{2} 
 \,\, \, 
-20\,{M}^{5} \, (M+1) \cdot \,  x
\, \,  \,  +5\,{M}^{6}, 
\end{eqnarray}
It is straightforward to calculate the next  $\, P_p(z, \, M)$ 
for $\, p\, = \, 7, \, 11, \, 13, \, \cdots$, but the 
expressions become quickly too large to be given here.

As expected, the two rational functions
(\ref{Aaprime}) {\em commute} for different primes $\, p$.  
The series expansion of these rational transformations 
read:
\begin{eqnarray}
\label{Aaprimeseries}
\hspace{-0.95in}&&   \quad \quad \quad  \quad \quad   
R_p(x) \, \, = \, \, \,  \,  \, 
p^2 \cdot \, x \, \, \, \,
- \,  {{ p^2 \cdot \, (p^2-1)}  \over {3}} 
 \cdot \, {{M\, +1} \over {M}} \cdot \, x^2
\,  \, \, \, +  \, \cdots  
\end{eqnarray}
When $\, p$ is not a prime the rational functions $\, R_p(x)$
corresponding to $\, \theta \, \rightarrow \, \, p \, \theta$,
are no longer of the form (\ref{Aaprime}) but they still have 
the series expansion (\ref{Aaprimeseries}). 

We have the following identity on a Heun function
where $\, R_p(x)$ are the previous rational functions 
(\ref{Aaprime}):
\begin{eqnarray}
\label{Fdoublingideapp}
\hspace{-0.95in}&&   \quad \quad \quad   \quad \quad   
 R_p(x) \cdot \, 
Heun\Bigl(M, \, {{M\, +1} \over {4}}, \,  {{1} \over {2}}, 
\, 1, \, {{3} \over {2}}, \,{{1} \over {2}}, \, R_p(x) \Bigr)^2
\nonumber \\
\hspace{-0.95in}&&   \quad \quad \quad  \quad \quad \quad \quad \quad   
 \, \, = \, \, \, 
p^2 \cdot \, x \cdot \, 
Heun\Bigl(M, \, {{M\, +1} \over {4}}, \,  {{1} \over {2}}, 
\, 1, \, {{3} \over {2}}, \,{{1} \over {2}}, \, x\Bigr)^2. 
\end{eqnarray}
Note that the Heun identity  (\ref{Fdoublingideapp}) is valid 
even when the integer $\, p$ is no longer a prime, $\, R_p(x)$ being
a rational function representation of 
$\,\,\, \theta \, \rightarrow \, \, p \cdot \, \theta$, and that
all these (commuting) rational transformations are solutions 
of the rank-two  condition.

\vskip .2cm
  
\section{The Schwarzian conditions are compatible with the composition of functions}
\label{Schwarzcomp}

We want to have 
\begin{eqnarray}
\label{condition1zy}
\hspace{-0.95in}&& \quad    \quad  \quad   \quad  \, 
 W(x) 
\, \, \,  \,-W(z(y(x))) \cdot  \, \Bigl({{d z(y(x))} \over { dx}}\Bigr)^2 
\, \, \,  \,+ \,  \{ z(y(x)), \, x\} 
\, \,\, \, = \,\, \, \,  \, 0, 
\end{eqnarray}
which reads using the derivative of composition of function and 
the previous chain rule (\ref{chainrule}):
\begin{eqnarray}
\label{condition1zy}
\hspace{-0.95in}&& \quad  \quad \quad    \quad  \quad \, 
 W(x) 
\, \, \,  \, -W(z(y(x))) \cdot  \,
 \Bigl({{d z(y)} \over { dy}} \Bigr)^2 \cdot \, y'(x)^2  
\nonumber \\ 
\hspace{-0.95in}&& \quad   \quad \quad \quad  \quad \quad \quad    \quad   \quad  
\, \, \,  \,+ \,   \{ z(y), \, y \} \cdot y'(x)^2 
\, \,\,\, + \, \, \{ y(x), \, x\}
\, \,\, \, = \,\, \, \,  \, 0, 
\end{eqnarray}
from 
\begin{eqnarray}
\label{condition1y}
\hspace{-0.95in}&& \quad  \quad   \quad  \quad   \quad   \, \, 
 W(x) 
\, \, \,  \,-W(y(x)) \cdot  \, y'(x)^2
\, \, \,  \,+ \,  \{ y(x), \, x\} 
\, \,\, \, = \,\, \, \,  \, 0, 
\end{eqnarray}
and 
\begin{eqnarray}
\label{condition1z}
\hspace{-0.95in}&& \quad   \quad   \quad  \quad  \quad  \, \, 
 W(y) 
\, \, \,  \,-W(z(y)) \cdot  \, z'(y)^2
\, \, \,  \,+ \,  \{ z(y), \, y\} 
\, \,\, \, = \,\, \, \,  \, 0.
\end{eqnarray}
Let us multiply the previous relation (\ref{condition1z}) 
by $\, y'(x)^2$ one gets:
\begin{eqnarray}
\label{condition1y2z}
\hspace{-0.95in}&& \quad      
 W(y) \cdot \,  y'(x)^2
\, \, \,  \,-W(z(y)) \cdot  \, z'(y)^2 \cdot \,  y'(x)^2
\, \, \,  \,+ \,  \{ z(y), \, y\} \cdot \,  y'(x)^2
\, \,\, \, = \,\, \, \,  \, 0.
\end{eqnarray}
Adding (\ref{condition1y}) to (\ref{condition1y2z})
one gets:
\begin{eqnarray}
\label{condition1y2z}
\hspace{-0.95in}&& \quad       \quad  \, \,  \, 
W(x) \, \,\,  + W(y) \cdot \,  y'(x)^2
\, \, \,  \,-W(z(y)) \cdot  \, z'(y)^2 \cdot \,  y'(x)^2
\, \, \,  \,+ \,  \{ z(y), \, y\} \cdot \,  y'(x)^2 
\nonumber \\ 
\hspace{-0.95in}&& \quad    \quad \quad \quad  \quad \quad \quad  
\, \, \,  \,-W(y(x)) \cdot  \, y'(x)^2
\, \, \,  \,+ \,  \{ y(x), \, x\} 
\, \,\, \, = \,\, \, \,  \, 0.
\end{eqnarray}
which gives after simplification nothing 
but (\ref{condition1zy}). Q. E. D. 

\vskip .1cm 

\section{Compatibility of the three Schwarzian conditions 
 (\ref{Harnad111}), (\ref{Harnad214}) and  (\ref{condition1})\\} 
\label{Schwarzmirror}

\vskip .1cm 

The Schwarzian equation on the $\, j$-invariant  are known to be invariant 
by the group of modular transformations (see for instance equation (1.26) 
in~\cite{HarnadHalphen} or (1.13) in~\cite{Harnad}). More remarkably 
(and less known) the Schwarzian equation (\ref{Harnad214}) on the 
nome\footnote[1]{In the Schwarzian equation (\ref{Harnad214}) the nome 
is seen as a function of the Hauptmodul.} is {\em invariant}
 under the transformations\footnote[2]{See the concept of 
{\em replicable functions}~\cite{Replicable}.}
 $\,\, q \,  \longrightarrow \, \, S(q) \, = \, \, e \cdot \, q^N$. Equation 
(\ref{Harnad214}) is clearly invariant under the rescaling 
$\, Q(x) \, \rightarrow \, e \cdot \, Q(x)$, and one can verify easily, 
using the chain rule for the Schwarzian derivative of a composition, that 
the sum of the first two terms in the LHS of (\ref{Harnad214}), namely 
$\,\{Q(x), \, x \} \, + \, Q'(x)^2/Q(x)^2/2\,\, $ is actually invariant by 
$\,   Q(x)\, \rightarrow \,  \,  Q(x)^N$. Therefore we also have the equation:
\begin{eqnarray}
\label{Harnad214bis}
\hspace{-0.95in}&& \quad \quad 
 \, \{S(Q(x)), \, x \} \, \, \, \, 
 + {{1} \over {2  \cdot \, S(Q(x))^2 }} 
 \cdot  \Bigl({{ d S(Q(x))} \over {d x}} \Bigr)^2
\,  \, +  \, W(x) \, \,  \, =  \, \, \, \, \, 0.
\end{eqnarray}
Equation (\ref{Harnad111}) yields 
\begin{eqnarray}
\label{Harnad111bis}
\hspace{-0.95in}&& 
\{X(S(Q(q))), \, S(Q(x)) \} \, \,   -{{1} \over { 2 \, S(Q(x))^2}} \,    \,
 - \, W(X(S(Q(q))))
\cdot \,  \Bigl( {{ d X(S(Q(x)))} \over {d S(Q(x))}}  \Bigr)^2
\,  = \,\, \, 0, 
\nonumber 
\end{eqnarray}
and thus: 
\begin{eqnarray}
\label{Harnad111ter}
\hspace{-0.95in}&& \quad  \quad   
\{X(S(Q(x))), \, S(Q(x)) \} \cdot \, \Bigl({{ d S(Q(x))} \over {d x}} \Bigr)^2
 \, \, \,\,   -{{1} \over { 2 \, S(Q(x))^2}} 
\cdot  \Bigl({{ d S(Q(x))} \over {d x}} \Bigr)^2
\nonumber \\
\hspace{-0.95in}&&  \quad  \quad  \quad  \quad  
 \, \, \,    \,
 - \, W(X(S(Q(x)))) 
\cdot \,  \Bigl( {{ d X(S(Q(x)))} \over {d S(Q(x))}}  \Bigr)^2
 \cdot  \Bigl({{ d S(Q(x))} \over {d x}} \Bigr)^2
\, \, \, = \,\,\, \, 0. 
\end{eqnarray}
Using the chain rule for Schwarzian derivative of 
the composition of functions
\begin{eqnarray}
\label{Harnad111chain}
\hspace{-0.95in}&&  \, 
\{X(S(Q(x))), \, x\}  \, \,  \, = \, \, \,  \, 
\{X(S(Q(x))), \, S(Q(x)) \} \cdot
  \Bigl({{ d S(Q(x))} \over {d x}} \Bigr)^2 \, + \, \, 
 \,  \, \{S(Q(x)), \, x \}, 
\nonumber 
\end{eqnarray}
we see immediately that the sum of (\ref{Harnad214bis}) 
and (\ref{Harnad111ter}) gives: 
\begin{eqnarray}
\label{condition1ter}
\hspace{-0.95in}&& \quad   \quad \quad  \quad \quad  \quad 
 W(x) 
\, \, \,  \,-W(y(x)) \cdot  \, y'(x)^2
\, \, \,  \,+ \,  \{ y(x), \, x\} 
\, \,\, \, = \,\, \, \,  \, 0, 
\end{eqnarray}

\vskip .1cm

\section{The $\, _2F_1([1/6,1/3],[1],x)$ hypergeometric function. } 
\label{16131app}

Let us consider the Schwarzian condition in the case of the 
$\, _2F_1([1/6,1/3],[1],x)$ hypergeometric function.
 
The one-parameter 
 family of  commuting series solution of the Schwarzian condition
reads:
\begin{eqnarray}
\label{361}
\hspace{-0.95in}&& \, \,         \quad   \quad    \quad       
y_1(e, \, x) \, \, \, = \, \, \,  \, 
e \cdot \, x   \, \,\, \, + \, e  \cdot \, (e-1) \cdot \, S_e(x),
 \quad \quad \quad     \quad  \quad \hbox{where:} 
\nonumber \\
\hspace{-0.95in}&& \, \,  \quad   \quad    \quad \quad   \quad   \quad        
 S_e(x)\, \, = \, \, \, \, 
-{\frac {7}{18}}  \cdot \,  {x}^{2} \, 
\, \, \, +{\frac { (109 \,e -283) }{1296}} \cdot \,  {x}^{3} 
\,  \, \, \,+ \, \,\, \cdots 
\end{eqnarray}
The series of the form $\, a \cdot \, x^2 \, + \, \cdots \,\,   $ reads
\begin{eqnarray}
\label{361y2}
\hspace{-0.95in}&&         
y_2(a, \, x) \, = \, \,   a \cdot x^{2} 
 + {{7 \, a} \over { 9}} \cdot  x^{3}  
-  a \cdot  {{84 a  -127  } \over {216 }} \cdot x^{4} 
 -  a \cdot  {{47628 a  - 36049} \over {78732}} \cdot  x^{5}  +  \cdots 
\end{eqnarray}
and is such that:
\begin{eqnarray}
\label{such}
\hspace{-0.95in}&&   \quad   \quad  \quad 
 y_1(e, \,y_2(a, \, x)) \, = \, \, y_2(a\, e, \, x), 
\quad   \quad  \quad 
 y_2(a, \,y_1(e, \, x)) \, = \, \, y_2(a\, e^2, \, x).
\end{eqnarray}
The series of the form $\, b \cdot \, x^3 \, + \, \cdots \,\,   $ 
reads
\begin{eqnarray}
\label{361y2}
\hspace{-0.95in}&&          \,  
y_3(b, \, x) \, \,  = \, \,\,  \,   b \cdot x^{3} \, \,
+{{7 \, b} \over {6}} \cdot x^{4}  \, \,
+{{479 \, b} \over {432}} \cdot x^{5} \, \,
+ b \cdot \, {{81648 b - 210031} \over {209952 }} \cdot x^{6} 
\, \,  \,   + \, \, \cdots 
\end{eqnarray}
and is such that
\begin{eqnarray}
\label{such2}
\hspace{-0.95in}&&   \quad   \quad \quad 
 y_1(e, \,y_3(b, \, x)) \, = \, \, y_3(b\, e, \, x), 
\quad   \quad  \quad 
 y_3(b, \,y_1(e, \, x)) \, = \, \, y_3(b\, e^3, \, x).
\end{eqnarray}
The two series $\, y_2(a, \, x)$ and $\, y_3(b, \, x)$ 
commute for $\, b \, = \, a^2$. For $\, a \, = \, 1/108$
the series (\ref{361y2}) becomes the series expansion 
\begin{eqnarray}
\label{such2}
\hspace{-0.95in}&&   
y \, \, = \, \, \, {\frac {{x}^{2}}{108}} \,\, \, +{\frac {7\,{x}^{3}}{972}} \, \, \,
+{\frac {71\,{x}^{4}}{13122}} \,\,  \,+{\frac {4451\,{x}^{5}}{1062882}} \,\,
\, +{\frac {63997\,{x}^{6}}{19131876}} \,\,  \,
+{\frac {1417505\,{x}^{7}}{516560652}} 
\,\, \, + \, \, \cdots 
\end{eqnarray}
which corresponds to the modular equation (A.3) in~\cite{IsingCalabi2}:
\begin{eqnarray}
\label{A3}
\hspace{-0.95in}&&   \quad   \quad  \quad  \quad \quad
4\,x^3\,y^3 \, \,\,  -12\,x^2\,y^2 \cdot \,(x+y) \,\, \,  \,  
 +3\, x \,y \cdot \,(4\,x^2-127\,x\,y+4\,y^2) \,
\nonumber \\
\hspace{-0.95in}&&   \quad   \quad   \quad  
 \quad   \quad  \quad   \quad   \quad 
 -4 \cdot \,(x+y)\cdot\,(x^2+83\,x\,y+y^2)\,  \,\,   +432\,x\,y 
\, \,\,  \, = \, \,  \, \, 0, 
\end{eqnarray}
This modular equation has a rational parametrization: it corresponds to the
relation between  two rational pullbacks in the hypergeometric
 identity (A.11) in~\cite{Christol}:
\begin{eqnarray}
\label{A11}
\hspace{-0.95in}&&   \quad   \quad  \quad   
_2F_1\Bigl([{{1} \over {6}}, \, {{1} \over {3}}], \, [1], \, \, 108 \, v^2 \cdot \, (1+4v)\Bigr) 
\nonumber \\
\hspace{-0.95in}&&   \quad   \quad  \quad   \quad \quad  \quad 
\,  \,= \, \, \, (1 \, -12\, v)^{-1/2} \cdot \, 
_2F_1\Bigl([{{1} \over {6}}, \, {{1} \over {3}}], \, [1], \, \,
 -{{ 108 \, v \cdot \, (1+4v)^2} \over { (1 \, -12\, v)^3 }} \Bigr).
\end{eqnarray}

\vskip .1cm

\section{The solutions $\, Q(x)$ of the non-linear conditions  (\ref{QDAfirst}) 
seen  as  solutions of the Schwarzian conditions on the mirror maps} 
\label{subcase}

In all the cases recalled in section (\ref{recalls}), the differentially algebraic 
function $\, Q(x)$ was of the form\footnote[1]{The constant $\,N$ being 
a positive integer $\, Q(x)$ was, in fact, holonomic.} $\, Y(x)^N$. From 
$\, Q(x) \, = \, \, Y(x)^N$ or even
$\, Q(x) \, = \, \, \alpha \cdot \, Y(x)^N$, one can rewrite the Schwarzian 
derivative on $\, Q(x)$ with respect to $\, x$:
\begin{eqnarray}
\label{rewritewithresp}
\hspace{-0.95in}&&   \quad  \quad
\{ Q(x), \, x\} \, \, = \, \,\, \{ Y(x)^N, \, x\} \, \, = \, \,\,
 -{{N^2\, -1} \over {2}} \cdot \, \Bigl({{Y'(x)} \over {Y(x)}}\Bigr)^2 
\, + \,\,\{ Y(x), \, x\}. 
\end{eqnarray}
Since $\, Y(x)$ is a solution of the operator $\, \Omega$, 
the ratio $\, Z(x) \, = \, \, Y"(x)/Y'(x)$ (log-derivative of $\, Y'(x)$) 
is in fact a rational function,
namely $\, -A_R(x)$. The Schwarzian derivative $\, \{ Y(x), \, x\}$
can also be written as:
\begin{eqnarray}
\label{rewritewithresp}
\hspace{-0.95in}&&   \quad  \quad \quad   
 \{ Y(x), \, x\} \, \, = \, \,
  Z'(x) \,\, -\, {{Z(x)^2} \over {2}} \, \, = \, \, 
-\, A'_R(x) \, -\, {{A_R(x)^2} \over {2}}
 \, \, = \, \, -W(x).
\end{eqnarray}
From $\, Q(x) \, = \, \, \alpha \cdot \, Y(x)^N$ one deduces immediately
the relation between the log-derivative of $\, Q(x)$ and $\, Y(x)$, 
namely $\, Q'(x)/Q(x) \, = \, N \cdot  Y'(x)/Y(x)$. 
Equation (\ref{rewritewithresp}) can be rewritten
using $\, Q'(x)/Q(x) \, = \, N \cdot  Y'(x)/Y(x)$ 
and (\ref{rewritewithresp}), 
as\footnote[2]{One recovers the Schwarzian condition 
(\ref{Harnad214}) in the $\, N \, \rightarrow \, \infty$ limit.}:
\begin{eqnarray}
\label{rewritewithrespN}
\hspace{-0.95in}&&   \quad  \quad \quad  \quad
\{ Q(x), \, x\} \, \, \,
 +{{N^2\, -1} \over {2 \, N^2}} \cdot \, \Bigl({{Q'(x)} \over {Q(x)}}\Bigr)^2 
\,\, \,  + \,\, W(x) \, \, \, = \,\, \,  \, 0. 
\end{eqnarray}
For instance, one verifies immediately that $\, Q(x)$ given 
by $\, Q(x) \, = \, Y(x)^N$ and $\, Y(x)$ given by 
(\ref{vid}),  (\ref{YM3first}),  (\ref{YM6first}),  (\ref{zero}), 
(\ref{more1}) (which identifies with (\ref{NEWident})) and (\ref{QQQ8first})
are actually solutions of the Schwarzian condition 
(\ref{rewritewithrespN}) 
for the corresponding  $\, A_R(x)$ given in (\ref{respecAR}) 
for respectively $\, N= \, 4, \, 3, \, 6, \, 2, \, 4, \, 6$.
Note that the higher-genus case hypergeometric function (\ref{Ygenus})
is {\em also such that} $\, Q(x) \, = \, Y(x)^6$ 
{\em is solution of  the Schwarzian condition} 
(\ref{rewritewithrespN}) with $\, N\, = \, 6$ and $ \, W(x)$ deduced from 
$\, A_R(x)$ given by (\ref{aAgenus}).

\vskip .1cm
\vskip .1cm

 One gets immediately the Schwarzian condition for the composition 
inverse $\, P(x) \, = \, Q^{-1}(x)$, namely: 
\begin{eqnarray}
\label{rewritewithrespNrev}
\hspace{-0.95in}&&   \quad  \quad \quad  \quad
\{ X(q), \, q\} \, \, \,
 -{{N^2\, -1} \over {2 \, N^2}} \cdot \, {{1} \over { q^2}}
\,\, \,  - \,\, W(X(q)) \cdot \,   \Bigl( {{ d X(q)} \over {d q}}  \Bigr)^2
 \, \, \, = \,\, \,  \, 0. 
\end{eqnarray}

\vskip .1cm

 {\bf Remark:} One verifies straightforwardly for the Heun function 
example of section (\ref{moreHeun}) that
\begin{eqnarray}
\label{onefindsfi}
\hspace{-0.95in}&&   \quad  \quad \quad  \quad
Q(x) \, = \, \, Y(x)^2 \, = \, \, x \cdot \, 
Heun\Bigl(M, \, {{M\, +1} \over {4}}, \,  {{1} \over {2}}, 
\, 1, \, {{3} \over {2}}, \,{{1} \over {2}}, \, x\Bigr)^2, 
\end{eqnarray}
is {\em actually solution of the  Schwarzian condition} 
(\ref{rewritewithrespN})  with $\, N \, = \, \, 2$,
with $\, W(x)$ given by (\ref{AadoublingbisW}). The composition 
inverse of the holonomic function $\, Q(x)$ 
given by (\ref{onefindsfi}) is 
\begin{eqnarray}
\label{onefindsinv}
\hspace{-0.95in}&&   \quad  \quad \quad  \, \, 
P(x) \,\, = \, \, \, sn\Bigl(x^{1/2}, \, {{1} \over {M^{1/2}}}\Bigr)^2
\\
\hspace{-0.95in}&&   \quad  \quad \quad \quad  \quad   \quad
 \,\, \,\, = \, \, \, \,  x \, \,   \,  \,  \, 
- {{1} \over {3}} \,{\frac { M \, +1}{M}} \cdot \, x^2 \, \, \,   \,  
+\, {{1} \over {45}} \,{\frac { 2\,{M}^{2}+13\,M+2}{{M}^{2}}} \cdot \, x^3
\,\, \,  \,     +\, \,  \,   \cdots
 \nonumber 
\end{eqnarray}
It is solution of (\ref{rewritewithrespNrev}) with $\, N\, = \, 2$: 
\begin{eqnarray}
\label{rewritewithrespNrevP}
\hspace{-0.95in}&&   \quad  \quad \quad  \quad \quad  \quad  
\{ P(x), \, x\} \, \, \,\,
 -{{3} \over {8}} \cdot \, {{1} \over { x^2}}
\,\, \,  \,- \,\, W(P(x)) \cdot \,  P'(x)^2
 \, \, \, = \,\, \,  \, 0. 
\end{eqnarray}

\vskip .2cm

\subsection{From the Schwarzian condition (\ref{rewritewithrespN}) 
back to the differential algebraic condition (\ref{simplerelation61first})  on $\, Q(x)$\\} 
\label{backto}

If one compares the Schwarzian condition (\ref{rewritewithrespN}) with 
the differentially algebraic condition (\ref{simplerelation61first})  
on $\, Q(x)$, one finds that they both have a third derivative $\, Q'''(x)$
but one condition depends on a constant $\, N$, while the other one 
is ``universal'': let us try to understand the compatibility between 
these two conditions. If one eliminates the third derivative $\, Q'''(x)$
between these two equations one finds a remarkably factorized condition 
$\, \,  E_{+} \cdot \, E_{-} \, = \, \, 0\, \, $ where:
\begin{eqnarray}
\label{factorised}
\hspace{-0.95in}&&   \, \,   \, \,  \, 
E_{\pm} \, = \, \,  {\cal F}'(x) \,  \,  
 - \, A_R(x) \cdot \,  \,  {\cal F}(x) \, \, \pm \, {{1} \over { N}} 
 \quad  \quad \quad  \hbox{with:} \, \, \quad \quad   
{\cal F}(x) \, = \, \,  {{Q(x)} \over {Q'(x)}}. 
\end{eqnarray}
Recalling (\ref{QF}) we see that $\, {\cal F}(x)$ is nothing but $\, F(x)$.  
The holonomic function $\, F(x)$ is known to be solution of 
$\, \Omega^{*}$, which can be rewritten, after one integration step, 
as $\, F'(x) \,  - \, A_R(x) \cdot \, F(x) \, = \, \, Cst$, 
which is actually (\ref{factorised}). The compatibility of  
the Schwarzian condition (\ref{rewritewithrespN}) 
with the differentially algebraic condition (\ref{simplerelation61first})
thus corresponds to  $\, F(x)$ being annihilated by  $\, \Omega^{*}$.

\vskip .1cm

\section{Reduction of $\, _3F_2$ identities to $\, _2F_1$ Schwarzian conditions}
\label{Reduc3F2}

Performing the derivative of the Schwarzian condition
(\ref{condition3F2}) one can eliminate this fourth derivative of $\, y(x)$, 
and then, in a second step, eliminate the third  derivative of $\, y(x)$
between the previous result and  the Schwarzian condition
(\ref{condition3F2}), and so on.  One finally gets the following 
relation that can be seen as the compatibility condition between 
the two previous conditions:
\begin{eqnarray}
\label{compat}
\hspace{-0.95in}&& \quad    \quad \quad  
 x^3 \cdot \, (1\, -x)^3 \cdot \, Q(y(x))
 \cdot \, y'(x)^3 
\, \, = \, \, \,  \, 
y(x)^3 \cdot \, (1\, -y(x))^3 \cdot \, Q(x), 
\end{eqnarray}
where the polynomial $\, Q(x)$ reads:
\begin{eqnarray}
\label{compatQ}
\hspace{-0.95in}&&  \, \,   \,    
Q(x) \, \, = \, \, \,  \, 
-2\, \cdot \, (b+c -2\,a)  \, (a+ c -2\,b) 
 \, (a+b -2\,c) \cdot \, {x}^{3} 
\nonumber \\
\hspace{-0.95in}&& \quad  \quad    \,  \, 
+3 \cdot \, q_2 \cdot \,  {x}^{2} \,  \,  \, +3 \cdot \, q_1 \cdot \, x 
\, \,  \,  -2\, \cdot \, (1+d-2\,e)  \, (d+e-2)  \, (2\,d -e-1), 
\end{eqnarray}
where 
\begin{eqnarray}
\label{compatQ2}
\hspace{-0.95in}&&  \quad  
q_2 \, = \, \, \, 6\,{a}^{2}b\, +6\,{a}^{2}c\, -4\,{a}^{2}d\,\,  -4\,{a}^{2}e \,\, 
 +6\,  a {b}^{2} \, \, -18\, a bc \,\,  -2\, a bd \, 
-2\, a be \,\,  +6\, a  {c}^{2} \, 
\nonumber \\
\hspace{-0.95in}&& \quad \quad   
-2\, a cd \,\,  -2\,  a  ce \, \, 
+6\, a de \,\,  +6\,{b}^{2}c \,\,  -4\,{b}^{2}d \, \, 
-4\,{b}^{2}e \,\,  +6\,b{c}^{2} \,\,  -2\,bcd \,\,  -2\,bce 
\nonumber \\
\hspace{-0.95in}&& \quad \, \quad  
+6\,bde \,\,  -4\,{c}^{2}d\, -4\,{c}^{2}e \, \, 
+6\,cde \,\,  +2\,{a}^{2} \,\,  + a b \,\,  
+ \, a c\, +2\,{b}^{2}\, +bc\, +2\,{c}^{2}
\nonumber \\
\hspace{-0.95in}&& \quad \quad   
\, -9\,de \, \, -3\,a\, -3\,b\, -3\,c\, +6\,d\, +6\,e \,\,  -3, 
\end{eqnarray}
\begin{eqnarray}
\label{compatQ1}
\hspace{-0.97in}&& 
q_1 \, = \, \,18\, a bc \, -6\, a bd \, -6\, a be \,
 -6\,  a  cd  \, -6\,  a ce \, +4\,  a {d}^{2}
\, +2\,  a de \, +4\, a {e}^{2} \,-6\,bcd \, -6\,bce \, 
\nonumber \\
\hspace{-0.97in}&& \quad  
+4\,b{d}^{2} \, +2\,bde+4\,b{e}^{2}+4\,c{d}^{2}+2\,cde
+4\,c{e}^{2} \,-6\,{d}^{2}e \,
 -6\,d{e}^{2} +3\, a  b+3\, a c\, -4\, a d
\nonumber \\
\hspace{-0.97in}&& \quad  
 \, -4\, a  e \, +3\,bc \, -4\,bd \, -4\,be-4\,cd-4\,ce+
21\,de \, +a+b+c \, -6\,d-6\,e+3.
\end{eqnarray}
The condition (\ref{compat}) can be seen as a an equality on 
a one-form and the same one-form where $\, x$ 
has been changed into:
\begin{eqnarray}
\label{compatQ1}
\hspace{-0.95in}&& 
 \quad \quad \quad \quad 
Q(x)^{1/3} \cdot \, {{  dx} \over { x \cdot \, (1\, -x) }} 
\, \, = \, \, \,  {{ dx} \over { u}} 
\, \, = \, \, \,  \, 
Q(y)^{1/3} \cdot \, {{  dy} \over { y \cdot \, (1\, -y) }}. 
\end{eqnarray}
This one-form is clearly associated with the algebraic curve:
\begin{eqnarray}
\label{curveAA}
\hspace{-0.95in}&& \quad \quad \quad \quad \quad \quad \quad \quad 
Q(x)\cdot u^3 \, \, =  \, \,\, \,  x \cdot \, (1\, -x). 
\end{eqnarray}
One actually finds that this algebraic curve (\ref{curveAA})
is a {\em  genus-one curve}.

One can go a step further by eliminating all
the derivatives $\, y'(x)$,$\, y''(x)$,$\, y'''(x)$, 
from the confrontation of the Schwarzian condition
 (\ref{condition3F2}) with the compatibility condition 
(\ref{compat}). One gets that way (after some calculation) 
a condition reading
\begin{eqnarray}
\label{curveone}
\hspace{-0.95in}&& \quad \quad \quad \quad 
{\cal I}(x) \, = \, \, \, {\cal I}(y(x))
 \quad \quad  \quad \hbox{where:} 
\quad \quad \quad \quad  {\cal I}(x)
 \, = \, \, \, {{Q(x)^8 } \over { P_8(x)^3 }}, 
\end{eqnarray}
where $\, P_8(x)$ is a (quite large) polynomial 
of degree $\, 8$ in $\, x$, sum of 4724 terms. 

We are seeking for non-trivial pullbacks $\, y(x)$ being 
different from  the obvious solution $\, y(x) \, = \, \, x$.
The interesting cases for physics are the one where 
$\, x \, \, \rightarrow \, \, y(x)$ is an infinite order
transformation. In such cases one has 
\begin{eqnarray}
\label{curvetwo}
\hspace{-0.95in}&& \quad \quad \quad \quad 
{\cal I}(x) \, = \, \, \, {\cal I}(y(x))\, = \, \, \,{\cal I}(y(y(x))) 
\, = \, \, \,{\cal I}(y(y(y(x)))) \,  \, = \, \, \, \cdots 
\end{eqnarray}
which amounts to saying that $\, {\cal I}(x)$ must be a constant.
The cases where $\, Q(x)^8 \, = \, \, \, \lambda \cdot \, P_8(x)^3$
correspond to a set of extremely large conditions on the 
parameters $\, a, \, b, \, c, \, d, \, e$ of the $\, _3F_2$ 
hypergeometric function, that is difficult to study
because of the size of polynomial $\, P_8(x)$.
However a simple case can fortunately be analyzed, namely 
$ \, {\cal I}(x) \, = \, \, \, 0$, which corresponds to 
$\, Q(x)  \, = \, \, \, 0$. 
In such a case the two conditions are compatible, and one just has one 
condition:  the Schwarzian condition (\ref{condition3F2}) with
the additional condition being automatically verified (see (\ref{compat})).

One finds that all the 
conditions on the parameters $\, a, \, b, \, c, \, d, \, e$ 
of the $\, _3F_2$ hypergeometric function associated with 
$\, Q(x)  \, = \,  0\, $
in fact correspond to cases where the order-three operator 
is exactly the symmetric 
power of a second order operator have $\, _2F_1$ solutions.
In other words this situation corresponds to
 the {\em Clausen identity}, the $\, _3F_2$ hypergeometric function
reducing to the square of a $\, _2F_1$ hypergeometric function:
\begin{eqnarray}
\label{Clausen}
\hspace{-0.95in}&& \quad \quad \quad \quad \quad  
 _3F_2\Bigl([2\,a, \, a\, +b, \, 2\, b], \, 
[a\, +b\, +{{1} \over {2}}, \, 2\,a \, +2 \, b], \, x  \Bigr) 
\nonumber \\
\hspace{-0.95in}&& \quad \quad \quad \quad  
\quad \quad \quad \quad \quad  \, \, \, 
\, = \, \, \, 
 _2F_1\Bigl([a, \, b], \, [a\, +b\, +{{1} \over {2}}], \, x  \Bigr)^2.
\end{eqnarray}
In this Clausen identity case, we found that the Schwarzian 
condition (\ref{condition3F2}) for $\, _3F_2$ is nothing but the 
Schwarzian condition for the underlying $\, _2F_1$. 

If one considers the other case $\,  P_8(x)\, = \,\, 0$ the
vanishing condition of the $\, x^8$ coefficient
and the
vanishing condition of the constant coefficient in $\, x$ read
respectively: 
\begin{eqnarray}
\label{C8}
\hspace{-0.95in}&&  \quad   
 ({a}^{2}-ab-ac+{b}^{2}-bc+{c}^{2})
 \cdot \, (a \, +c -2\,b)^{2} \cdot \, (a+b\, -2\,c)^{2} 
\cdot \,  (2\,a \, -c-b)^{2} \, \, = \, \, \, 0,
\nonumber \\ 
\label{C0}
\hspace{-0.95in}&& \quad   
({d}^{2}-de+{e}^{2}-d-e+1)  \cdot \, 
(1+d \,-2\,e)^{2} \cdot \,  \, (d +e \, -2)^{2} \cdot \,  (2\,d -e-1)^{2}
 \, \, = \, \, \, 0.
\nonumber \
\end{eqnarray}
These two conditions are, respectively, very similar to the 
vanishing condition of the $\, x^3$ and constant coefficient 
of $\, Q(x)$, the other coefficients of $\,  P_8(x)$ being
more involved. The vanishing condition of all the 
$\, x^n$ coefficients of $\, P_8(x)$ yields more relations on the 
$\, a, \, b, \, c, \, d, \, d, \, e$ parameters. All these 
miscellaneous cases correspond to cases where the order-three
linear differential operator reduces to the symmetric square 
of an order-two operator, and to the Clausen identities 
of the form (\ref{Clausen}). More simply on can verify that 
for parameters such that $\, Q(x) \, = \, \, 0$ 
(for which a Clausen reduction take place  (\ref{Clausen}))
are also such that  $\,  P_8(x)\, = \,\, 0$ 
(the invariant $\, {\cal I}(x)$
in (\ref{curveone}) is thus of the form $\, 0/0$).

\vskip .1cm
 
\section{Schwarzian condition and other generalised hypergeometric functions} 
\label{Schwarz4F32F2}

\vskip .1cm
 
\subsection{Schwarzian condition and $\, _4F_3$ hypergeometric functions} 
\label{Schwarz4F3}

\vskip .1cm
 
Let us consider a $\, _4F_3$ hypergeometric known to correspond 
to a Calabi-Yau ODE~\cite{IsingCalabi,IsingCalabi2} and seek
 an identity of the form:  
\begin{eqnarray}
\label{modularform4F3}
\hspace{-0.95in}&& \quad \quad  \quad \quad    \quad \quad  
  {\cal A}(x) \cdot \, _4F_3\Bigl(
[{{1} \over {2}}, \, {{1} \over {2}},  \,  {{1} \over {2}},  \,  {{1} \over {2}}], 
\, [1, \, 1, \, 1], \,  x  \Bigr)
\nonumber \\
\hspace{-0.95in}&& \quad \quad  
\quad \quad   \quad \quad  \quad \quad \quad \quad 
\, = \, \, \, \,\,  
 _4F_3\Bigl(
[{{1} \over {2}}, \, {{1} \over {2}},  \,  {{1} \over {2}},  \,  {{1} \over {2}}], 
\, [1, \, 1, \, 1], \,  y(x)  \Bigr)
\end{eqnarray}
where $\, {\cal A}(x)$ is an algebraic function.
Again we introduce the order-four linear differential operator 
annihilating the LHS and RHS of identity (\ref{modularform4F3}).
The equality of the wronskians of these two linear differential 
operators enables us to get the expression of $\, {\cal A}(x)$ 
in terms of $\, y(x)$, namely: 
\begin{eqnarray}
\label{A4F3}
\hspace{-0.95in}&& \quad \quad  \quad \quad \quad \quad  \quad 
 {\cal A}(x) \, \, = \, \, \, 
\Bigl( {{ (1-y(x))\cdot \, y(x)^3 } \over { 
(1\, -x) \cdot \, x^3  \cdot \, y'(x)^3}}   \Bigr)^{1/2}. 
\end{eqnarray}
After eliminating $\, {\cal A}(x)$ from (\ref{A4F3}), the identification
of the $\, D_x^2$ coefficients for these two linear differential operators 
of order four gives the Schwarzian condition
\begin{eqnarray}
\label{condition4F3}
\hspace{-0.95in}&& \quad   \quad \quad  \quad \quad  \quad 
 W(x) 
\, \, \,  \,-W(y(x)) \cdot  \, y'(x)^2
\, \, \,  \,+ \,  \{ y(x), \, x\} 
\, \,\, \, = \,\, \, \,  \, 0, 
\end{eqnarray}
where $\, W(x)$ reads:
\begin{eqnarray}
\label{condition4F3W}
\hspace{-0.95in}&& \quad    \quad \quad  \quad \quad  \quad 
W(x)\, \, = \, \, \, 
-\, {{1} \over { 10}} \cdot \,
{\frac {5\,{x}^{2} \, -7\,x\, +5}{{x}^{2} \cdot \, (1\, - x)^{2}}}.
\end{eqnarray}
The condition corresponding to the identification
of the $\, D_x$ coefficient can be seen to be compatible with the previous 
Schwarzian condition: it can be seen to be a consequence of  
condition (\ref{condition4F3}), corresponding to
a combination of (\ref{condition4F3})
with the derivative of (\ref{condition4F3}). One 
finds, unfortunately, that the only solution is the trivial solution
$\, y(x) \, = \, \, x$, the other solutions being spurious solutions
$\, y \, +5 \, = \, 0$, $\, 5\,y \, +1 \, = \, 0$, 
$\, \,y \, -1 \, = \, 0$, etc ...

\vskip .1cm

\vskip .1cm

{\bf Remark:} Similarly to what has been performed 
in section (\ref{Schwarz4F3}) one can imagine 
to  seek for an identity (\ref{modularform4F3}) but, now, 
for the general $\, _4F_3$ hypergeometric function 
$\, _4F_3([a,\, b, \, c, \, d], \, [e, \, f, \, g], \, x)$. These 
calculations  are really too large. 

\vskip .1cm

\subsection{Schwarzian condition and hypergeometric functions with irregular singularities} 
\label{Schwarz2F2}

\vskip .1cm

The $\, n$-fold integrals emerging in
lattice statistical mechanics or enumerative combinatorics 
are naturally diagonal of rational functions~\cite{Christol},
the corresponding linear differential operators being 
globally nilpotent, and in particular Fuchsian. In such a lattice 
framework only $\, _nF_{n-1}$ hypergeometric functions~\cite{Heckman} 
with {\em regular} singularities occur. Of course {\em irregular} 
singularities can also occur in physics~\cite{penson1,penson2,penson3},
in particular in the {\em scaling limit} of lattice models~\cite{scaling,Holo} 
(modified Bessel functions, etc ...).

Let us consider a hypergeometric function with an {\em irregular} 
singularity, namely
a simple $\, _2F_2$ hypergeometric function
 solution of an order-three linear differential operator.
We seek an identity of the form:  
\begin{eqnarray}
\label{modularform4F3bis}
\hspace{-0.95in}&& \quad  \quad  \quad  \quad 
  {\cal A}(x) \cdot \, _2F_2\Bigl(
[{{1} \over {2}}, \, {{1} \over {2}}], 
\, [1, \, 1], \,  x  \Bigr)
\,\,\, = \, \, \, \,\,  
 _2F_2\Bigl(
[{{1} \over {2}}, \, {{1} \over {2}}], 
\, [1, \, 1], \,  y(x)  \Bigr).
\end{eqnarray}
The calculations are the same as in section (\ref{Schwarz3F2}). The identification
of the wronskians of the two operators (the pullbacked  order-three linear 
differential operator and the conjugated one) gives
\begin{eqnarray}
\label{A2F2}
\hspace{-0.95in}&& \quad \quad  \quad \quad \quad \quad  \quad 
 {\cal A}(x) \, \, = \, \, \, \exp\Bigl( {{x\, -y(x)} \over {3}} \Bigr) \cdot \, 
 {{ y(x) } \over { 
 \, x  \cdot \, y'(x)}}, 
\end{eqnarray}
and the Schwarzian equation: 
\begin{eqnarray}
\label{condition2F2bis}
\hspace{-0.95in}&&   
 W(x) 
\,  \,-W(y(x)) \cdot  \, y'(x)^2
\,   \,+ \,  \{ y(x), \, x\} 
\, \, = \,\, \,  0, \quad \hbox{with:}\quad 
 W(x)\, = \,\,  {{1} \over { 6}} \cdot \, {{x^2 \, -3 } \over {x^2}}.
\end{eqnarray}
However, combining equation (\ref{condition2F2bis}) with the last
condition emerging from the identification of the terms with no $\, D_x$ 
in the two operators, one finds that there is no pullback $\, y(x)$ 
for (\ref{modularform4F3bis}) except the trivial solution 
$\, y(x) \, = \, x$.

\vskip .2cm

\end{document}